\PassOptionsToPackage{square,comma,numbers,sort&compress,super}{natbib}
\documentclass[reprint,superscriptaddress,amsmath,amssymb,aps,longbibliography,nofootinbib]{revtex4-1}
\usepackage[version=3]{mhchem}
\usepackage[english]{babel}
\usepackage{placeins}

\usepackage{amssymb,amsmath,amsthm,mathrsfs,mathtools}
\usepackage{comment,afterpage,accents,subcaption,physics,booktabs,gensymb,graphicx,epstopdf,float,xcolor,multirow}
\usepackage{dcolumn}
\usepackage{bm}
\usepackage[colorlinks]{hyperref}
\hypersetup{colorlinks=blue,
            breaklinks,
            urlcolor=teal,
            linkcolor=red,
            citecolor=blue}
\usepackage[normalem]{ulem}

\newcommand{\matr}[1]{\mathbf{#1}}
\newcommand{\opr}[1]{\hat{#1}}

\newcommand{\eq}[1]{Eq.~\hyperref[eq:#1]{(\ref*{eq:#1})}}
\renewcommand{\sec}[1]{\hyperref[sec:#1]{Section~\ref*{sec:#1}}}
\newcommand{\app}[1]{\hyperref[app:#1]{Appendix~\ref*{app:#1}}}
\newcommand{\tab}[1]{\hyperref[tab:#1]{Table~\ref*{tab:#1}}}
\newcommand{\fig}[1]{\hyperref[fig:#1]{Figure~\ref*{fig:#1}}}
\newcommand{\figa}[2]{\hyperref[fig:#1]{Figure~\ref*{fig:#1}#2}}
\newcommand{\figx}[2]{\hyperref[fig:#1]{Figure~\ref*{fig:#1}(#2)}}
\newcommand{\thm}[1]{\hyperref[thm:#1]{Theorem~\ref*{thm:#1}}}
\newcommand{\lem}[1]{\hyperref[lem:#1]{Lemma~\ref*{lem:#1}}}
\newcommand{\cor}[1]{\hyperref[cor:#1]{Corollary~\ref*{cor:#1}}}
\newcommand{\defn}[1]{\hyperref[def:#1]{Definition~\ref*{def:#1}}}
\newcommand{\alg}[1]{\hyperref[alg:#1]{Algorithm~\ref*{alg:#1}}}

\newcommand{\twori}[2]{\left\langle#1 || #2\right\rangle}

\definecolor{CadetBlue}{RGB}{95, 158, 160}

\begin{document}

\title{Say NO to Optimization: A Non-Orthogonal Quantum Eigensolver}

\author{Unpil Baek}
\email{baek.un@berkeley.edu}
\affiliation{Department of Physics, University of California, Berkeley}
\affiliation{Berkeley Quantum Information and Computation Center, University of California, Berkeley, CA 94720, USA}
\affiliation{Challenge Institute for Quantum Computation, University of California, Berkeley, CA  94720}
\author{Diptarka Hait}
\affiliation{Department of Chemistry, University of California, Berkeley}
\affiliation{%
 Chemical Sciences Division, Lawrence Berkeley National Laboratory, Berkeley, CA, 94720, USA
}
\author{James Shee}
\affiliation{Department of Chemistry, University of California, Berkeley}
\author{Oskar Leimkuhler}
\affiliation{Berkeley Quantum Information and Computation Center, University of California, Berkeley, CA 94720, USA}
\affiliation{Department of Chemistry, University of California, Berkeley}
\affiliation{Challenge Institute for Quantum Computation, University of California, Berkeley, CA  94720}
\author{William J. Huggins}
\thanks{current address: Google Quantum AI, San Francisco, CA 90291, United States}
\affiliation{Department of Chemistry, University of California, Berkeley}
\author{Torin F. Stetina}
\affiliation{Simons Institute for the Theory of Computing, Berkeley, CA, 94704}
\affiliation{Berkeley Quantum Information and Computation Center, University of California, Berkeley, CA 94720, USA}
\affiliation{Challenge Institute for Quantum Computation, University of California, Berkeley, CA  94720}
\author{Martin Head-Gordon}
\affiliation{Department of Chemistry, University of California, Berkeley}
\affiliation{%
 Chemical Sciences Division, Lawrence Berkeley National Laboratory, Berkeley, CA, 94720, USA
}
\affiliation{Challenge Institute for Quantum Computation, University of California, Berkeley, CA  94720}
\author{K. Birgitta Whaley}
\email{whaley@berkeley.edu}
\affiliation{Department of Chemistry, University of California, Berkeley}
\email{whaley@berkeley.edu}
\affiliation{Berkeley Quantum Information and Computation Center, University of California, Berkeley, CA 94720, USA}
\affiliation{Challenge Institute for Quantum Computation, University of California, Berkeley, CA  94720}

\date{\today}

\begin{abstract}
A balanced description of both static and dynamic correlations in electronic systems with nearly degenerate low-lying states presents a challenge for multi-configurational methods on classical computers. We present here a quantum algorithm utilizing the action of correlating cluster operators to provide high-quality wavefunction ans{\"a}tze employing a non-orthogonal multireference basis that captures a significant portion of the exact wavefunction in a highly compact manner, and that allows computation of the resulting energies and wavefunctions at polynomial cost with a quantum computer. This enables a significant improvement over the corresponding classical non-orthogonal solver, which incurs an exponential cost when evaluating off-diagonal matrix elements between the ansatz states, and is therefore intractable. We implement the non-orthogonal quantum eigensolver (NOQE) here with an efficient ansatz parameterization inspired by classical quantum chemistry methods that succeed in capturing significant amounts of electronic correlation accurately. By taking advantage of classical methods for chemistry, NOQE provides a flexible, compact, and rigorous description of both static and dynamic electronic correlation, making it an attractive method for the calculation of electronic states of a wide range of molecular systems.
\end{abstract}

\maketitle

\section{Introduction}\label{sec:intro}

The computation of electronic states and energies for molecular and extended systems, known as ``the electronic structure problem", has emerged as one of the most prominent practical problems for which quantum computers might show an advantage over their classical counterparts. The worst-case quantum complexity of the electronic structure problem is known to be QMA-Complete~\cite{o2021electronic}, i.e., the quantum analog of NP-hard. Although this implies that solving the general electronic structure problem to arbitrary accuracy may be intractable for quantum computers, a significant advantage over existing classical methods for systems of interest may still be achievable. 

Focusing on molecular electronic systems, often referred to as ``quantum chemistry", there are several facts that make the search for quantum algorithms in this area both interesting and suggestive of promise for demonstrating a quantum advantage. First, the level of accuracy required to make quantitative chemical predictions is 1 kcal/mol ($\sim 1.6$ mHa = 43 meV) or less (which corresponds to the reported uncertainties of typical experimental measurements for thermochemistry). This quantity is referred to as ``chemical accuracy''~\cite{boys1969calculation}. Therefore, achieving arbitrary accuracy is not necessary for this high-impact domain-specific application of quantum algorithms. Second, energy splittings and relative orderings, which involve low-lying excited states in addition to the ground state, are of paramount importance in many problems of interest. This is especially true in so-called ``strongly correlated'' electronic systems~\cite{small2011post,shee2021revealing}, wherein multiple spin-states or phases are separated by small energy gaps. Such states can only be described by multi-configurational wavefunctions composed of many Slater determinants, and chemically-accurate \emph{ab initio} predictions remain a challenge for existing classical computational methods. 

A proper description of strong electronic correlation is required to describe a wide range of interesting physical phenomena, ranging from materials such as cuprates which can exhibit long-range order and high-temperature superconductivity~\cite{imada1998metal}, to bond-breaking chemical reactions and the intricate electronic processes found in many biological and synthetic catalysts that contain magnetically-coupled transition metals~\cite{witzke2020bimetallic} or f-block atoms~\cite{gould2022ultrahard}. One very well-known system with multiple spin states is the oxygen-evolving complex (OEC) of Photosystem II in green plants, which plays a critical role in the oxidation or ``splitting" of water molecules to generate protons and free molecular oxygen~\cite{askerka2017o2}.  This complex has four manganese atoms in different oxidation states~\cite{umena2011crystal}, which are involved in a complex series of electron transfer reactions that catalyze the oxidation of water~\cite{raymond2008origin}. Understanding the mechanism of this critical step in photosynthesis of green plants requires that the spin states involving the transition metal atoms and the electronic states of coordinated reactants be well characterized, which presents a major challenge for quantum chemistry today. Efforts to perform electronic structure calculations with quantum computers started with an application of the quantum phase estimation (QPE) algorithm~\cite{aspuru2005simulated} that requires fault-tolerant quantum hardware, and later shifted to the variational quantum eigensolver (VQE)~\cite{peruzzo2014variational,mcclean2016theory}. VQE is a hybrid quantum-classical algorithm, in which the quantum computer is used to generate a wavefunction ansatz with a parametrized quantum circuit, the expected value of the Hamiltonian terms are measured with this circuit, and the variational parameters are then optimized by a classical solver in an iterative manner~\cite{mcclean2016theory}. By avoiding the large circuit depth necessary for QPE, VQE is suitable for running on Noisy Intermediate-Scale Quantum (NISQ) hardware. 

Considerable efforts have gone into developing a broad range of different quantum circuit ans{\"a}tze that approximate the ground states of Hamiltonians. The Unitary Coupled-Cluster (UCC) ansatz, a unitary generalization of the coupled-cluster ansatz used in quantum chemistry, has been popular due to its variational and size-consistent nature~\cite{mcclean2016theory,peruzzo2014variational,taube2006new}. However, the generic UCC ansatz requires a circuit depth that increases as higher-order terms in the coupled-cluster expansion are included, rendering hybrid calculations on near-term machines  very sensitive to noise and limited by current constraints on qubit coherence.  Furthermore, the classical optimization loop often requires many circuit evaluations and is also susceptible to the ``barren plateau'' phenomenon in high dimensions~\cite{mcclean2018barren,uvarov2020variational,arrasmith2021effect}. Hence there are two generic limitations of VQE: (i) the qubits must have coherence times long enough to generate ansatz states with complex structures, and (ii) the required number of measurements should not exceed the wall-clock time available for the near-term devices with consistent calibration. 

An extension of VQE was recently proposed by some of the authors to allow greater expressivity of the final wavefunction without increasing the circuit depth~\cite{huggins2020non}. This is the non-orthogonal VQE (termed NOVQE) that uses the quantum computer to generate a set of non-orthogonal ansatz states, on which the Hamiltonian and overlap matrices are measured using a modified Hadamard test. The resulting generalized eigenvalue problem is solved classically to provide an estimate of the ground-state energy, which is then optimized with respect to the circuit parameters for the set of non-orthogonal ansatz states. Results from NOVQE in Ref.~\cite{huggins2020non} demonstrated a systematic increase in wavefunction complexity and a greater fidelity with the true ground state, relative to that obtained with VQE using a single reference state ansatz. But this comes at the cost of a greater number of measurements. Therefore, NOVQE presents a tradeoff between decreased requirements on qubit coherence time for an increased number of circuit repetitions, and consequently also an increase in the measurement cost. 

To break out of this inevitable tradeoff for variational hybrid quantum algorithms, we present here a novel quantum algorithmic approach to electronic structure calculations that does not use variational optimization. The non-orthogonal quantum eigensolver proposed in the current work, which we refer to as NOQE, takes advantage of domain-specific knowledge available in classical quantum chemistry to construct high-quality wavefunctions at a low cost. Specifically, our protocol uses spin-unrestricted methods that optimize spin-symmetry-broken Slater determinants (i.e., with different spatial distributions for the up and down spins) to yield more accurate energies and electronic densities in the strongly correlated limit. Explicit diagonalization within the subspace spanned by all qualitatively relevant unrestricted single determinants, known in the classical quantum chemical literature as non-orthogonal configuration interaction or NOCI, offers a straightforward way to approximate spin-purification of unrestricted solutions~\cite{broer1981broken,thom2009hartree,sundstrom2014non}. However, while strong correlations are reasonably well addressed by the classical NOCI approach~\cite{sundstrom2014non,yost2016size}, taken alone this protocol cannot recover the weak (dynamical) correlation present in the system. In the current work, we show that the effects of dynamical correlation can be introduced via the application of UCC-like operators to the subspace states with perturbative parameters. This ``perturb and then diagonalize" approach confers the advantage of computing at once both the ground and low-lying excited states of the system through subspace diagonalization, with approximately correct spin quantum numbers. Consequently, the energy gaps between the different low-lying spin states of a strongly correlated electronic system can be easily computed. The application of UCC-like operators can be efficiently implemented on quantum computers. Furthermore, by making a subspace diagonalization on the resulting states, NOQE also avoids the long time evolution (and consequent deep circuits) required by other optimization-free hybrid quantum algorithms~\cite{Parrish2019-dv, Kyriienko2020-yc, Stair2020-co, klymko2022real, Cohn2021-mq}. 

The quantum computational advantage of the NOQE approach comes from the efficient calculation of off-diagonal matrix elements on a quantum computer. Classical evaluation of a single NOCI matrix off-diagonal element has only $\mathcal{O}(N^2)$ asymptotic cost with the basis size $N$. However, the inclusion of dynamic correlation with CC operators (perturbative or otherwise) results in each off-diagonal element having a classical cost that scales $\mathcal{O}(\exp (N))$. On the other hand, evaluation of these matrix elements via the modified Hadamard test of Ref.~\cite{huggins2020non} scales with $\mathcal{O}(\text{poly}(N))$. In principle, this allows for an exponential separation between quantum and classical non-orthogonal quantum chemistry calculations. In this work, we explore the quality of the NOQE approach using both a UCC-derived ansatz and a cluster Jastrow modification thereof. We make a resource estimate for NOQE with these ans{\"a}tze and give a prognosis for the extent to which classically-intractable NOCI calculations including dynamic correlation may be efficiently realized by our quantum NOQE approach.

The remainder of the paper is structured as follows. \sec{background} discusses the background of current classical quantum chemistry methods available for electronic structure calculations for systems with strong correlation. \sec{theory} introduces the NOQE protocol after presenting an essential preliminary analysis of the modified UCC wavefunction ans{\"a}tze used in this work. \sec{results} presents numerical results for two model systems, the potential energy surface of molecular hydrogen H$_2$ and the square H$_4$ system which is a well-known benchmark for calibration of electronic structure algorithms for strongly correlated systems. In \sec{circuit-resource-estimation}, we analyze the scaling of quantum circuit resources needed to run the NOQE routine on a quantum computer for molecular systems. Finally, in \sec{discussion-and-outlook}, we summarize and provide an outlook for both theoretical and experimental developments suggested by this work. 

\section{Background}\label{sec:background}

Quantum chemistry aims to develop computationally affordable approximations for modeling the quantum many-body problem. The simplest quantum chemistry method is the Hartree-Fock (HF) approximation~\cite{szabo2012modern}, which provides a mean-field description of the many-electron ground state wavefunction by variationally optimizing a single Slater determinant (an antisymmetrized product of single-particle orbitals~\cite{szabo2012modern}). The computational cost of Hartree-Fock calculations scales as $\mathcal{O}(N^4)$. While global minimization of the Hartree-Fock procedure is formally an NP-complete problem~\cite{whitfield2013computational}, in most cases the physically motivated heuristics involved in Hartree-Fock calculations nevertheless allow for efficient convergence to local minima. The energy not captured by this independent particle approximation is typically referred to as the electron correlation energy, which is often broadly categorized into ``weak/dynamic” and ``strong/static” (although no rigorous boundary exists between the two limits).

Nevertheless, a distinction between weak and strong correlations can be made in terms of limiting behavior. The Hartree-Fock approximation works best when the true wavefunction is well-approximated by a single Slater determinant, wherein the orbitals are filled in ascending order of energy and are doubly occupied (with one spin-up electron and one spin-down electron) when possible. However, the strong correlation between electrons makes it energetically unfavorable to maintain electron pairs, leading to a (partial) separation of paired electrons into distinct orbitals occupying different spatial regions.  Strong correlation, therefore, entails a significant level of electron pair breaking, which requires several Slater determinants to have significant weights in the wavefunction. On the other hand, weak electron correlation does not require electron unpairing and largely stems from small but significant contributions of many electronic configurations with individually small amplitudes in the overall wavefunction. Traditional quantum chemistry has been quite successful at modeling weak correlation for systems where the Hartree-Fock approximation is qualitatively valid (`single-reference'), through perturbative~\cite{cremer2011moller} and projected coupled-cluster~\cite{bartlett2007coupled} approaches that have only polynomial scaling with $N$. Systems with strong correlation are more challenging and generally require solving the exact diagonalization or full CI (FCI) problem within some subspace of orbitals~\cite{helgaker2014molecular}. This has an exponential cost, and  exact treatment is only feasible for small subspaces up to 24 spatial orbitals~\cite{vogiatzis2017pushing} at present (selected CI solvers permit substantially larger orbital spaces with soft exponential scaling~\cite{holmes2016heat,levine2020casscf,tubman2020modern}). In addition, weak correlation outside the subspace is essential for chemical accuracy~\cite{andersson1990second,andersson1992second,angeli2001introduction}, and considerably adds to computational cost.

One possible route to describing strong correlations starting from mean-field theory is via spin-symmetry breaking. It is possible to converge Slater determinants where the up and down spin orbitals have different spatial components, and can thus accommodate a level of electron unpairing. The downside of such spin-unrestricted Hartree-Fock calculations is that the resulting determinant is no longer a pure spin-eigenstate, and therefore cannot be used to estimate the spin-spectrum. Restoration of spin-symmetry is also quite challenging for single-reference wavefunction methods, often leading to suboptimal performance of such approaches utilizing an unrestricted reference~\cite{gill1988does}. However, unrestricted determinants have much better energies and total electron densities~\cite{szabo2012modern}, making them attractive candidates for modeling strongly correlated systems if spin-purity can be realized. 

Classically, the non-orthogonal configuration interaction (NOCI) approach offers a reasonably straightforward route to an approximate spin-purification of such unrestricted solutions~\cite{sundstrom2014non}. Explicit diagonalization within the subspace spanned by all unrestricted configurations corresponding to different permutations of all (partially or completely) unpaired spins should therefore lead to pure spin states, via indirectly coupling the angular momenta. In practice, the resulting states are not always quite spin-pure, due to differences in orbitals corresponding to stationary states for each permuted configuration. However they are much better approximations to spin-eigenstates than are the untreated unrestricted single determinants, and furthermore, they also permit computation of spin-gap energies. 

In practice, NOCI with single determinants is not very accurate, because the CI protocol only recovers the strong correlation between the unpaired electrons~\cite{sundstrom2014non,yost2016size,mayhall2014spin}. Weak correlation can be included via a `perturb and then diagonalize' approach, in which NOCI is performed within a subspace of unrestricted wavefunctions that already include a measure of dynamic correlation. NOCI using unitary coupled-cluster (UCC) wavefunctions, where the UCC states are constructed from unrestricted Hartree-Fock solutions falls into the latter category, and this is the route that we shall explore in this work. However, for classical computers, the computational demands of UCC amplitude optimization are considerable, particularly for non-orthogonal problems. It is important to note that the classical cost of computing the UCC operator for a single reference determinant scales exponentially with the number of spin-orbitals. For this reason, the projected coupled-cluster approaches~\cite{bartlett2007coupled} widely used in classical quantum chemistry do not evaluate energies as the expectation value of the Hamiltonian over the full coupled-cluster wavefunction, since that task also has an exponential classical cost. Instead, a formally non-variational single reference energy is defined via projection equations that can be solved at an increasing high-scaling polynomial cost as more correlations are included in the CC operator~\cite{shavitt2009many}. For example, the commonly used CCSD approach (single and double excitations from a Hartree-Fock reference state) scales as $\mathcal{O}(N^6)$, and the CCSDT approach (single, double, and triple excitations) scales as $\mathcal{O}(N^8)$, and so on (so-called local correlation approximations can reduce this scaling at the expense of some numerical error~\cite{schutz1999low,riplinger2016sparse,subotnik2006near}). However, this classical approach does not readily generalize to off-diagonal matrix elements between multiple coupled-cluster states. Indeed, any use of coupled-cluster wavefunctions for direct evaluation of off-diagonal elements would incur an exponential cost for even orthogonal reference states. The situation is even more challenging for the non-orthogonal case because the Slater-Condon rules~\cite{szabo2012modern} cannot be applied to simplify many terms to zero, incurring an even higher cost. Therefore, the advantage of using a quantum processor in NOQE is not only due to the simple implementation of the UCC operator within a quantum algorithm for generating the reference states of a NOCI problem, but also because the standard projected coupled-cluster theory is computationally intractable for the NOCI problem on classical computers. This provides strong motivation for the introduction of NOQE in this work as a quantum algorithm yielding a substantial quantum advantage relative to NOCI.

In this work, we therefore explore the use of amplitudes derived from the perturbative analysis in NOQE calculations without further optimization. This is expected to be a reasonable assumption because the amplitudes from second-order many-body perturbation theory (known as MP2) are first-order approximations to actual UCC amplitudes, as is shown explicitly in \sec{firstorder}. We denote this version of NOQE using UCC ans{\"a}tze together with MP2 amplitudes, as NOUCC(2). We note that NOUCC(2) ground state energies are variational in the sense that they are bounded from below by the exact FCI energy which provides a lower bound on diagonalization of the Hamiltonian within an NO subspace of wavefunctions. As an alternative, we also examine the benefits of adding a Cluster Jastrow correlator to a UCC ansatz, i.e., using the UCJ ansatz of~\cite{matsuzawa2020jastrow}. We denote the version of NOQE using this ansatz as NOUCJ. 

\section{Theory}\label{sec:theory}

We first define the notation employed in this work and then provide in \sec{prelim} some preliminary analysis for the construction of the NOUCC(2) ansatz reference states, as well as the alternative NOUCJ ansatz based on a Cluster Jastrow decomposition.  The general NOQE approach is then presented in \sec{noqe_circuit}, together with details of the quantum circuit implementation.

\subsection*{Notation}

We define $N$ as the number of spin-orbitals and make use of the Jordan-Wigner transformation to map electronic states (determinantal wavefunctions) constructed from these to a qubit representation~\cite{mcardle2020quantum}. Then $N$ is the number of qubits required to represent the quantum state. We shall use $\eta$ to represent the number of electrons and $r$ to denote the number of radical sites involved in a specific molecular calculation (\sec{circuit-resource-estimation}). All references to orbitals in this work are to molecular orbitals (MOs) unless specified otherwise; the MOs are assumed to be real. The MOs are generally expressed as linear combinations of Gaussian basis functions in quantum chemistry, for ease of computation. Indices $\{i,j\}$ refer to occupied spin-orbitals, indices $\{a,b\}$ to  virtual spin-orbitals, and indices $\{p,q,r,s\ldots\}$ are employed to represent general spin-orbitals. Capitalized indices $\{I,J\}$ are used to index the reference states for the generalized eigenvalue equation of NOQE, with a total number of $M$ reference states (determinants).  We follow the chemistry convention that spin-up electrons are referred to as $\alpha$-electrons and spin-down electrons as $\beta$-electrons.

The two-electron repulsion integrals (ERIs) between MOs are abbreviated according to standard chemical notation as
\begin{align}
    \braket{ij}{ab}&=\displaystyle\int d\vec{r}\displaystyle\int d\vec{r}\,' \dfrac{\phi_{i}\left(\vec{r}\,\right)\phi_{j}\left(\vec{r}\,'\right)\phi_{a}\left(\vec{r}\,\right)\phi_{b}\left(\vec{r}\,'\right)}{|\vec{r}-\vec{r}\,'|}\\
\twori{ij}{ab}&=\braket{ij}{ab}-\braket{ij}{ba}
\end{align}

The unitary coupled-cluster doubles (UCCD) ansatz is defined, in this work, relative to a Hartree-Fock reference state $\ket{\Phi_\text{HF}}$ as $e^{\hat\tau} |\Phi_\text{HF}\rangle$, with
\begin{align}
    \opr{\tau} &\equiv \opr{T}-\opr{T}^\dag, \label{eq:ucc1_theory}\\
    \opr{T} &= \sum_{pqrs=1}^N t_{ps,qr} \opr{a}_p^\dag \opr{a}_q^\dag \opr{a}_r \opr{a}_s \\ &= \sum_{pqrs=1}^N t_{ps,qr} \opr{a}_p^\dag \opr{a}_s \opr{a}_q^\dag \opr{a}_r,
    \label{eq:ucc2_theory}
\end{align}
where the $N^2\times N^2$ supermatrix $\matr{T}$ is defined by its grouped two-index matrix elements
\begin{equation}
    t_{ps,qr} \equiv \begin{cases} 
    t_{pqrs} & p<q;s<r;r,s\in\text{occ};p,q\in\text{virt} \\
    0 & \text{otherwise}.
    \end{cases}
    \label{eq:t_pqrs_theory}
\end{equation}
In the above, the four-index $t_{pqrs}$ refer to the standard UCCD amplitudes. The primary focus of the present work is the use of MP2 amplitudes for $t_{pqrs}$ (generating the NOUCC(2) ansatz), which we shall denote as $t^\text{MP2}_{pqrs}$.  However, we shall also consider the use of Jastrow-correlated amplitudes (generating the NOUCJ ansatz), which we shall denote as $t^\text{J}_{pqrs}$.  We shall refer to the cluster operator with MP2 amplitudes as UCC-MP2 and to the cluster operator with Jastrow-correlated amplitudes as UCC-J.

\subsection{Preliminaries}\label{sec:prelim}

\subsubsection{MP2 amplitudes as first-order approximation to UCCD amplitudes}\label{sec:firstorder}

Here we derive a first-order approximation to the UCCD amplitudes $t_{abij}$ in terms of MP2 amplitudes obtained from perturbation theory. Suppose we have a UCC wavefunction given by:
\begin{align}
    \ket{\Psi(\{t_{abij}\})}&=e^{\opr{\tau}}\ket{\Phi_\text{HF}}
\end{align}
with $\hat\tau$ given by equations \eq{ucc2_theory} and \eq{t_pqrs_theory}. We approximate the amplitudes $\{t_{abij}\}$ via gradient descent in the space of $t$ amplitudes. For example, starting from the case where all $t_{abij}=0$, the gradient of the energy with respect to $t_{abij}$ is
\begin{align}
    \left(\dfrac{\partial E}{\partial t_{abij}}\right)_{\{t_{abij}\} =0}&=\bra{\Phi_\text{HF}}[\opr{H},\opr{a}_a^\dag \opr{a}_b^\dag \opr{a}_j  \opr{a}_i]\ket{\Phi_\text{HF}}\\
    &=\twori{ij}{ab},
\end{align}
and the diagonal elements of the Hessian are readily evaluated as
\begin{align}
    \left(\dfrac{\partial^2 E}{\partial t_{abij}^2}\right)_{\{t_{abij}\}=0}=&\bra{\Phi_\text{HF}} \left(\opr{a}_a^\dag \opr{a}_b^\dag \opr{a}_j \opr{a}_i\right)^\dag \opr{H} \opr{a}_a^\dag \opr{a}_b^\dag  \opr{a}_j \opr{a}_i\ket{\Phi_\text{HF}}\notag\\&-\bra{\Phi_\text{HF}} \opr{H} \ket{\Phi_\text{HF}}\label{eq:approxhess}\\
    \approx &\,\, \epsilon_a+\epsilon_b-\epsilon_j-\epsilon_i,
\end{align}
where $\epsilon_i$ denotes the single-particle energy of orbital $i$. A first-order Newton-Raphson approximation to $t_{abij}$ is then given by:
\begin{align}
    t_{abij}&=-\dfrac{\left(\dfrac{\partial E}{\partial t_{abij}}\right)_{\{t_{abij}\}=0}}{\left(\dfrac{\partial^2 E}{\partial t^2_{abij}}\right)_{\{t_{abij}\}=0}}\approx -\dfrac{\twori{ij}{ab}}{\epsilon_a+\epsilon_b-\epsilon_j-\epsilon_i},
\end{align}
which is seen to be identical to the $t$ amplitudes that would be obtained from second-order perturbation theory, i.e., from MP2. In fact, the energy of the wavefunction $|\Psi(\{t^\text{MP2}_{abij}\})\rangle$ is correct not only to the second order in MP perturbation theory, but to the third. We also note this protocol predicts zero single excitation amplitudes from unrestricted Hartree-Fock reference determinants, consistent with Brillouin's theorem~\cite{szabo2012modern}.

We note that using the exact form of \eq{approxhess} would lead to Epstein-Nesbet perturbation theory~\cite{epstein1926,nesbet1955} instead, which is known to be inferior to MP2 for single Slater determinants~\cite{murray1992different} and is therefore not explored here. We do however note that there are some other routes to improve upon using bare MP2 $\{t_{abij}\}$. Perhaps the simplest route is via scaling $t_{abij}$, in the spirit of a line-search. Scaling MP2 parameters has precedence in classical quantum  chemistry, with Ref.~\cite{grimme2003improved} proposed that empirically scaling the same spin amplitudes (i.e., all of $a,b,i,j$ have same spin) by 0.33 and opposite spin amplitudes (i.e., $i,j$ have different spin, and so do $a,b$) by 1.2 would lead to better results. Ref.~\cite{jung2004scaled} went even further and neglected same spin amplitudes entirely, scaling opposite spin amplitudes by 1.3. These spin-component-scaled (SCS) and scaled-opposite-spin (SOS) MP2 methods yield superior quantitative performance to normal MP2~\cite{grimme2003improved,jung2004scaled}, indicating that unscaled MP2 appears to underestimate opposite spin correlation. Furthermore, the very slow convergence of MP theory for spin-contaminated systems~\cite{gill1988does} indicates the possibility of unscaled MP2 amplitudes being too small in the spin-polarized limit. Scaling up the amplitudes thus has the potential to be more effective.  We will therefore explore whether SCS/SOS-style scalings affect NOUCC(2) results, as well investigate if uniformly scaling the amplitudes is also effective.

\subsubsection{Single-reference state generation: Quantum circuit ansatz based on the low-rank decomposition}\label{sec:singleref}

To simulate the action of the unitary cluster operator we employ the technique of low-rank tensor decomposition~\cite{motta2021low,rubin_compressing_2021}, which has been previously introduced and utilized in classical contexts~\cite{peng2017highly,motta2019efficient}. In this approach we decompose the rank-4 doubles cluster tensor $\hat{\tau}=\hat{T}-\hat{T}^\dag$, Eqs.~(\ref{eq:ucc1_theory})-(\ref{eq:ucc2_theory}), into a sum-of-squares of one-electron normal operators. This can be done using either a singular value decomposition (SVD) or a Takagi factorization~\cite{motta2021low,rubin_compressing_2021}, to yield
\begin{align}
    e^{\hat{\tau}} = \exp(-i\sum_{l=1}^L\sum_{\mu=1}^{m} {\hat{Y
    }_{l,\mu}}^2),
    \label{eq:sum_of_squares}
\end{align}
where the number of nonzero singular values (or Takagi diagonals) $L\leq N^2$, and $m=4$ if the decomposition is by SVD, or $m=2$ if the Takagi factorization has been used (this may increase $L$ by up to a factor of two, see \app{details-low-rank}). The (normal) operators $\hat{Y}_{l,\mu}$ are then further diagonalized and the resulting unitary is Trotterized to obtain the final low-rank form of the state preparation ansatz as
\begin{align}
    e^{\hat\tau} \approx \opr{\mathcal{U}}_B^{(1,1)\dag}\prod_{l=1}^L\prod_{\mu=1}^m \exp\left(-i \sum_{pq}^{\rho_l}\lambda_p^{(l,\mu)}\lambda_q^{(l,\mu)} \opr{n}_p\opr{n}_q\right) \tilde{\mathcal{U}}_B^{(l,\mu)}.
    \label{eq:lowrank_final}
\end{align}
Here $\lambda_p^{(l,\mu)}$ are eigenvalues of the $\opr{Y}_{l,\mu}$ operators, and the total number of nonzero eigenvalues is $\rho_l\leq N$. The unitary operators $\opr{\mathcal{U}}_B^{(1,1)\dag}$ are single basis rotations and $\tilde{\mathcal{U}}_B^{(l,\mu)}$ are sequences of neighboring basis rotations (see \app{details-low-rank}). The approximation in the above decomposition is entirely due to Trotter error. The effect of this error on the energy expectation values can be made arbitrarily small by increasing the order of the Trotter expansion (see \app{error-estimation}). 

The double decomposition of \eq{lowrank_final} results in a circuit structure of blocks of alternating unitary basis rotations (${\tilde{\mathcal{U}}}_B^{(l,\mu)}$) requiring up to $2 {N/2 \choose 2}$ nearest-neighbor Givens rotations~\cite{motta2021low,kivlichan2018quantum}, and sets of exponentiated number operator pairs that require at most $N \choose 2$ two-qubit CZ gates on a fully connected architecture, which may be applied in $N$ layers of parallel gates~\cite{motta2021low} (see \sec{circuit-resource-estimation} for further details of the resource estimation). We note that a singles excitation term can easily be included in the ansatz, requiring only an additional single basis rotation in front of the product in \eq{lowrank_final} and resulting in an additional cost of up to $2 {N/2 \choose 2}$ Givens rotations. 

The representation in \eq{lowrank_final} is advantageous for two reasons. Firstly the low-rank decomposition allows us to systematically truncate the rank of the MP2 $\hat\tau$-tensor by thresholding the singular values, reducing the number of circuit blocks and hence the overall circuit depth, while preserving the desired level of accuracy in the ansatz. Secondly, the Jordan-Wigner mapping from orbitals to qubits requires lengthy strings of $Z$ gates to encode fermionic anti-commutation relations in excitations ($\opr{a}^\dag_p\opr{a}_q$) between geometrically distant qubits $p$ and $q$, which results in multiple additional layers of CNOT gates per excitation term~\cite{whitfield2011simulation,hastings2015improving}. By rotating into bases where the fermionic excitations are represented by number operators ($\opr{n}_p=\opr{a}^\dag_p\opr{a}_p$), these Jordan-Wigner strings are entirely avoided, significantly reducing the circuit depth.

\subsubsection{Single-reference state generation: adding Cluster-Jastrow correlators}\label{sec:Jastrow}

In this work we also explore the addition of Jastrow correlations to the MP2 approximation to the UCCD ansatz wavefunctions, leading to the NOUCJ modification of NOQE.  We show below that incorporating Jastrow correlations allows for a more expressive but equally compact representation, suggesting that this approach is compatible with a lower truncation of the tensor decomposition and showing potential for implementation with shorter quantum circuits than the NOUCC(2) reference ansatz.

The idea of applying a two- or more-electron Jastrow correlator to a Slater determinant, i.e., $e^{\hat{J}} | \phi \rangle$, was first introduced to efficiently satisfy the well-known cusp condition~\cite{kato1957eigenfunctions} and forms the basis of classical quantum chemistry methods such as variational and diffusion Monte Carlo~\cite{hammond1994monte,reynolds1982fixed}. In such methods, the Jastrow term is typically represented in real space, although Neuscamman and others have recently developed promising approaches in which a Cluster Jastrow (CJ) correlator in orbital space appears in front of the antisymmetrized geminal power (pairing) ansatz~\cite{neuscamman2013communication,neuscamman2013jastrow,neuscamman2016improved}. A unitary variant of the CJ operator, denoted UCJ, was recently proposed in Ref.~\cite{matsuzawa2020jastrow}, where the cluster operator takes the form:
\begin{equation}
    e^{\hat{\tau}} = \prod_{l=0}^L e^{\hat{\mathcal{K}}_l} e^{\hat{\mathcal{J}}_l} e^{\hat{\mathcal{K}}_l}
    \label{eq:JastrowAnsatz}
\end{equation}
with $\hat{\mathcal{K}}_l= \sum_{pq} \mathcal{K}_{pq}^l \opr{a}^\dag_p \opr{a}_q$ and $\hat{J}_l = \sum_{pq} \mathcal{J}_{pq}^l \opr{n}_p \opr{n}_q$. 
The $\bm{\mathcal{K}}$ matrices are complex and anti-Hermitian, while the $\bm{\mathcal{J}}$ matrices are symmetric and purely imaginary.  As a result, the (generally complex) amplitudes of the UCCD operator can then be represented as: 
\begin{equation}
    t^{\text J}_{pqrs} = \sum_l\sum_{jk} U_{pj}^l U_{qj}^{l,*}\mathcal{J}_{jk}^l U_{rk}^l U_{sk}^{l,*}.
\end{equation}
As noted in Ref.~\cite{matsuzawa2020jastrow}, this form is identical to the double-SVD decomposition shown in \eq{lowrank_final} when $\lambda_i \lambda_j \rightarrow \mathcal{J}_{ij}$.  As such, the UCJ ansatz has the potential to be slightly more expressive.  We note also that while the unitaries and $\lambda$ eigenvalues that result from double factorization are determined by the cluster amplitudes (which for NOUCC(2) are taken from a classical MP2 calculation), in the UCJ ansatz (cf. \eq{JastrowAnsatz}) the matrix elements of $\bm{\mathcal{K}}$ and $\bm{\mathcal{J}}$ are classically optimized variationally.  Furthermore, it has been demonstrated that significantly fewer $l$ terms are required to recover high accuracy comparable to exact results~\cite{matsuzawa2020jastrow}, thereby providing a promising avenue toward reduced depth quantum circuits.  

In this work, we illustrate the use of the UCJ ansatz with $L=1$ in a NOQE calculation, denoted NOUCJ($L=1$), for the STO-3G H$_2$ system.  Accuracy comparable to untruncated (i.e., $L=N^2$) NOUCC(2) is achieved, suggesting that variational optimization of the Jastrow ansatz parameters can effectively compensate for the relatively small number of these imposed by truncation.  We note that by construction, the $L=1$ UCJ ansatz does not require further Trotterization, which is not the case for doubly-factorized reference ans{\"a}tze in the NOUCC(2) procedure. 

\begin{figure*}[thb!]
\centering
\includegraphics[width=0.6\textwidth]{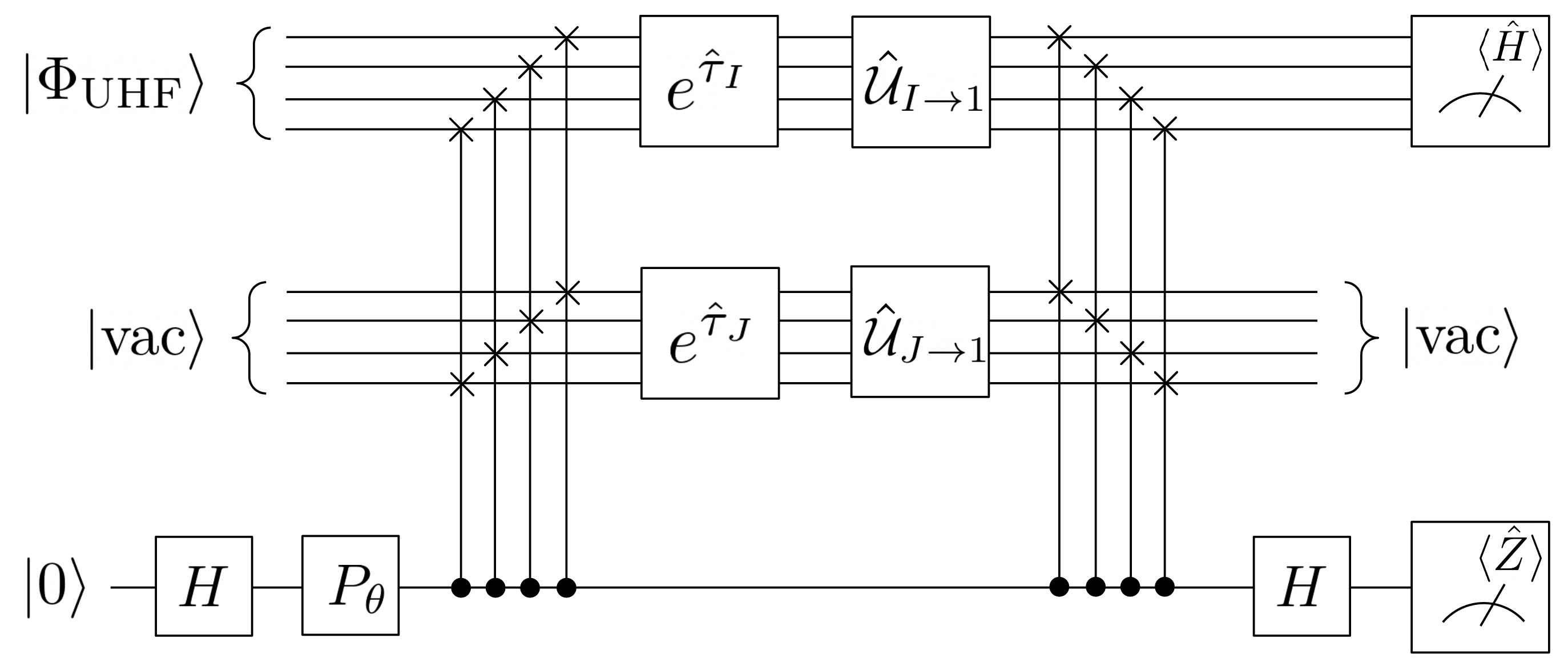}
\caption{NOQE circuit for the evaluation of off-diagonal matrix elements of the Hamiltonian ($H_{IJ}$) and overlap matrix ($S_{IJ}$) between NOUCC(2) ansatz states $\ket{\phi_I}$ and  $\ket{\phi_J}$. 
For the phase gate $P_\theta$, we set $\theta=0$ ($P_\theta=I$) when measuring $\text{Re}(H_{IJ})$, and $\theta=\pi/2$ when measuring $\text{Im}(H_{IJ})$  (see \app{noqe-circuit}).
}
\label{fig:noqe-circuit}
\end{figure*}

\begin{figure*}[thb!]
\includegraphics[width=0.7\textwidth]{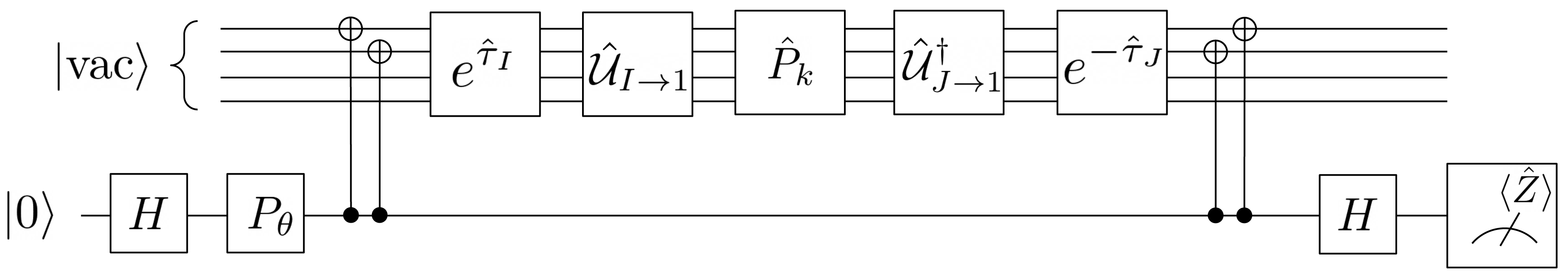}
\caption{NOQE circuit for evaluation of the off-diagonal matrix elements with fewer qubits. Here \(\hat{P}_k\) arises from a decomposition of the Hamiltonian (expressed in the \(I\)th UHF basis) in terms of Pauli operators. For the phase gate $P_\theta$, we set $\theta=0$ ($P_\theta=I$) when measuring $\text{Re}(H_{IJ})$, and $\theta=\pi/2$ when measuring $\text{Im}(H_{IJ})$  (see \app{noqe-circuit}).
}
\label{fig:noqe-circuit_moreefficient}
\end{figure*}

\subsection{Non-Orthogonal Quantum Eigensolver, NOQE}\label{sec:noqe_circuit}

We now come to the main theoretical exposition of this work, which is the construction of a non-orthogonal quantum eigensolver (NOQE) that does not require variational optimization. We construct multi-reference ansatz state as a linear combination of $M$ UHF base reference states:
\begin{equation}
    \ket{\Psi_\text{NOUCC(2)}} = \sum_{J=1}^Mc_J\ket{\phi_J} = \sum_{J=1}^Mc_Je^{\hat{\tau}_J}\ket{\Phi_J}.
\end{equation}
Here $\ket{\Phi_J}$ and $\hat{\tau}_J$ are, respectively, the $J$th UHF base reference state and the corresponding $J$th doubles MP2 tensor, which is constructed in the single-particle basis of $\ket{\Phi_J}$. Key features of this expansion, which also distinguish this work from the previous NOVQE work~\cite{huggins2020non}, are first that more than one base reference state is employed, and, secondly, that the base reference states are unrestricted Hartree-Fock states rather than restricted Hartree-Fock state, i.e., UHF rather than RHF. Since all base reference states are described by their expansion in the underlying common atomic orbital basis, the first feature introduces the need to track the unitary transformations between these expansions (see~\app{orbs-coeffs}). The second feature introduces significantly greater flexibility for the description of strongly correlated systems.

The NOQE coefficients $c_J$ are determined by classically diagonalizing the Hamiltonian matrix in the subspace of these non-orthogonal reference states. This requires solving the generalized eigenvalue problem
\begin{equation}
\matr{H}\vec{c} = E\matr{S}\vec{c},
\label{eq:generalized_eval_prob}
\end{equation}
with Hamiltonian and overlap matrix elements given by 
\begin{equation}
    H_{IJ} = \bra{\phi_I}\opr{H}\ket{\phi_J}, \qquad S_{IJ} = \braket{\phi_I}{\phi_J}.
    \label{eq:matrix_els}
\end{equation}
Provided that the $\matr{S}$ matrix is far from singular, and therefore the solutions to \eq{generalized_eval_prob} are numerically stable, this small eigenvalue problem is trivial for a classical computer. We note that even in the near-singular case it can be solved reliably using the standard procedure of thresholding the singular values of $\matr{S}$, even in the presence of relatively large perturbations to $\matr{H}$ and $\matr{S}$ that might arise due to device noise~\cite{epperly_theory_2021}.

We evaluate the off-diagonal elements of the Hamiltonian and overlap matrices  (\eq{matrix_els}) using a quantum circuit that is closely related to the modified Hadamard test protocol of Ref.~\cite{huggins2020non}. \fig{noqe-circuit} shows the general circuit for NOQE calculations using different NOUCC(2) ansatz states. The input to the circuit is an $N$-qubit reference state $\ket{\Phi_\text{UHF}}$, an ancilla register of size $N$, and a single control qubit, resulting in a total of $2N+1$ qubits. The $\eta$-electron UHF state in the first single-particle basis is prepared as $\ket{\Phi_\text{UHF}}=\ket{1}^{\otimes \eta}\otimes\ket{0}^{\otimes N-\eta}$. For the example of H$_2$ in the STO-3G basis with four spin-orbitals (see \tab{num-qubits-h2}), we would have $\ket{\Phi_\text{UHF}}=\ket{1100}$. The operation of the circuit in \fig{noqe-circuit} and its implementation of a modified Hadamard test for the evaluation of the off-diagonal Hamiltonian and overlap matrix elements of \eq{matrix_els} is described in detail in \app{noqe-circuit}.

Note that in contrast to the circuit employed in the non-orthogonal variational quantum eigensolver (NOVQE) of Ref.~\cite{huggins2020non}, which employed non-orthogonal reference states constructed with respect to a single (spin-restricted) Hartree-Fock state, the circuit of \fig{noqe-circuit} contains additional unitaries that rotate the atomic orbital bases on each register from that of UHF reference $I$ or $J$ into the (arbitrarily chosen) first single-particle basis, i.e., $\opr{\mathcal{U}}_{J \rightarrow 1}$. 
This accounts for the fact that the NOQE reference states are constructed here over different UHF basis sets, while the computation of the matrix elements and overlaps requires a consistent mapping of the orbital space onto the qubit register for all of the reference states, as well as the system Hamiltonian. 
We apply the unitary basis rotation following the preparation of the ansatz state in the default basis of the quantum register, which is equivalent under cancellation of unitaries to the transformation
\begin{align}
    \ket{\Phi_J}&\mapsto\opr{\mathcal{U}}_{J\rightarrow1}\ket{\Phi_\text{UHF}},\label{eq:basis_transform_line1}\\
    e^{\hat\tau_J}&\mapsto\opr{\mathcal{U}}_{J\rightarrow1}e^{\hat\tau_J}\opr{\mathcal{U}}_{J\rightarrow1}^\dag.\label{eq:basis_transform_line2}
\end{align}
Thus, on the right-hand side of \eq{basis_transform_line1} and \eq{basis_transform_line2}, $\ket{\Phi_\text{UHF}}$ and $e^{\opr{\tau}_J}$ are implemented in the default basis but are implicitly understood to be representations in the $J$th UHF basis, prior to the application of $\opr{\mathcal U}_{J\rightarrow1}$, after which everything is correctly expressed in the same (first) UHF basis. The basis rotation is constructed in terms of the $N\times N$ coefficient matrices $\matr{C}_J$ transforming the atomic orbital basis to the $J$th molecular orbital basis and the atomic orbital overlap matrix $\matr{S}_\text{A}$:
\begin{equation}
    \opr{\mathcal{U}}_{J \rightarrow 1} = \exp\left( \sum_{pq=1}^N[\ln(\matr{C}_J^T\matr{S}_\text{A}\matr{C}_1)]_{pq} (\opr{a}_p^\dag \opr{a}_q - \opr{a}_q^\dag \opr{a}_p) \right).
\end{equation}
This can be efficiently implemented with Givens rotations and single-qubit rotation gates, even over a device with linear connectivity~\cite{kivlichan2018quantum} (see \sec{circuit-resource-estimation}). We note that the circuit can be modified to reduce the gate cost by transforming one of the two reference ansatz states to the basis of the other one, rather than transforming both to the common basis $I=1$. This modification is employed in obtaining the resource estimate counts in \sec{circuit-resource-estimation}.

While the current work focuses on ground and low-lying electronic energy states, we point out that one may also replace the UHF reference state in the circuit diagram of \fig{noqe-circuit} with a UHF state representing an excited electronic configuration to find other higher lying states. For example, the excited determinant $\ket{\Phi_i^a}$, where $i\leq\eta$ and $a>\eta$, would be prepared by $X_i\ket{1}^{\otimes \eta}\otimes X_a\ket{0}^{\otimes N-\eta}$, where $X_p$ denotes a bit-flip gate acting on the $p$th qubit.

One additional benefit of performing the measurements using the modified Hadamard test is that we can thereby directly incorporate the reduced overhead measurement techniques of Ref.~\cite{Huggins2021-vu} without increasing the circuit depth. The basic idea is to apply a tensor factorization to the two-body part of the Hamiltonian, similar to the one described for the cluster tensor in \eq{lowrank_final}:
\begin{equation} \label{eq:h_tensordecomp}
        \hat{H} \approx \sum_{l=1}^L \opr{\mathcal{U}}_B^{l\dagger} \left( \sum_{pq}\lambda_p^{(l)}\lambda_q^{(l)} \opr{n}_p\opr{n}_q\right) \tilde{\mathcal{U}}_B^{l}.
\end{equation}
One can then measure all of the \(\opr{n}_p\opr{n}_q\) terms corresponding to a particular value of \(l\) simultaneously by explicitly applying the change of basis \(\opr{\mathcal{U}}_B^{l}\) on the quantum device before performing a standard measurement in the computational basis. In our case, this can be accomplished without any additional quantum resources because the product \(\opr{\mathcal{U}}_B^{l} \opr{\mathcal{U}}_{J\rightarrow1}\) amounts to a single change of basis and can be implemented using the same number of gates as \(\opr{\mathcal{U}}_{J\rightarrow1}\) alone. Empirically, it has been found that taking \(L = \mathcal{O}(N)\) is sufficient to obtain a fixed relative error in the energy due to the decay of the singular values in the first tensor factorization of the Hamiltonian~\cite{peng2017highly}. Ref.~\cite{Huggins2021-vu} found that taking advantage of the decomposition \eq{h_tensordecomp} reduced the variance of the energy estimator by orders of magnitude, even for small VQE calculations. More study in the context of the off-diagonal matrix element measurements considered here will be useful.

There are also variants of the NOQE circuit that can reduce the number of qubits required from $2N+1$ to $N+1$. In this case, however, the circuit depth will need to be approximately doubled to accommodate application of all four unitaries \(e^{\hat{\tau}_I},\) \(\opr{\mathcal{U}}_{I \rightarrow 1},\) $\opr{\mathcal{U}}_{J \rightarrow 1}^\dagger,$  and $e^{-\opr{\tau}_J}$ to the $N$ system qubits. Suppose that the Hamiltonian (expressed in the first UHF basis) is decomposed as a linear combination of Pauli operators, \(\opr{H} = \sum_k c_k \opr{P}_k\). One strategy for reducing the number of qubits would then be to use a single ancilla qubit to perform a Hadamard test on each of the unitary operators \(e^{-\opr{\tau}_J} \opr{\mathcal{U}}_{J \rightarrow 1}^\dagger \opr{P}_k \opr{\mathcal{U}}_{I \rightarrow 1} e^{\hat{\tau}_I}\), estimating their expectation value with respect to \(\ket{\Phi_\textrm{UHF}}\). Naively, this requires us to condition the application of the state preparation unitaries on the state of the ancilla qubit. 

However, as with the generic case of \fig{noqe-circuit} in which \(2N + 1\) qubits are used, we can take advantage of the fact that the state preparation unitaries conserve particle number. Rather than explicitly conditioning them on the state of the ancilla qubit, we can instead use the ancilla to control the preparation and uncomputation of \(\ket{\Phi_\textrm{UHF}}\) from the vacuum state, together with the operator \(\opr{P}_k\). Note that, as above, \(\ket{\Phi_\textrm{UHF}}\) is simply a computational basis state (although we implicitly work in two different bases). We can therefore make a controlled preparation of \(\ket{\Phi_\textrm{UHF}}\) from the vacuum state using a single CNOT gate for each occupied orbital, as shown in \fig{noqe-circuit_moreefficient}. Let \(C{\text -}U_\textrm{UHF}\) denote the set of CNOT gates which prepares \(\ket{\Phi_\textrm{UHF}}\) conditional on the state of an ancilla qubit, and let \(C{\text -}P_k\) denote the controlled-\(P_k\) gate. Due to the fact that \(e^{\opr{\tau}_I},\) \(\opr{\mathcal{U}}_{I \rightarrow 1},\) $\opr{\mathcal{U}}_{J \rightarrow 1}^\dagger,$  and $e^{-\opr{\tau}_J}$ all conserve particle number, we have the following equality:
\begin{align}
     & (C{\text -}U_\textrm{UHF})^\dagger e^{-\hat{\tau}_J} \opr{\mathcal{U}}_{J \rightarrow 1}^\dagger (C{\text -}P_k)  \opr{\mathcal{U}}_{I \rightarrow 1} e^{\hat{\tau}_I} (C{\text -}U_\textrm{UHF}) \ket{+}\ket{\textrm{vac}} \nonumber \\
     =& \frac{1}{\sqrt{2}} \big(\ket{0}\ket{\textrm{vac}} + U_\textrm{UHF}^\dagger e^{-\hat{\tau}_J} \opr{\mathcal{U}}_{J \rightarrow 1}^\dagger \opr{P}_k  \opr{\mathcal{U}}_{I \rightarrow 1} e^{\hat{\tau}_I} \ket{1}\ket{\Phi_\textrm{UHF}}\big).
\end{align}
A quick computation verifies that measuring the ancilla qubit in the \(X\) basis yields the quantity
\begin{equation}
    \textrm{Re}\big(\bra{\Phi_\textrm{UHF}} e^{-\hat{\tau}_J} \opr{\mathcal{U}}_{J \rightarrow 1}^\dagger \opr{P}_k  \opr{\mathcal{U}}_{I \rightarrow 1} e^{\hat{\tau}_I} \ket{1}\ket{\Phi_\textrm{UHF}}\big),
\end{equation}
as desired. The imaginary part can be estimated in the usual way, by beginning with the ancilla state in the \(-1\) eigenstate of the Pauli $Y$ operator (see \app{noqe-circuit}).

 On real quantum hardware, the matrix elements of the Hamiltonian and overlap matrix would be evaluated by repeatedly running the NOQE circuit of \fig{noqe-circuit} or \fig{noqe-circuit_moreefficient}, as in~\cite{huggins2020non}.  However, for this first accuracy benchmarking study of the NOQE approach on small molecules, we evaluate the matrix elements by use of a quantum simulator to generate representations of the ansatz states $\ket{\phi_J}=\opr{\mathcal{U}}_{J\rightarrow1}e^{\hat\tau_J}\ket{\Phi_J}$ in the $2^N$-dimensional vector space of FCI determinants. We then simulate the idealized (i.e., noiseless) circuit result classically, by directly evaluating the bitwise inner product $\bra{\phi_I}\opr{H}\ket{\phi_J}$.
 
\begin{figure*}[thb!]
\includegraphics[width=0.7\textwidth]{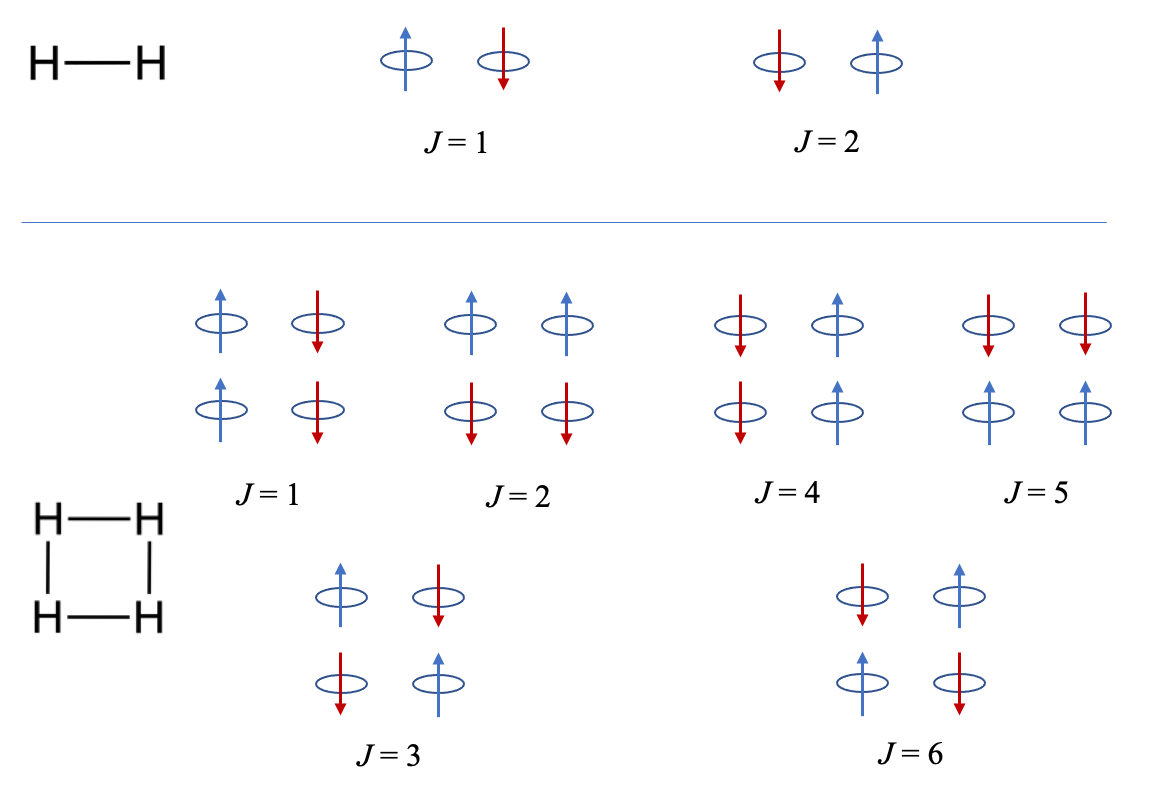}
\caption{UHF reference configurations for \ce{H2} and square \ce{H4} at the dissociation limit, showing spatial distribution of spins on individual H atoms. 
}
\label{fig:num_refs}
\end{figure*}

\subsection{Technical Details}\label{sec:technical}

We used OpenFermion~\cite{mcclean2020openfermion}, PySCF~\cite{sun2020recent}, and QChem~\cite{epifanovsky2021software} to generate the reference states used in the NOQE protocol. Both OpenFermion and OpenFermion-PySCF were augmented to allow unrestricted Hartree-Fock (UHF) states to be prepared. The UHF solutions themselves were obtained via the Q-Chem 5 software package~\cite{epifanovsky2021software}. These solutions to the Hartree-Fock equations were obtained with the following procedure:
\begin{enumerate}
    \item Optimize the restricted open-shell Hartree-Fock solution with all radical sites having unpaired electrons and all unpaired spins pointing in the same direction. For \ce{H2}, this is the triplet state; for \ce{H4}, it is the quintet state.
    \item Localize this first set of orbitals onto the radical sites~\cite{boys1960construction,edmiston1963localized}. Hartree-Fock equations can at times yield spurious results with spins in delocalized orbitals, especially for species like \ce{H4}~\cite{burton2020energy}, so this step removes that possibility. 
    \item Generate all possible permutations of up and down spins (on these radical sites) that have the desired total $m_s$ value (equal to zero in this work, unless specified otherwise). For \ce{H2} with $m_s=0$, there are only 2 radical sites possible for one up and one down spin, leading to 2 determinants. For $m_s=0$ \ce{H4}, there are 4 radical sites, on which 2 up and 2 down spins have to be placed, resulting in 6 possible spin configurations and hence 6 determinants. These spin arrangements for \ce{H2} and \ce{H4} are shown in \fig{num_refs}. 
    \item Optimize these permuted electronic configurations with the square gradient minimization~\cite{hait2020excited} algorithm, to ensure that the closest stationary state to the initial starting point is reached. The resulting Hartree-Fock solutions are used for NOQE applications. 
\end{enumerate}
The above procedure is similar to the manner in which reference determinants for the classical spin-flip NOCI method are generated~\cite{mayhall2014spin}, except no orbitals have to be held frozen, and full orbital relaxation can be carried out on account of advancements in classical algorithms for optimizing orbitals~\cite{hait2021orbital}.

In this work, we consider systems with $d$ radical sites and $\eta=d$ electrons within the total $m_s=0$ subspace.  As will be shown, using only the $m_s=0$ subspace to describe eigenvectors with different $\langle \opr{S}^2 \rangle$ (i.e., singlet, triplet, quintuplet, $\dots$) leads to advantageous cancellation of errors in the energy gaps between these states.  Therefore we have exactly $d/2$ up and $d/2$ down spin electrons. The total number of NOQE reference states $M$ is then equal to
\begin{align}\label{eq:NOQE_refs_paired}
    M=\dfrac{d!}{\left(\dfrac{\eta}{2}!\right)^2}
\end{align} 
determinants. For the general case with $d$ radical sites, $\eta_{\alpha}$ up spins and $\eta_{\beta}$ down spins (where $\eta_\alpha+\eta_\beta<d$, $m_s=\eta_\alpha-\eta_\beta$), the number of NOQE reference states is given by
\begin{align}\label{eq:NOQE_refs_general}
    M&= \dfrac{d!}{\eta_\alpha! \eta_\beta! (d-\eta_\alpha-\eta_\beta)!}
\end{align}

The low-rank circuits that prepare the NOUCC(2) reference states for NOQE were compiled with Cirq~\cite{The_Cirq_Developers2019-xo}. The full circuit for calculating the off-diagonal elements between NOQE reference states is shown in \sec{noqe_circuit}. However, as noted there, for the benchmark calculations presented in this work, we have evaluated the matrix elements directly in the computational basis and have not taken into the effect of the quantum measurement noise or circuit noise, which would be present in experimental evaluation of the off-diagonal matrix elements. This is consistent with our goal in this work of establishing the ideal values possible under a NOQE calculation. 

\section{Results}\label{sec:results} 

\begin{figure}[h!]
    \begin{minipage}{\columnwidth}
       \includegraphics[width=1\linewidth]{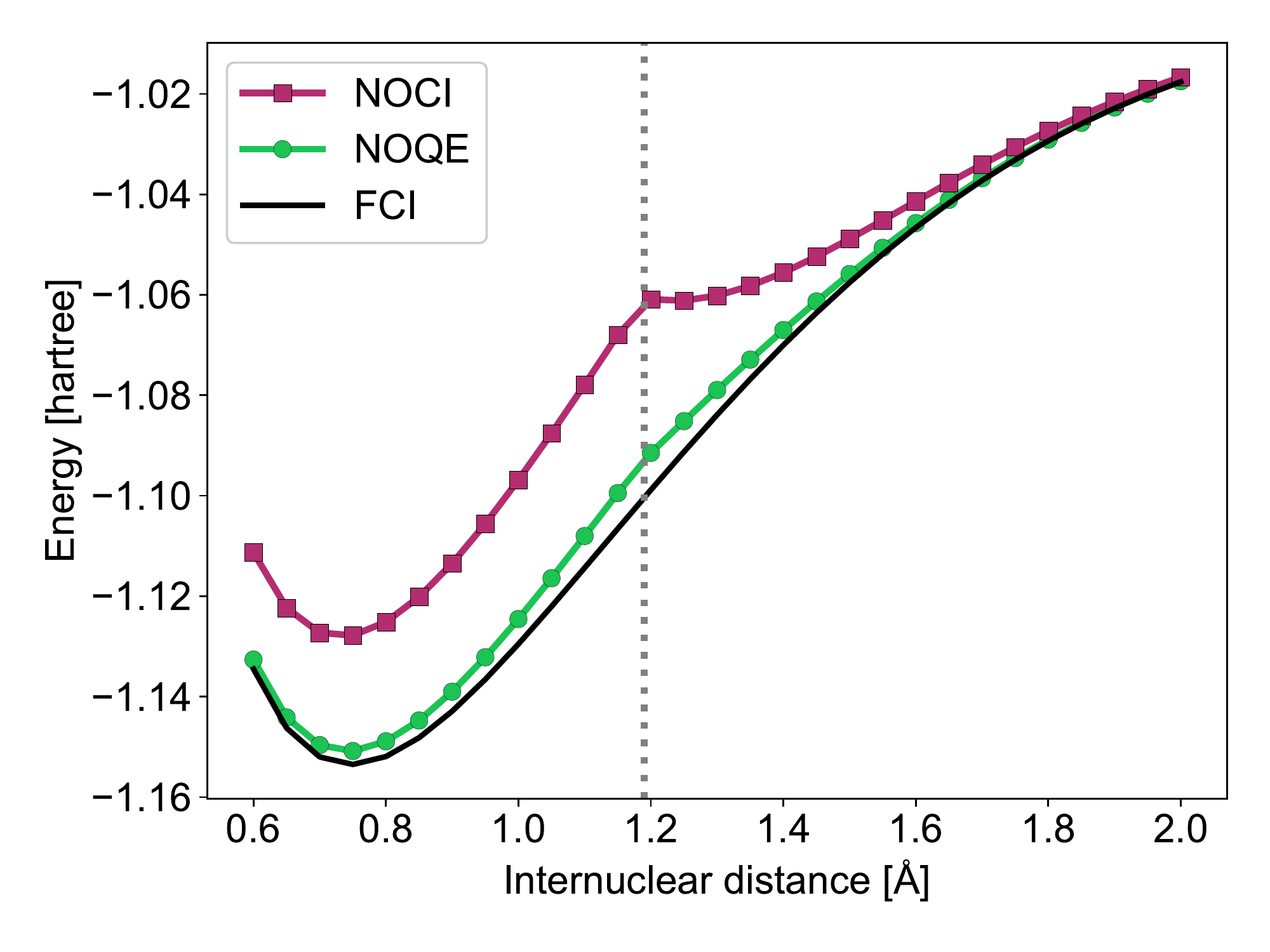}
       \subcaption{Lowest energy singlet (S$_0$) state.}
       \label{fig:h2-6311g-energies-S0} 
    \end{minipage}
    \begin{minipage}{\columnwidth}
       \includegraphics[width=1\linewidth]{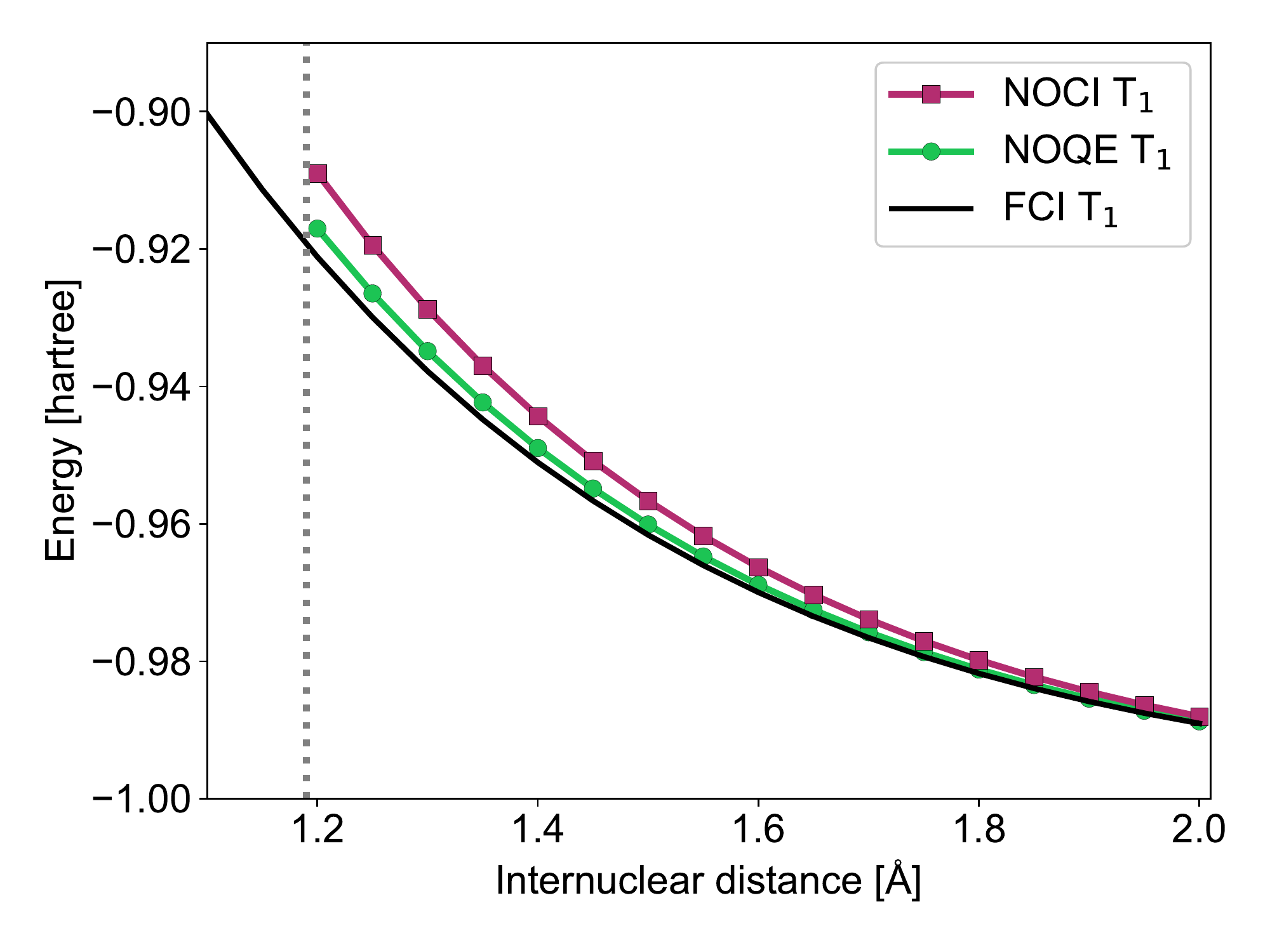}
       \subcaption{Lowest energy triplet (T$_1$) state.}
       \label{fig:h2-6311g-energies-T1}
    \end{minipage}
    \caption{Comparison of NOUCC(2) and FCI eigenenergies for \ce{H2}, with the 6-311G basis set. Classical NOCI results (without dynamic correlation) are also provided for comparison. The location of the CF point (1.19 {\AA} for HF/6-311G) is marked as a dotted gray line. Note that the T$_1$ state only appears beyond the CF point for NOCI and NOUCC(2) calculations.}
    \label{fig:h2-6311g-energies-absolute}
    \vspace{-10pt}
\end{figure}

\subsection{Hydrogen molecule}
At its minimum energy geometry, the \ce{H2} molecule has an internuclear separation (``bond length") of 0.74~\AA. The wavefunction at this geometry is well approximated by a single RHF Slater determinant in which both electrons occupy the same bonding spatial orbital. This RHF determinant is however incapable of describing the dissociation limit of two independent H atoms (see \app{cfpt}). Indeed, the optimal single determinant description for the dissociation limit of \ce{H2} consists of two independent atoms with one electron localized around each. There are thus two UHF states possible, corresponding to the two possible ways to arrange an up and a down spin on the two atoms without spin-pairing (as shown in \fig{num_refs}). 

The behavior at internuclear distances between these two limits is intermediate between them. The RHF determinant continues to be the minimal energy Hartree-Fock solution for internuclear separations smaller than a cutoff distance that is commonly referred to as the Coulson-Fischer (CF) point~\cite{coulson1949xxxiv}. Beyond the CF point, however, it is energetically favorable to (partially) unpair the two electrons. The minimum energy Hartree Fock solutions are then spin-symmetry broken UHF determinants wherein the individual atoms have non-zero net spin. There are two such energetically degenerate UHF states because the net spin on a particular atom can point either up or down (with the net spin on the other atom being in the opposite direction to preserve total $m_s=0$). The spin-symmetry breaking in these UHF states prevents them from being eigenstates of the $\opr{S^2}$ operator and instead causes them to be a mixture of singlet and triplet. The minimum energy Hartree Fock solution thus branches from a single RHF determinant to two (overlapping) UHF determinants at the CF point. This branching is continuous in both the energy and its first derivatives, with discontinuities arising only in the second derivatives of the energy, at the CF point~\cite{hait2019wellbehaved}.  The two spin-symmetry broken UHF states, therefore, have an overlap that is close to $1$ just beyond the CF point (having barely branched from the same parent RHF state) but equal to $0$ at an infinite internuclear distance.  

These branching UHF states constitute our NOUCC(2) reference states at distances beyond the CF point. They recouple to yield the lowest energy singlet (S$_0$) and lowest energy triplet (T$_1$) states in an approximately spin-pure manner. Potential singularities in the $\matr{S}$ matrix are removed by discarding singular values less than $10^{-4}$. At distances shorter than the CF point, the RHF state is the only possible reference state. This has the consequence that the T$_1$ state appears only beyond the CF point where there are two reference states, while the S$_0$ state can be described by the RHF state and is thus also found at shorter distances. We note that the inability of this chosen set of reference states to model the T$_1$ state at small internuclear distances is not particularly limiting, because the T$_1$ state is very high in energy relative to the S$_0$ ground state in this regime, and the ordering of the spin states is not in doubt here. Indeed, electronic structure in this ``single reference" regime is well described by classical methods~\cite{mardirossian2017thirty}. The description of the ground and excited states beyond the CF point is more challenging due to spin-symmetry breaking~\cite{szabo2012modern,hait2019wellbehaved,hait2019beyond}.

Since there are (at most) two reference states in this description of H$_2$, only a single off-diagonal matrix element energy evaluation is needed.  Given a single-particle basis size of $N$ spin-orbitals, we note that $2N+1$ qubits are needed to construct the NOQE circuit in \fig{noqe-circuit}, or $N+1$ qubits for the circuit shown in \fig{noqe-circuit_moreefficient}. We have carried out calculations with three basis sets, namely STO-3G~\cite{hehre1969self}, 6-31G~\cite{ditchfield1971self}, and 6-311G~\cite{krishnan1980self}. The following discussion will be based on results obtained with the 6-311G basis. Results obtained with the other two basis sets show comparable results and are presented in \app{h2-other-results}. The number of basis functions per hydrogen atom, spin-orbitals ($N$), and the total number of qubits required to construct the NOQE circuit of \fig{noqe-circuit} are listed in \tab{num-qubits-h2}. The resulting circuit depths (number of layers of parallel gate operations) for these calculations employing the low-rank factorization of the doubles tensor operator $\hat{\tau}$ are listed in \tab{circ-depth-h2}.

\begin{table}[h!]
\begin{tabular}{| c |c|c|c|c| } 
\hline
System & Basis set & Basis functions & Spin-orbitals  & Qubits   \\
& & per H atom & ($N$) & (2$N$+1)  \\
\hline
\multirow{3}{*}{\ce{H2}} & STO-3G  & 1& 4 & 9 \\ 
&6-31G & 2 & 8 & 17 \\ 
&6-311G & 3 & 12 & 25 \\ 
\hline
\multirow{1}{*}{\ce{H4}} & STO-3G  & 1 & 8 & 17 \\ 
\hline
\end{tabular}
\caption{\label{tab:num-qubits-h2} Size resources required for evaluation of matrix elements of the Hamiltonian and overlap for H$_2$ and H$_4$ with different basis sets (see \app{orbitals} for definition of these). The last column shows the number of qubits required to construct the NOQE circuit of \fig{noqe-circuit}.}
\end{table}

\begin{center}
\begin{table}[h!]
\begin{tabular}{ |c|c|c|c|c| } 
\hline
Basis set & Circuit depth & Circuit depth\\
& (SVD) & (Takagi) \\
\hline
STO-3G  & 53 & 38\\ 
6-31G   & 386 & 255\\
6-311G  & 1061 & 864\\
\hline
\end{tabular}
\caption{\label{tab:circ-depth-h2} 
Circuit depth of the (full rank) ansatz preparation unitary operator $\opr{\mathcal{U}}_{J\rightarrow 1}e^{\hat\tau_J}$ with MP2 amplitudes, i.e., NOUCC(2), for H$_2$ with increasing basis set size. These ansatz preparations may be performed in parallel, as in \fig{noqe-circuit}, or in sequence, as in \fig{noqe-circuit_moreefficient}.}
\end{table}
\end{center}

\fig{h2-6311g-energies-absolute} presents the computed NOQE absolute energies for the S$_0$ (\fig{h2-6311g-energies-S0}) and the T$_1$ (\fig{h2-6311g-energies-T1}) eigenstates as a function of internuclear separation, for calculations using the NOUCC(2) ansatz with the 6-311G basis set. Classical NOCI results without dynamic correlation, i.e., without MP2 amplitudes, and the exact FCI results are also provided for comparison.  We observe that NOQE with NOUCC(2) provides a significant improvement on classical NOCI at all distances for both the S$_0$ and T$_1$ eigenstates, yielding a much closer approximation to FCI. In particular, the NOQE S$_0$ result smoothly traverses the CF point and is successful in removing the spurious second local minimum just beyond the CF point that is evident in the classical NOCI S$_0$ state. Indeed, the results from NOQE with the NOUCC(2) ansatz are qualitatively in good agreement with FCI at all internuclear distances. 

\begin{figure}[htb!]
    \begin{minipage}{\columnwidth}
       \includegraphics[width=1\linewidth]{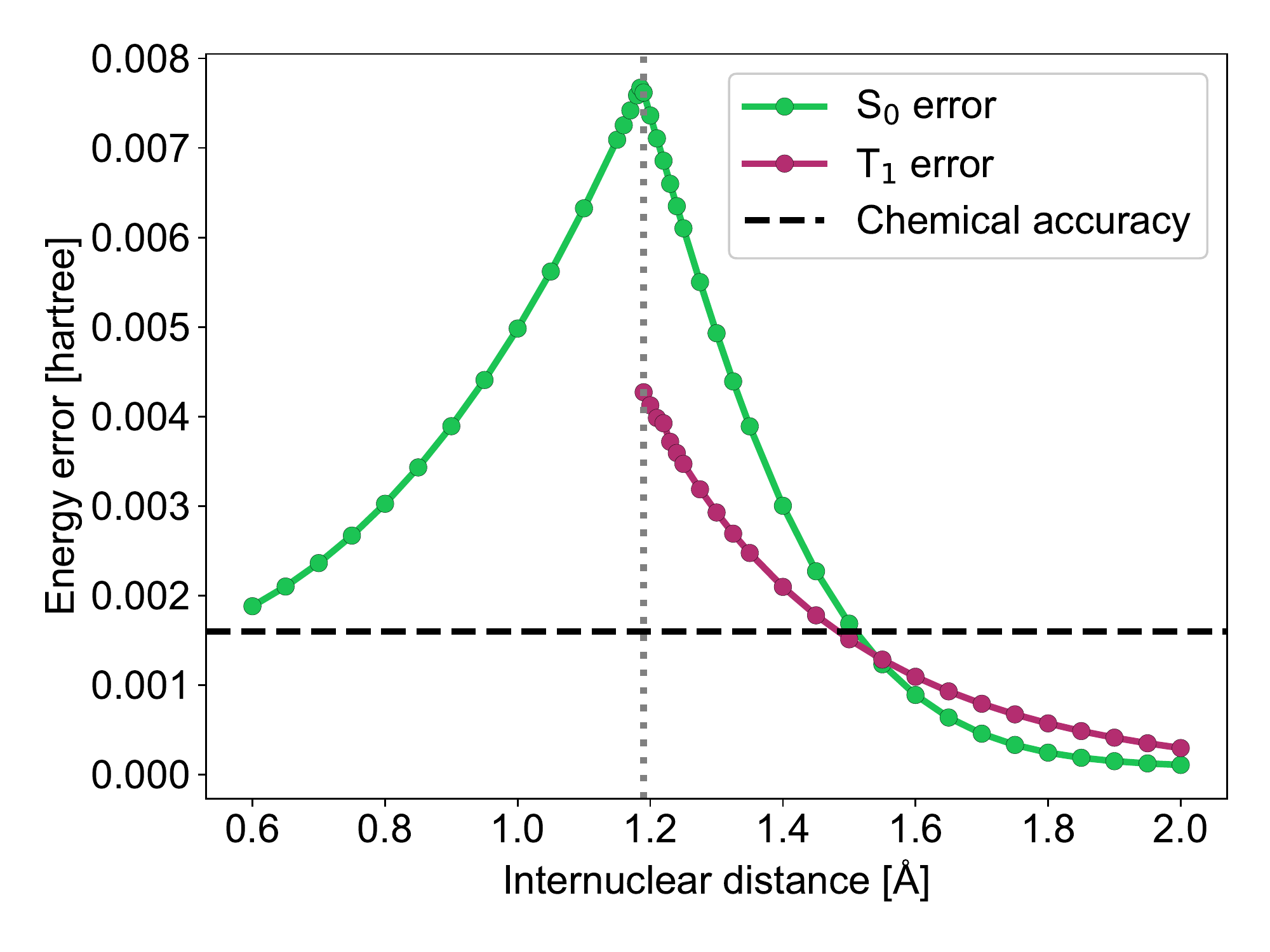}
       \subcaption{Error in the S$_0$ and T$_1$ eigenenergies with 6-311G basis.}
       \label{fig:h2-6311g-abserror} 
    \end{minipage}
    \begin{minipage}{\columnwidth}
       \includegraphics[width=1\linewidth]{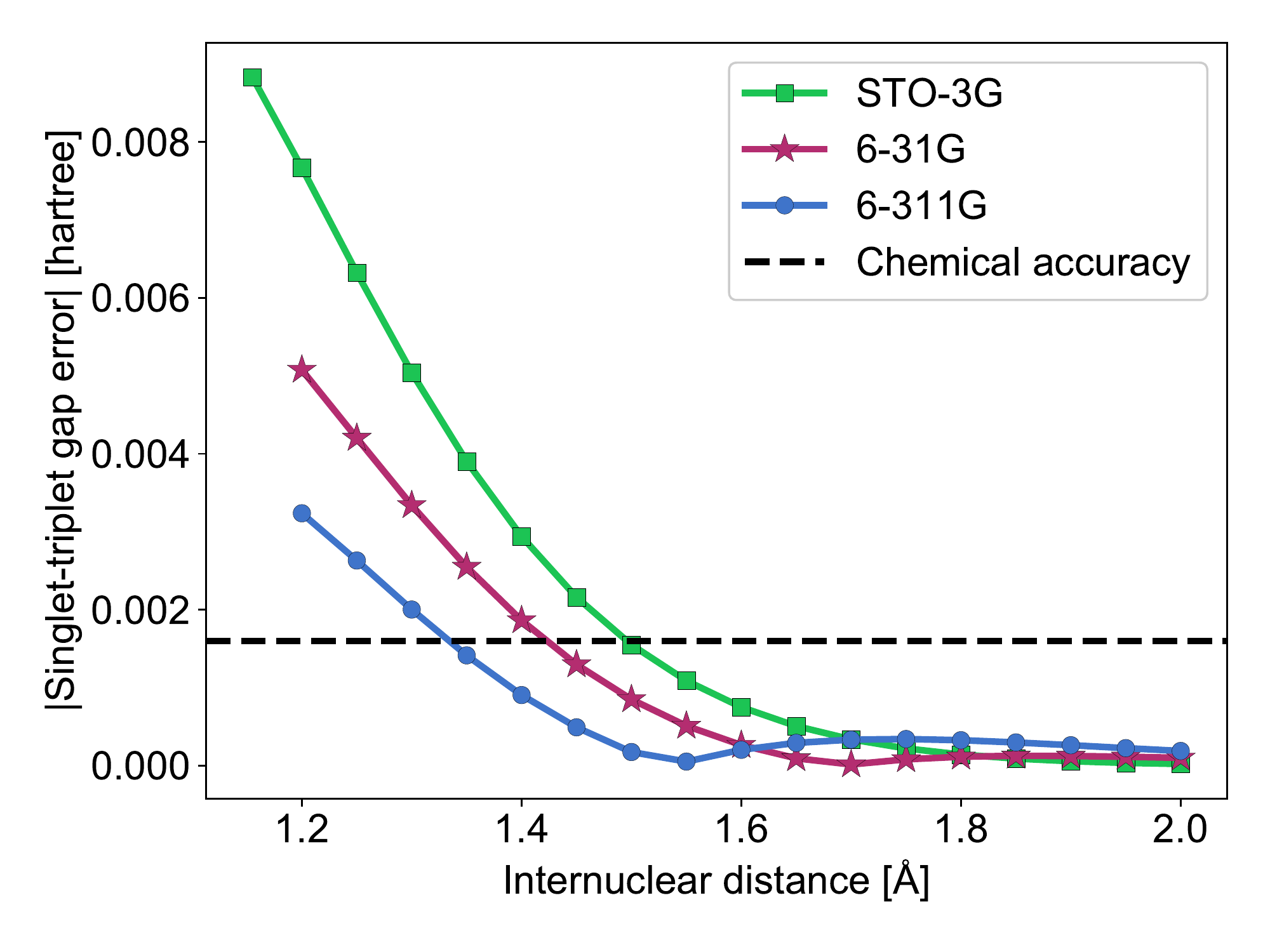}
       \subcaption{Singlet-triplet gap with different basis sets.}
       \label{fig:h2-stgaperror-basis}
    \end{minipage}
    \caption{Errors in energies relative to FCI for H$_2$ from NOQE with the NOUCC(2) ansatz states. The location of the CF point is marked with a dotted gray line in \fig{h2-6311g-abserror}, but not in \fig{h2-stgaperror-basis} since the precise location of this point depends slightly on the basis used.}
    \label{fig:h2-unscaled-errors}
     \vspace{-10pt}
\end{figure}

We now consider quantitative accuracy, showing first the energy error relative to the FCI results in \fig{h2-6311g-abserror}. We see that the NOQE S$_0$ and T$_1$ state energies differ from FCI by more than chemical accuracy at most distances and that the error is below chemical accuracy only at larger internuclear separations, where the system becomes increasingly well approximated by two independent atoms. The maximum error for the S$_0$ state is about 8 mHa, right around the CF point.

In practice, however, quantum chemists are seldom interested in absolute energies and instead prefer to look at energy differences between states (commonly referred to as ``relative energies"). The most chemically relevant quantity for H$_2$ is therefore the energy difference between the S$_0$ and T$_1$ states (i.e., the ``singlet-triplet gap"). Since NOQE with the NOUCC(2) ansatz overestimates the absolute energies of both the S$_0$ and T$_1$ states (see \fig{h2-6311g-energies-absolute}), this can lead to some cancellation of systematic error for the singlet-triplet gap. This is shown in \fig{h2-stgaperror-basis}, which shows that for the 6-311G basis calculation the singlet-triplet gap error is lower than the absolute energy errors at all internuclear distances. Around the CF point, it is less than 4 mHa, about half of the absolute error in S$_0$. Furthermore, the maximum singlet-triplet gap error decreases as the basis set size of the calculation is increased from STO-3G to 6-31G to 6-311G, suggesting that larger, more physically accurate basis sets like cc-pVDZ (with five basis functions per H atom, resulting in 20 spin-orbitals) could yield even lower errors.

\begin{figure}[h!]
    \begin{minipage}{\columnwidth}
       \includegraphics[width=1\linewidth]{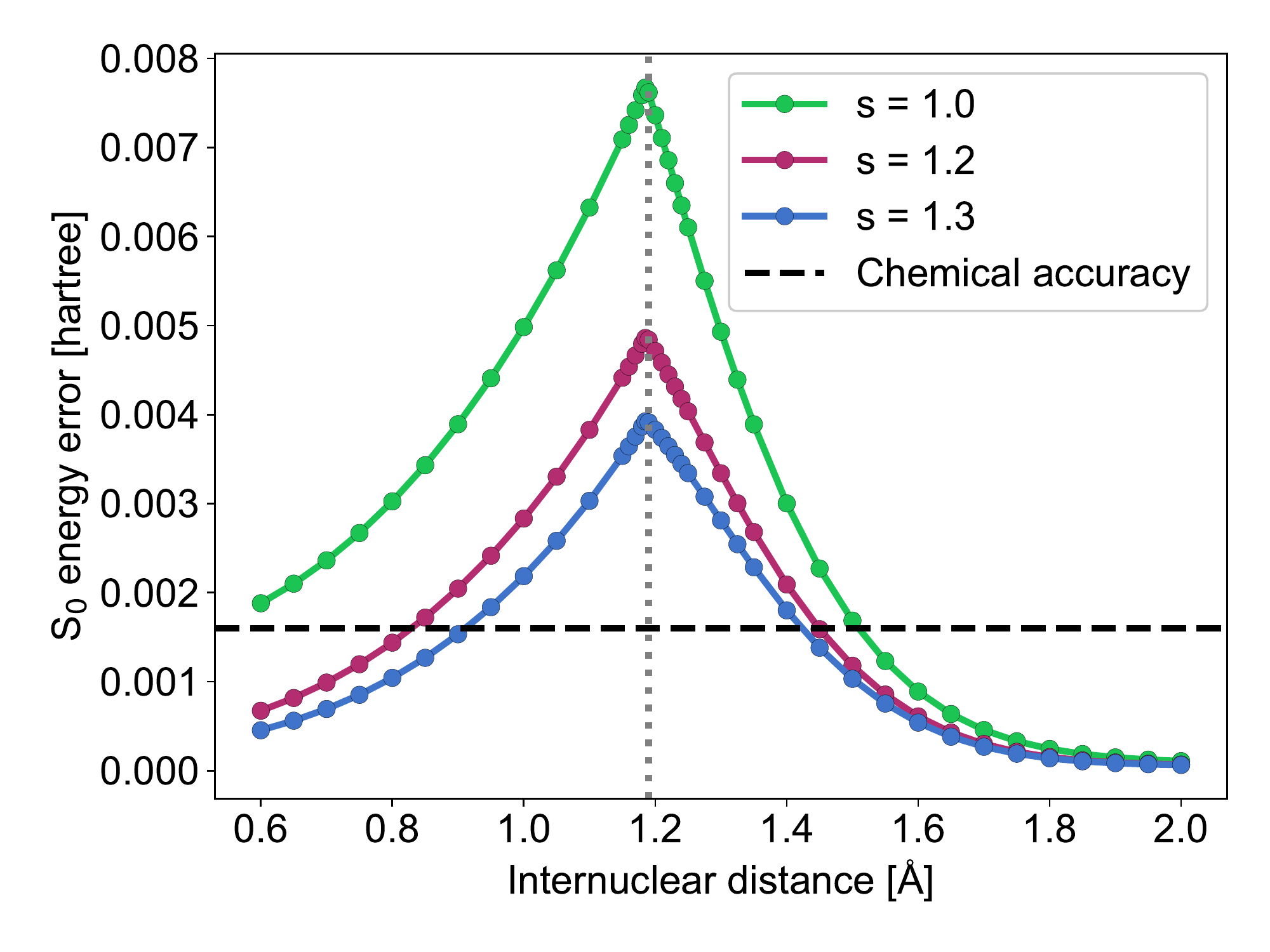}
       \subcaption{S$_0$ energy error relative to FCI.}
       \label{fig:h2-6311g-S0scale}
    \end{minipage}
    \begin{minipage}{\columnwidth}
       \includegraphics[width=1\linewidth]{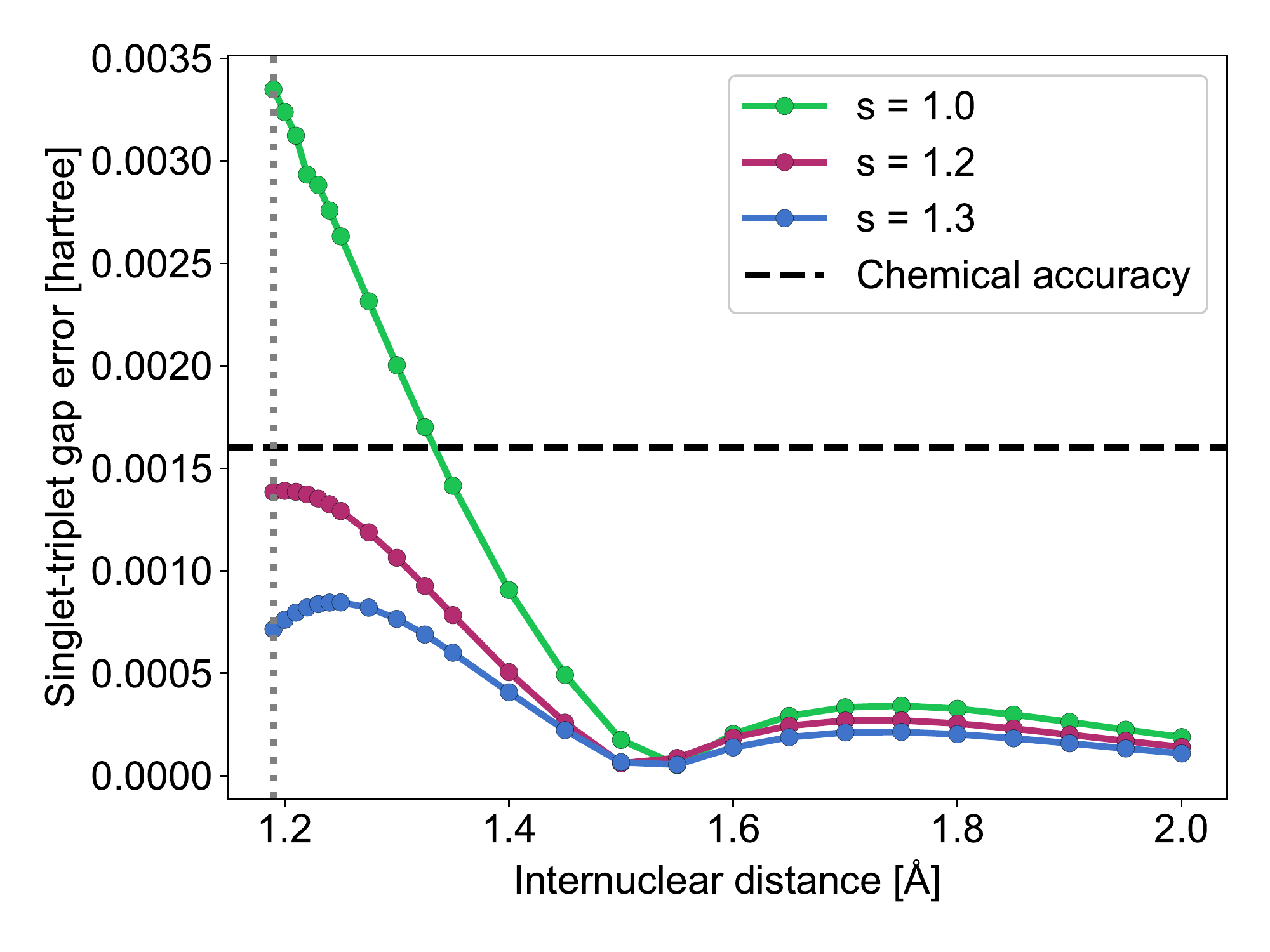}
       \subcaption{Singlet-triplet gap error relative to FCI.}
       \label{fig:h2-6311g-stgapscale} 
    \end{minipage}
    \caption{Energy errors relative to FCI for singlet and triplet states of H$_2$ from NOQE with NOUCC(2) ansatz states in which the MP2 amplitudes are scaled by $s$ (calculations made with the 6-311G basis). The location of the CF point is marked with a dotted gray line. }
    \label{fig:h2-6311g-scaled-errors}
     \vspace{-10pt}
\end{figure}

We now consider the impact of scaling the MP2 parameters on these results, which has been found to be beneficial in classical quantum chemistry~\cite{grimme2003improved,jung2004scaled}. Therefore, we investigate the effect of uniformly scaling MP2 amplitudes by scaling parameters $s=1.2$ and 1.3, as suggested classically by SCS- and SOS-MP2 (\sec{firstorder}). \fig{h2-6311g-scaled-errors} shows that the use of scaled MP2 amplitudes can significantly lower the errors in both absolute energies and the singlet-triplet gap. Indeed, the scaled MP2 amplitudes for scalings $s = 1.2$ and $s=1.3$ now yield singlet-triplet gaps within chemical accuracy from FCI at all internuclear distances (\fig{h2-6311g-stgapscale}).  Improvement of the S$_0$ absolute energies is less dramatic, but still quite significant, with calculations for $s=1.3$ almost halving the maximum error. 

\begin{figure}[h!]
    \begin{minipage}{\columnwidth}
       \includegraphics[width=1\linewidth]{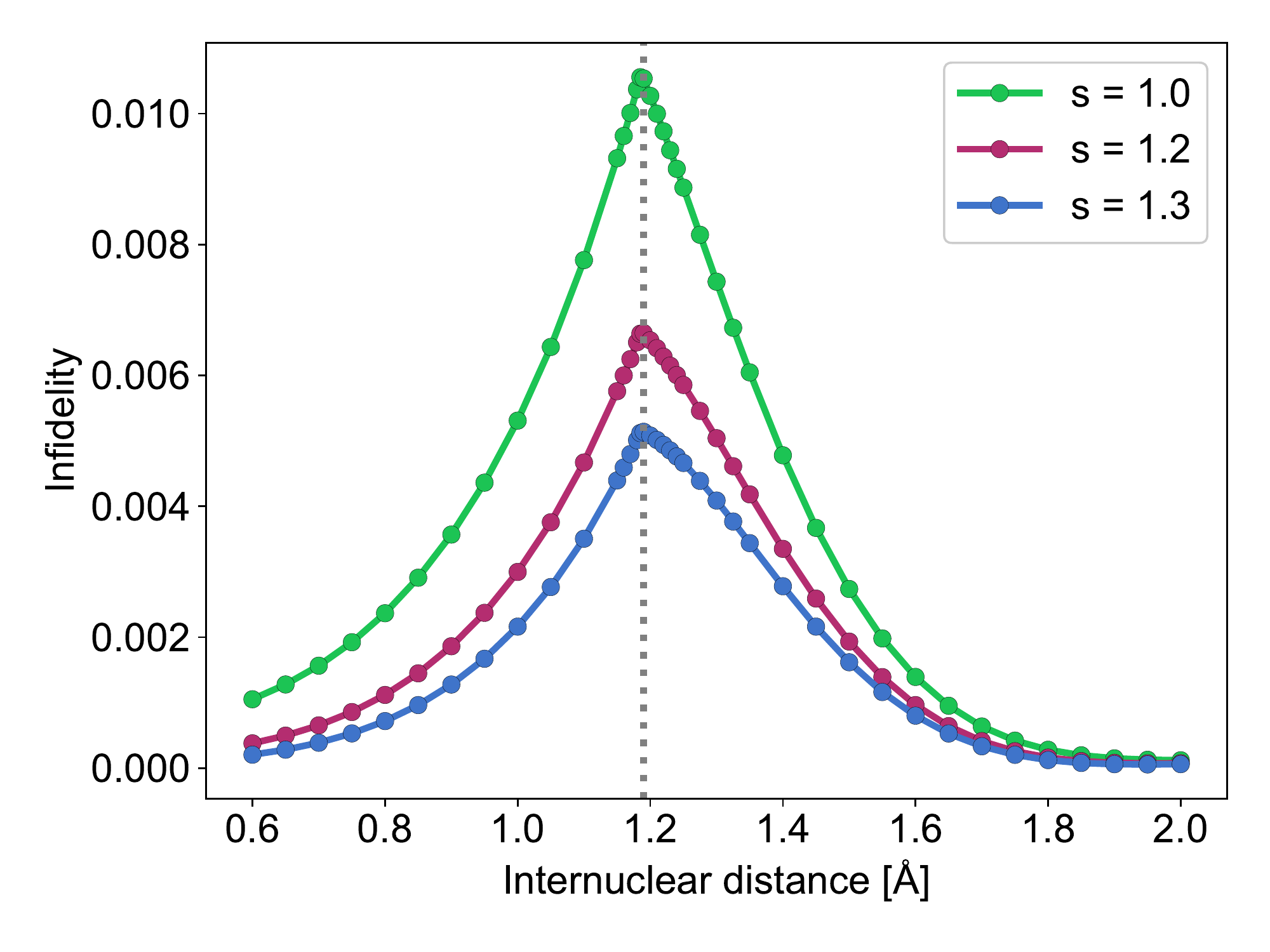}
       \subcaption{Infidelity of NOQE S$_0$ state.}
       \label{fig:h2-6311g-S0scaleinfidelity}
    \end{minipage}
    \begin{minipage}{\columnwidth}
       \includegraphics[width=1\linewidth]{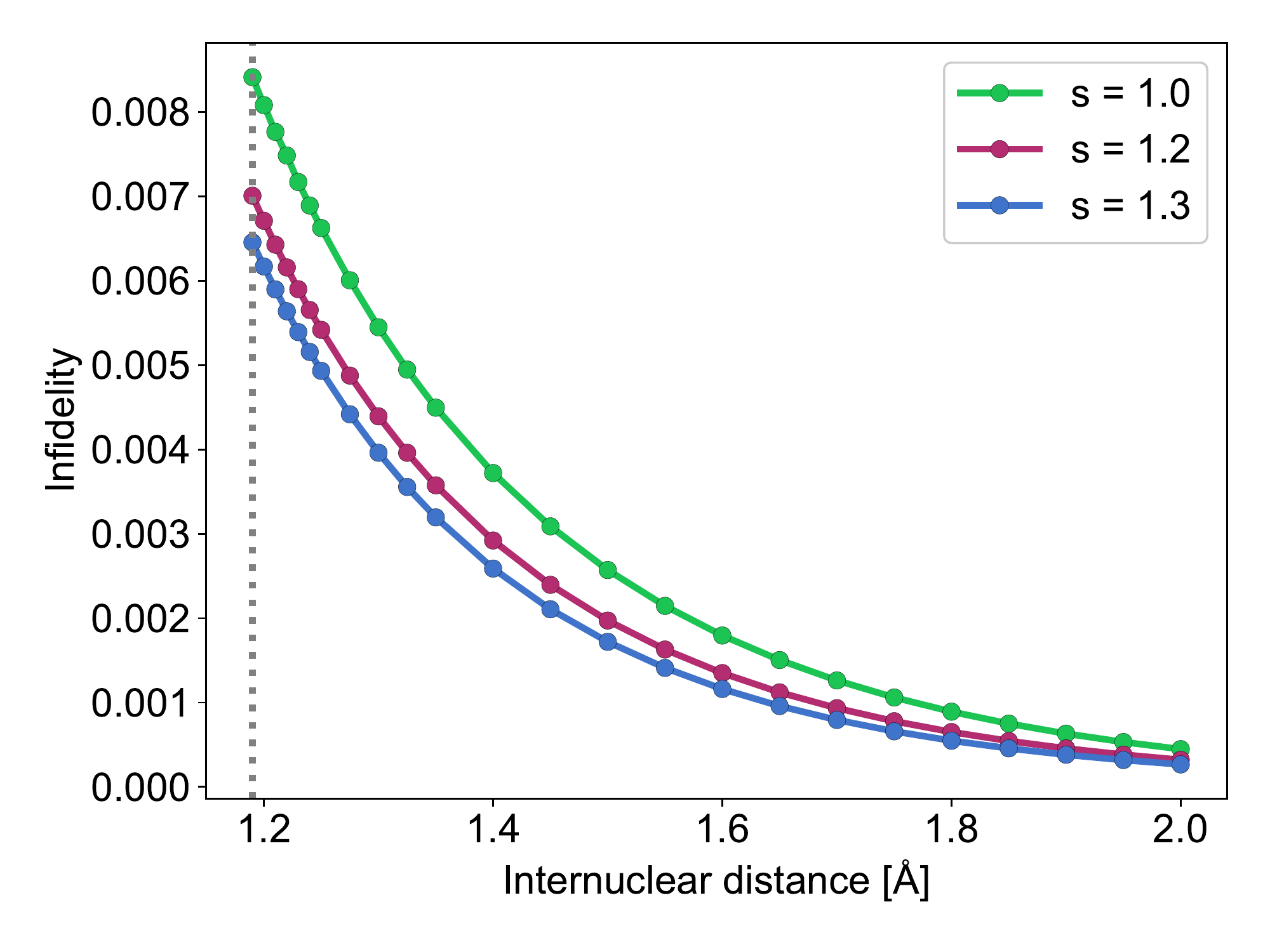}
       \subcaption{Infidelity of NOQE T$_1$ state.}
       \label{fig:h2-6311g-T1scaleinfidelity} 
    \end{minipage}
    \caption{Infidelity $I_\text{FCI}$ relative to the FCI states of the amplitude scaled NOQE states derived from the NOUCC(2) ansatz, for H$_2$ with the 6-311G basis. See \eq{Infidelity_FCI}. The location of the CF point is marked with a dotted gray line.}
    \label{fig:h2-6311g-infidelities}
\end{figure}

We are also interested to ascertain the quality of approximation of the NOQE states to the true FCI states. To this end, we define the state infidelity
\begin{equation} \label{eq:Infidelity_FCI}
    I_\text{FCI} = 1-\abs{\braket{\Psi_\text{NOQE}}{\Psi_\text{FCI}}}^2,
\end{equation}
which is the fraction of the FCI state not captured by NOQE.  \fig{h2-6311g-infidelities} shows the infidelities $I_\text{FCI}$ of the singlet and triplet NOQE states as a function of internuclear distance.  The maximum infidelity over all distances  is $\sim 1\%$ (at the CF point) without any amplitude scaling and is further reduced on scaling the amplitudes with $s > 1$. This indicates that both the S$_0$ and T$_1$ states are reproduced to $\sim99\%$ or better fidelity by NOQE with the NOUCC(2) ansatz. We also note that the NOQE states also show a high degree of spin purity, with a maximum error of $3\times 10^{-7}$ in $\langle \opr{S}^2 \rangle$ against FCI over the same range of internuclear distances.

\begin{figure}[h!]
    \begin{minipage}{\columnwidth}
        \includegraphics[width=1\linewidth]{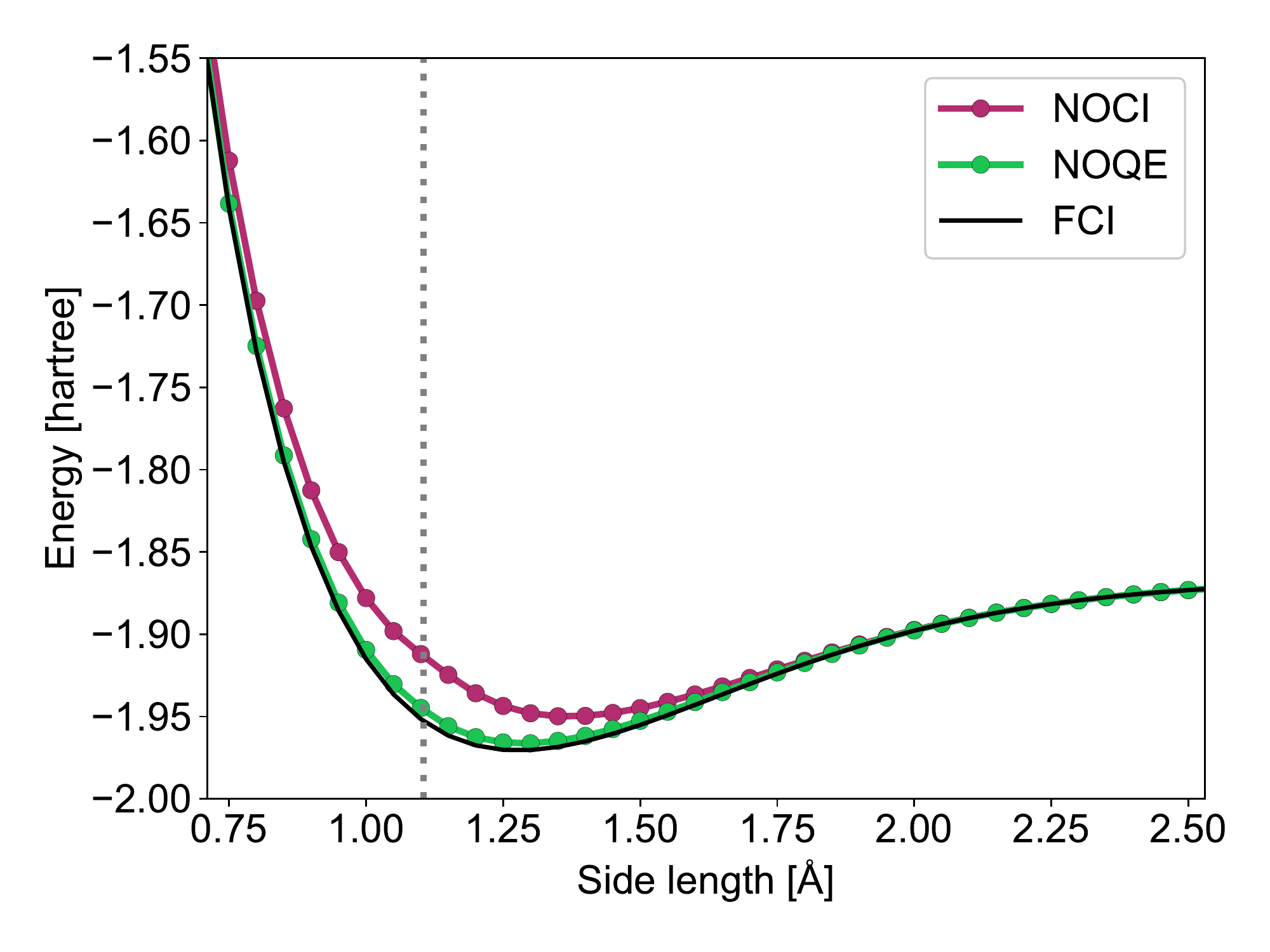}
       \subcaption{Lowest energy singlet (S$_0$) state.}
       \label{fig:h4-sto3g-energies-S0} 
    \end{minipage}
    \begin{minipage}{\columnwidth}
        \includegraphics[width=1\columnwidth]{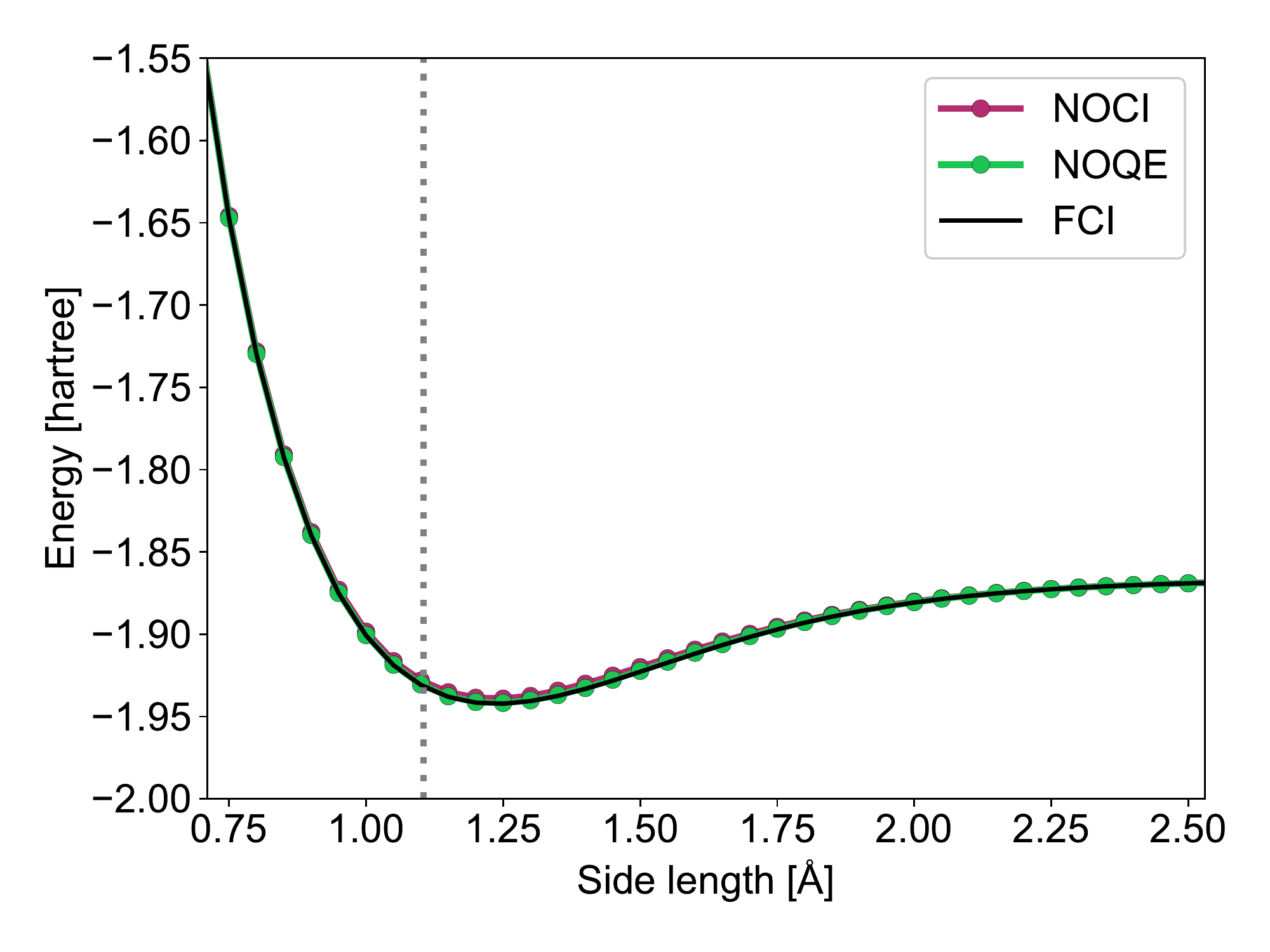}
      \subcaption{Lowest energy triplet (T$_1$) state.}
       \label{fig:h4-sto3g-energies-T1} 
    \end{minipage}
    \caption{Comparison of NOUCC(2) and FCI eigenenergies for \ce{H4}, with the STO-3G basis set. Classical NOCI results (without dynamic correlation) are also provided for comparison. The location of the CF point (1.105 {\AA} for HF/STO-3G) is marked as a dotted gray line. Note that the T$_1$ state only appears beyond the CF point for NOCI and NOUCC(2) calculations.}
   \label{fig:h4-sto3g-pes}
\end{figure}

\subsection{Square \ce{H4} system}

We now consider a system that is considerably more challenging for classical methods, namely the square geometry \ce{H4} species in which all four H atoms are equivalent. While the formation of \ce{H4} from separate \ce{H2} molecules is energetically unfavorable, the square \ce{H4} molecule represents a benchmark system for quantum chemical studies of strong electron correlation. In particular, for short side lengths, square \ce{H4} is a model for more complex antiaromatic molecules such as cyclobutadiene, which possesses a triplet ground state. However, longer side lengths lead to singlet ground states with four strongly correlated electrons. The switch from a triplet ground state to a singlet ground state occurs at $0.82$ {\AA} in the STO-3G basis, although the two states remain fairly close in energy in the neighborhood of the crossover point (with the singlet-triplet gap changing from $-5$ mHa at 0.76 {\AA} side length, to 5 mHa at $0.88$ {\AA}).

\begin{figure}[h!]
    \begin{minipage}{\columnwidth}
       \includegraphics[width=1\linewidth]{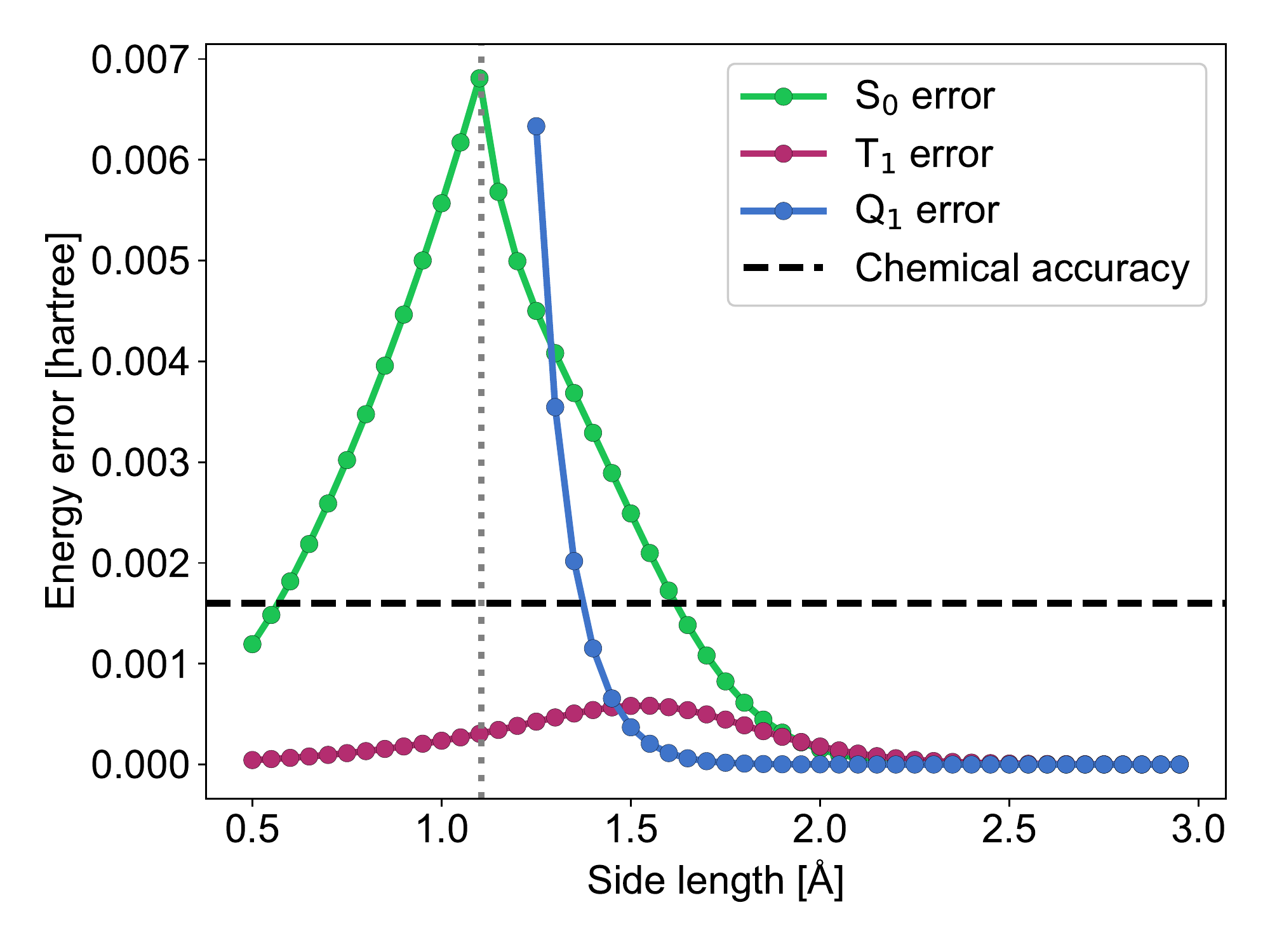}
       \subcaption{Errors in S$_0$, T$_1$ and Q$_1$ NOQE energies with NOUCC(2) ansatz.}
       \label{fig:h4-sto3g-abserrors}
    \end{minipage}
    \begin{minipage}{\columnwidth}
       \includegraphics[width=1\linewidth]{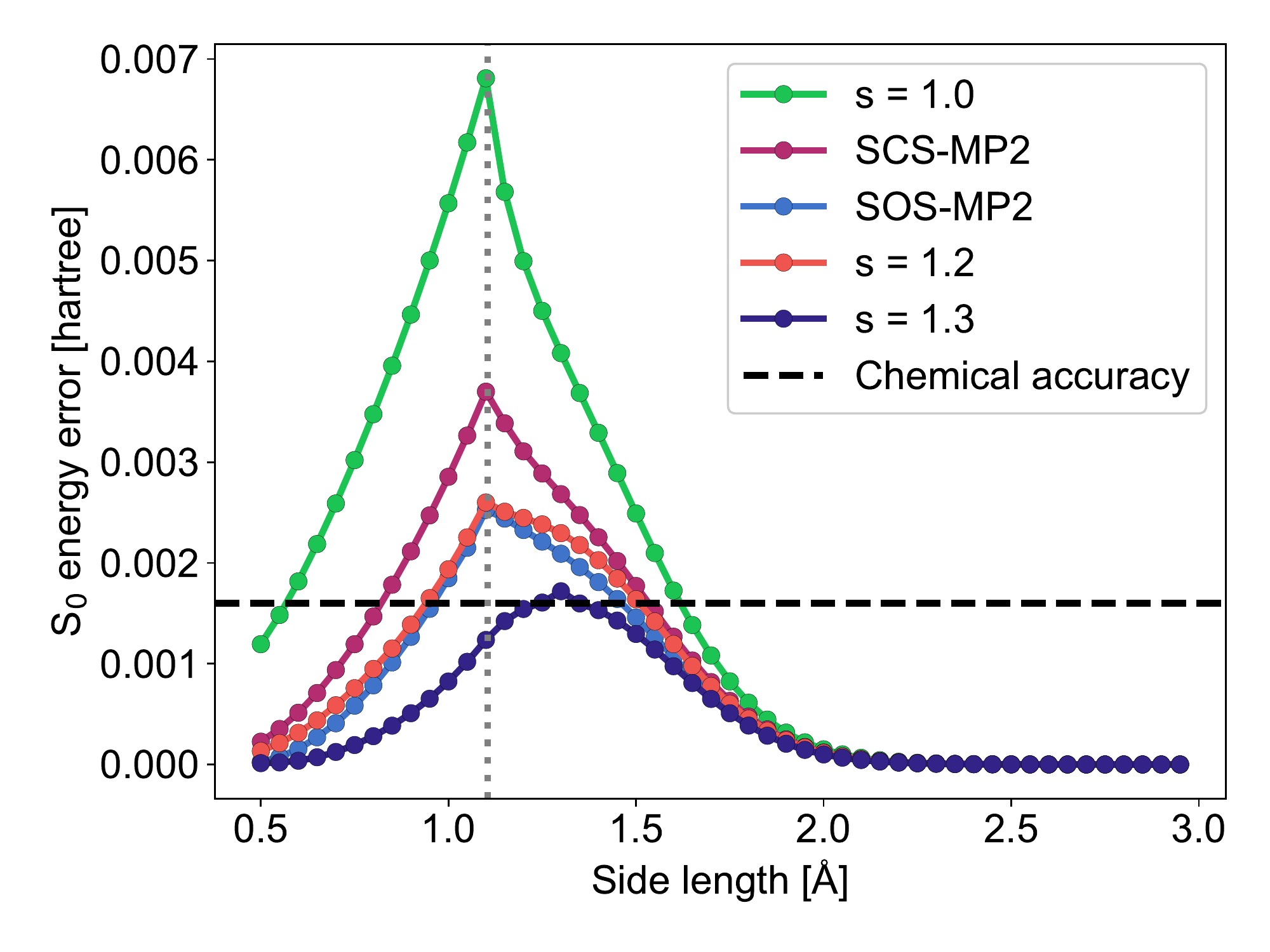}
       \subcaption{NOQE S$_0$ energy errors for the NOUCC(2) ansatz with scaled MP2 $\{t_{abij}\}$.}
       \label{fig:h4-sto3g-scaledabserrors} 
    \end{minipage}
    \caption{Errors in energies relative to FCI for square \ce{H4} from NOQE with the NOUCC(2) ansatz states. The location of the CF point is marked with a dotted gray line. }
    \label{fig:h4-sto3g-energy}
\end{figure}

At the dissociation limit for total spin $m_s=0$, there are four independent H atoms, two of which have up spins and two down spins, leading to six possible arrangements, each of which corresponds to a separate UHF state. These six UHF states are depicted in \fig{num_refs} and correspond to all possible spin arrangements without pairing.  At shorter side lengths, an analogous set of UHF solutions is targeted and constructed using the protocol of \sec{technical}. Unlike the situation for \ce{H2} where the two possible UHF states are always degenerate for all internuclear distances, for \ce{H4} the  six possible UHF states for \ce{H4}, that are shown in \fig{num_refs}, are not degenerate for all values of the side length. The $J=3$ and $J=6$ states form a doubly degenerate set corresponding to the UHF global minimum, while the $J=1,2,4,5$ states form another degenerate set of higher energy. Detailed analysis shows that in the STO-3G basis, for side length less than 1.105 {\AA}, the $J=1,2,4,5$ UHF states collapse to two closed-shell RHF states, reflecting a spin-pairing symmetry analogous to that found at the CF point for \ce{H2}, while for larger distances all six UHF states are linearly independent. The side length of 1.105 {\AA} is not strictly speaking a CF point, because not all determinants show the transition in this case. However, it is a close analogue to this and we shall refer to it as the CF point for \ce{H4}. 

In contrast to the situation for \ce{H2}, the minimum energy point for \ce{H4} now lies well within the 6 determinant, spin-symmetry broken regime, rather than in the spin-paired regime (see \fig{h4-sto3g-pes}). There are at most 6 UHF states to consider, and thus no more than 15 off-diagonal sets of matrix elements need to be made for all values of side length. Since the NOQE calculations remove linear dependencies around the CF point by discarding the overlap matrix singular values of $10^{-4}$ or less, in practice this number can be smaller for some distances.  The number of basis functions per hydrogen atom, spin-orbitals ($N$), and the total number of qubits required to construct the NOQE circuit of \fig{noqe-circuit} for \ce{H4} with the STO-3G basis are listed in \tab{num-qubits-h2}. The corresponding circuit depths employing the low-rank factorization of the doubles tensor operator $\hat{\tau}$ are listed in \tab{circ-depth-h4}.

\begin{center}
\begin{table}[h!]
\begin{tabular}{ |c|c|c|c|c| } 
\hline
Ref. state & Circuit depth & Circuit depth\\
($J$) & (SVD) & (Takagi) \\
\hline
1  & 555 & 371\\ 
2   & 594 & 386\\
3  & 283 & 183\\
\hline
4  & 593 & 399\\
5  & 583 & 390\\
6  & 281 & 183\\
\hline
\end{tabular}
\caption{\label{tab:circ-depth-h4} 
Circuit depth of the (full rank) ansatz preparation unitary operator $\opr{\mathcal{U}}_{J\rightarrow 1}e^{\hat\tau_J}$ with MP2 amplitudes, i.e., UCC-MP2, for the six reference states of H$_4$ shown in \fig{num_refs}, in the STO-3G basis. The $J=3$ and $J=6$ states require lower circuit depths, as the MP2 amplitudes corresponding to correlation between spins of the same sign are zero from symmetry. These ansatz preparations may be performed in parallel, as in \fig{noqe-circuit}, or in sequence, as in \fig{noqe-circuit_moreefficient}.}
\end{table}
\end{center}

\fig{h4-sto3g-pes} shows the NOQE potential energy surface for the $S_0$  and T$_1$ states of square \ce{H4} using the NOUCC(2) ansatz with the STO-3G basis, and \fig{h4-sto3g-abserrors} shows the corresponding energy errors for both states. We see that NOQE and FCI agree very well for these states, while classical NOCI proves to be quite inadequate at shorter side lengths for the S$_0$ state. \fig{h4-sto3g-abserrors} reveals that the energy difference between NOQE and FCI for the S$_0$ state is similar to that observed for \ce{H2}, with a maximum deviation around $7$ mHa near the CF point analogue for H$_4$. However, we note that the T$_1$ state has a very low deviation from FCI, being within chemical accuracy at all side lengths. 

The lowest energy quintet (Q$_1$) state shows very interesting behavior. \fig{h4-sto3g-abserrors}  shows that the error against FCI is quite low at longer bond lengths, but in contrast to the behavior of the S$_0$ and T$_1$ errors, the quintet Q$_1$ state energy error rises dramatically as the CF point is approached. This behavior has a physical interpretation, namely the consequence of increasing spin-contamination as the side length shortens. At side lengths shorter than the CF point at 1.105 {\AA} side length, the NOUCC(2) subspace of reference states provides an inadequate description of the Q$_1$ state, because the former is targeting the low-energy states, while the Q$_1$ state is very high in energy relative to the T$_1$ and S$_0$ states. Indeed, the state with the greatest quintet character deriving from the NOQE calculation with the NOUCC(2) ansatz in this regime is a heavily spin-contaminated state with $\langle \opr{S}^2\rangle \sim 4-5$, which is a poor approximation to the true Q$_1$ state for which $\langle \opr{S}^2\rangle=6$. This heavily spin-contaminated state is best viewed not as a proper eigenstate, but rather as the residue left in the NOQE subspace after the lower energy S$_0$ and T$_1$ states have been solved for, with the high degree of spin-contamination being a mark of its poor quality. However, when the side length is stretched, the Q$_1$ energy is lowered and the state is then better captured by the NOUCC(2) ansatz. Thus, for side length greater than 1.25 {\AA}, the absolute spin-contamination in the highest energy NOQE state is reduced to 0.05 or less, i.e., $\langle \opr{S}^2\rangle \ge 5.95$. The state can be reasonably labeled as Q$_1$ from that point on, but not at shorter distances due to greater spin-contamination and therefore we do not plot Q$_1$ at the shorter bond-lengths in \fig{h4-sto3g-abserrors} or elsewhere. This example indicates that the extent of spin-contamination provides a useful internal check on the accuracy of the NOQE energies.

The energy errors in the S$_0$ state can be further reduced by scaling the MP2 amplitudes, as shown in  \fig{h4-sto3g-scaledabserrors}. Since both same-spin and opposite-spin MP2 amplitudes are present for \ce{H4} with $m_s = 0$, we investigate the behavior of both the SCS- and SOS-MP2 models, as well as their uniform-scaled analogue. \fig{h4-sto3g-scaledabserrors} shows that uniform scaling of the MP2 amplitudes yields the lowest error, with a value of $s=1.3$ bringing the NOQE ground state singlet S$_0$ energy error below chemical accuracy for all internuclear distances. We do not separately examine the behavior of the singlet-triplet gap, as the much lower error in the T$_1$ energy for this basis set (as shown in \fig{h4-sto3g-abserrors}) indicates that it would look very similar to the S$_0$ energy error plot.

\begin{figure}[h!]
    \begin{minipage}{\columnwidth}
       \includegraphics[width=1\linewidth]{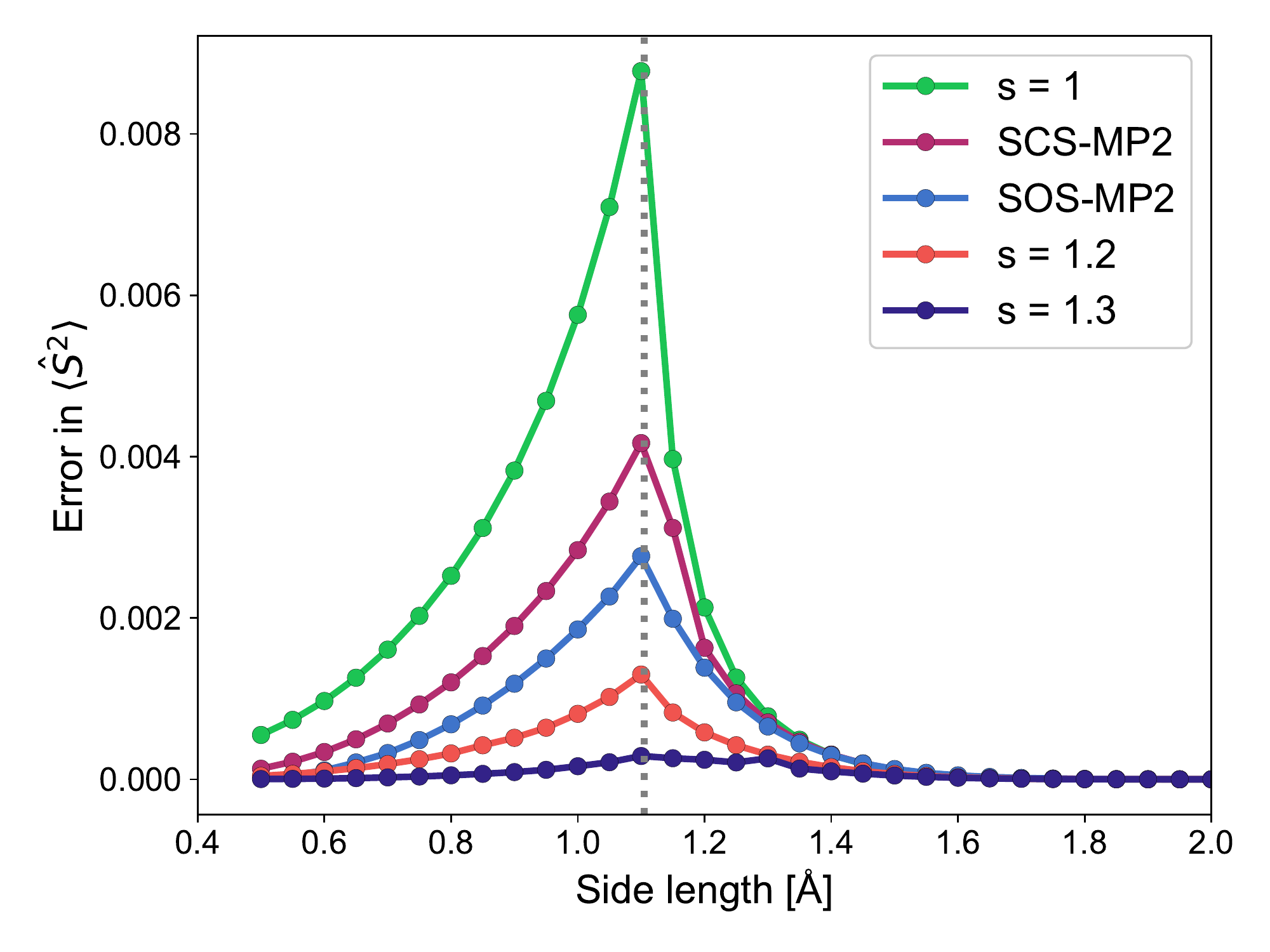}
       \subcaption{Error in $\langle \opr{S}^2 \rangle$ for S$_0$ state.}
       \label{fig:h4-sto3g-ssqerrors}
    \end{minipage}
    \begin{minipage}{\columnwidth}
       \includegraphics[width=1\linewidth]{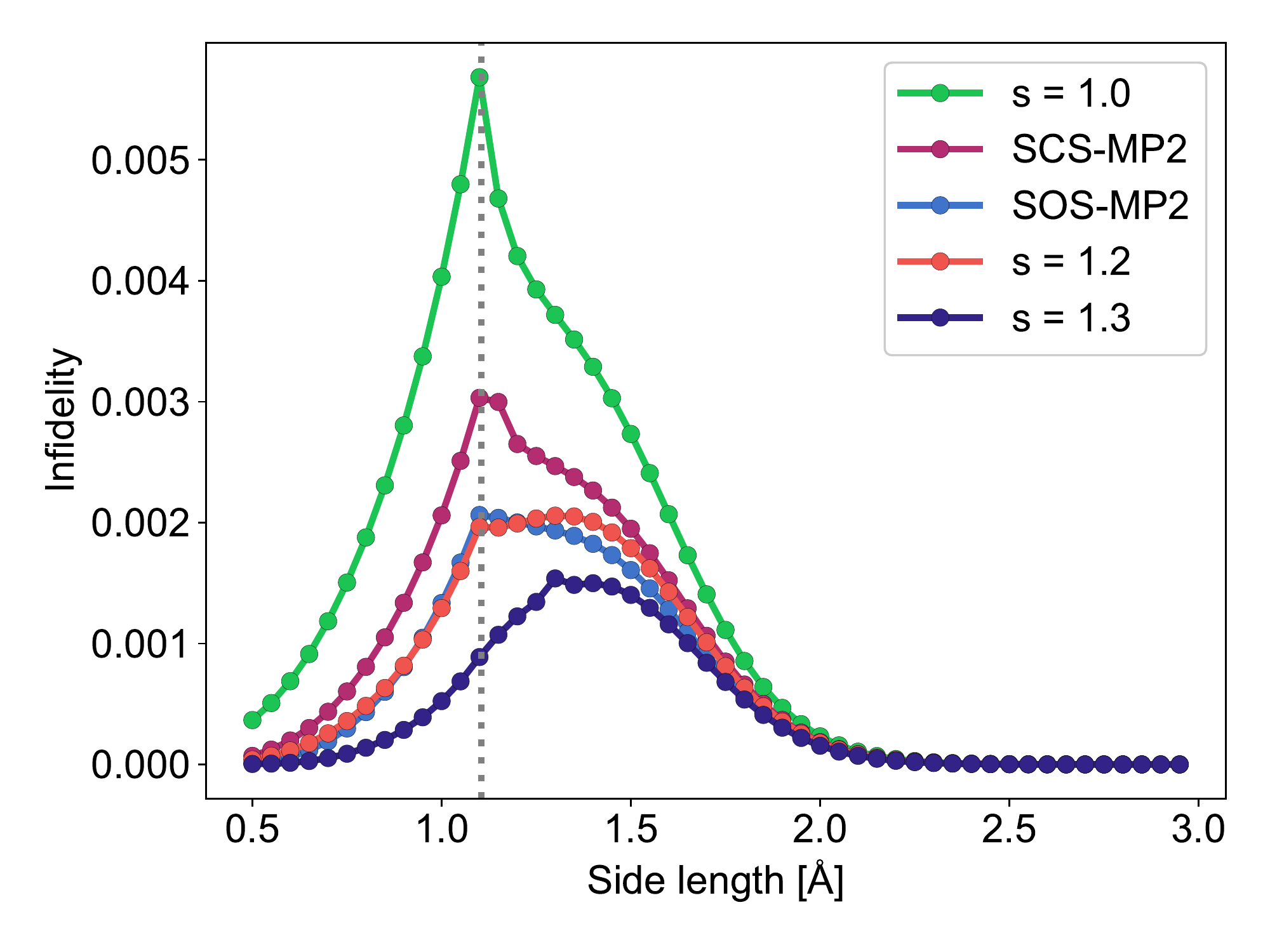}
       \subcaption{Infidelity in S$_0$ state.}
       \label{fig:h4-sto3g-infidelity} 
    \end{minipage}
    \caption{Spin contamination error and state infidelity ($I_\text{FCI}$, \eq{Infidelity_FCI}) relative to the FCI states of the amplitude scaled NOQE states derived from the NOUCC(2) ansatz, for \ce{H4} with the STO-3G basis. The location of the CF point is marked with a dotted gray line.}
    \label{fig:h4-sto3g-noten}
\end{figure}

We also evaluate the extent of spin-contamination in the S$_0$ state, shown in \fig{h4-sto3g-ssqerrors}. $\langle \opr{S}^2 \rangle$ is below $0.01$ with even the unscaled approach, indicating quite low error vs. the exact value of $\langle \opr{S}^2\rangle=0$. Scaling reduces the error further. The error in $\langle \opr{S}^2\rangle$ for the  T$_1$ state (not shown) is considerably lower (maximum deviation of $5\times10^{-6}$), showing that the low energy S$_0$ and T$_1$ states are modeled in a nearly spin-pure fashion by NOQE with the NOUCC(2) ansatz. On the other hand, the higher energy Q$_1$ state shows greater levels of spin-contamination, having $\langle \opr{S}^2\rangle < 5.94$ (in the unscaled case) at distances shorter than 1.25 {\AA} against the exact value of $\langle \opr{S}^2\rangle =6$. Indeed, $\langle \opr{S}^2\rangle$ for the closest analog to the Q$_1$ state reaches $4-5$ at side lengths shorter than the CF point, as previously noted. Interestingly, the SCS- and SOS-MP2 amplitudes worsen spin-contamination for the Q$_1$ state at shorter side lengths (having error $>0.01$ at 1.45 {\AA}), as compared to unscaled or uniformly scaled results (which have $\sim 0.006$ error at by 1.45 {\AA}).

We note that while spin-purity is a necessary measure of the quality of the final NOQE states, it is not a sufficient one. \fig{h4-sto3g-infidelity} shows the infidelity of the S$_0$ state for \ce{H4}, which reveals that NOQE with unscaled MP2 amplitudes in the NOUCC(2) reference states attains over 99\% fidelity at all values of the side length of the molecule, with the scaled amplitude ansatz states showing even greater overlaps with the FCI eigenstate. Lower infidelity values are obtained for the T$_1$ state, being $0.08\%$ or less in all cases. Even the Q$_1$ state shows infidelities $\sim 1\%$ or less for side lengths longer than 1.45 {\AA} with the unscaled and uniformly scaled ans{\"a}tze, although SCS- and SOS-MP2 amplitudes lead to greater infidelity (reaching $\sim 2\%$) at 1.45 {\AA}.

\begin{figure}[h!]
       \includegraphics[width=1\columnwidth]{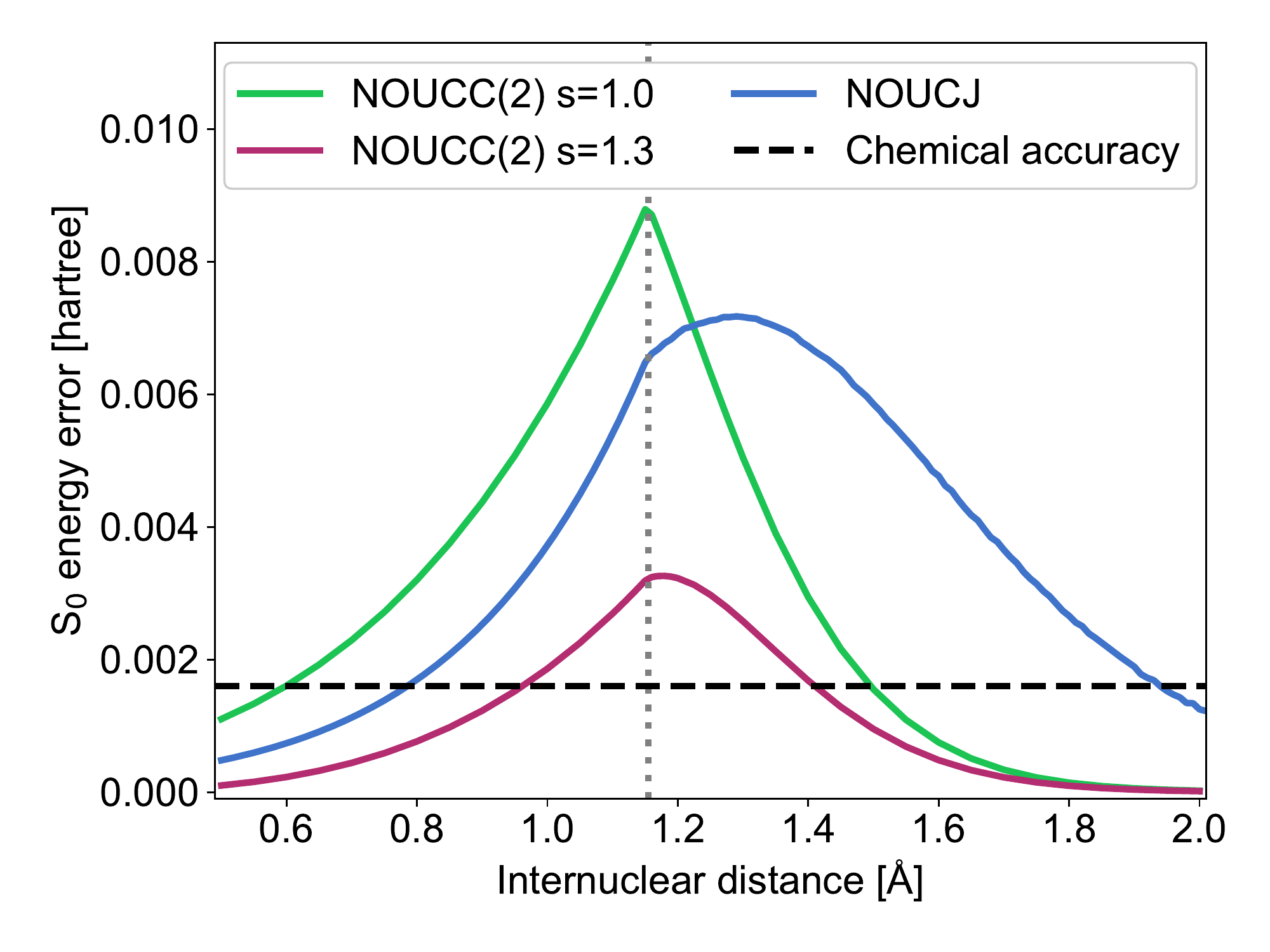}
    \caption{Comparison of S$_0$ energy errors with NOUCC(2) and NOUCJ ($L=1$) against FCI for \ce{H2}/STO-3G. The location of the CF point (1.155 {\AA} for \ce{H2}/STO-3G) is marked with a dotted gray line.}
    \label{fig:h2-sto3g-jastrow}
\end{figure}

\subsection{NOUCJ for \ce{H2}/STO-3G}

We also undertook a preliminary investigation of NOQE with the NOUCJ ($L=1$) ansatz (for the remainder of this section the $L=1$ label will be suppressed) by examining its performance for \ce{H2}/STO-3G. A comparison of the errors in the S$_0$ state (vs. FCI) for the NOUCJ and NOUCC(2) ans{\"a}tze is shown in \fig{h2-sto3g-jastrow}. The T$_1$ state in this basis has zero correlation and is thus not considered. It appears that the maximum energy error is lower for NOQE with the NOUCJ ansatz than the unscaled NOUCC(2) ansatz, although scaling NOUCC(2) amplitudes by 1.3 leads to even better performance. However, the error in the S$_0$ NOUCJ energy decays rather slowly with distance and remains above chemical accuracy until $\sim $ 2 {\AA}. In contrast, NOQE with NOUCC(2) shows a rapid decrease in error with increasing side length, going below the chemical accuracy threshold $\sim 1.5$ {\AA}. That said, increasing $L$ beyond 1 (up to $N^2$) in the NOUCJ ansatz offers a route toward systematic improvement.  As will be discussed further in \sec{circuit-resource-estimation}, the lower resource cost associated with the NOUCJ ansatz (particularly for the choice of $L=1$) makes it very appealing for use in the NOQE framework, meriting further investigation.

\subsection{Low-rank truncation for H$_2$ and H$_4$}

\begin{figure}[h!]
    \begin{minipage}{\columnwidth}
       \includegraphics[width=1\linewidth]{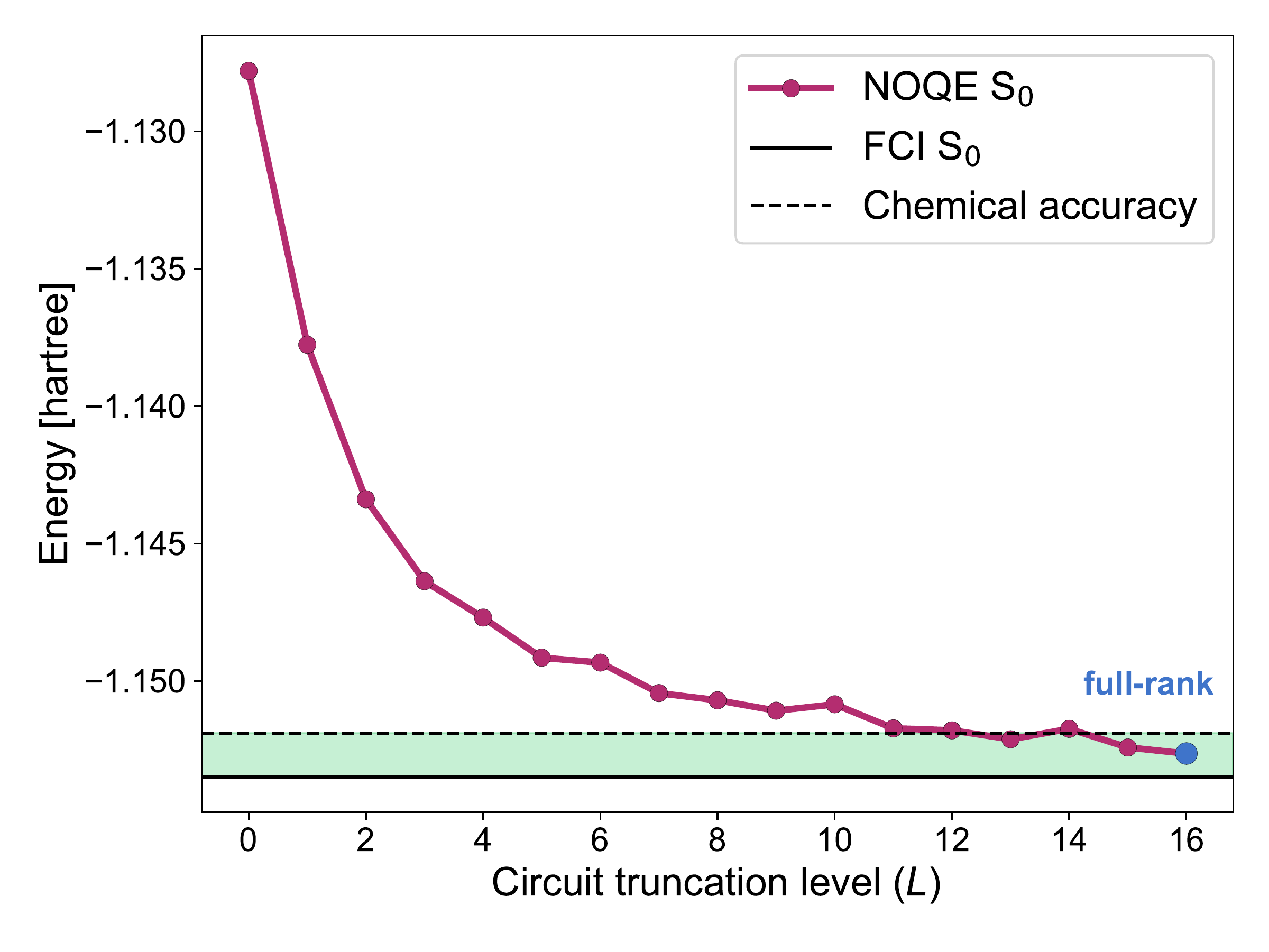}
       \subcaption{NOQE energy vs. $L$ at 0.75~\AA,~the equilibrium bond distance of H$_2$.}
       \label{fig:h2-0p75-ranktruncationerrors}
    \end{minipage}
    \begin{minipage}{\columnwidth}
       \includegraphics[width=1\linewidth]{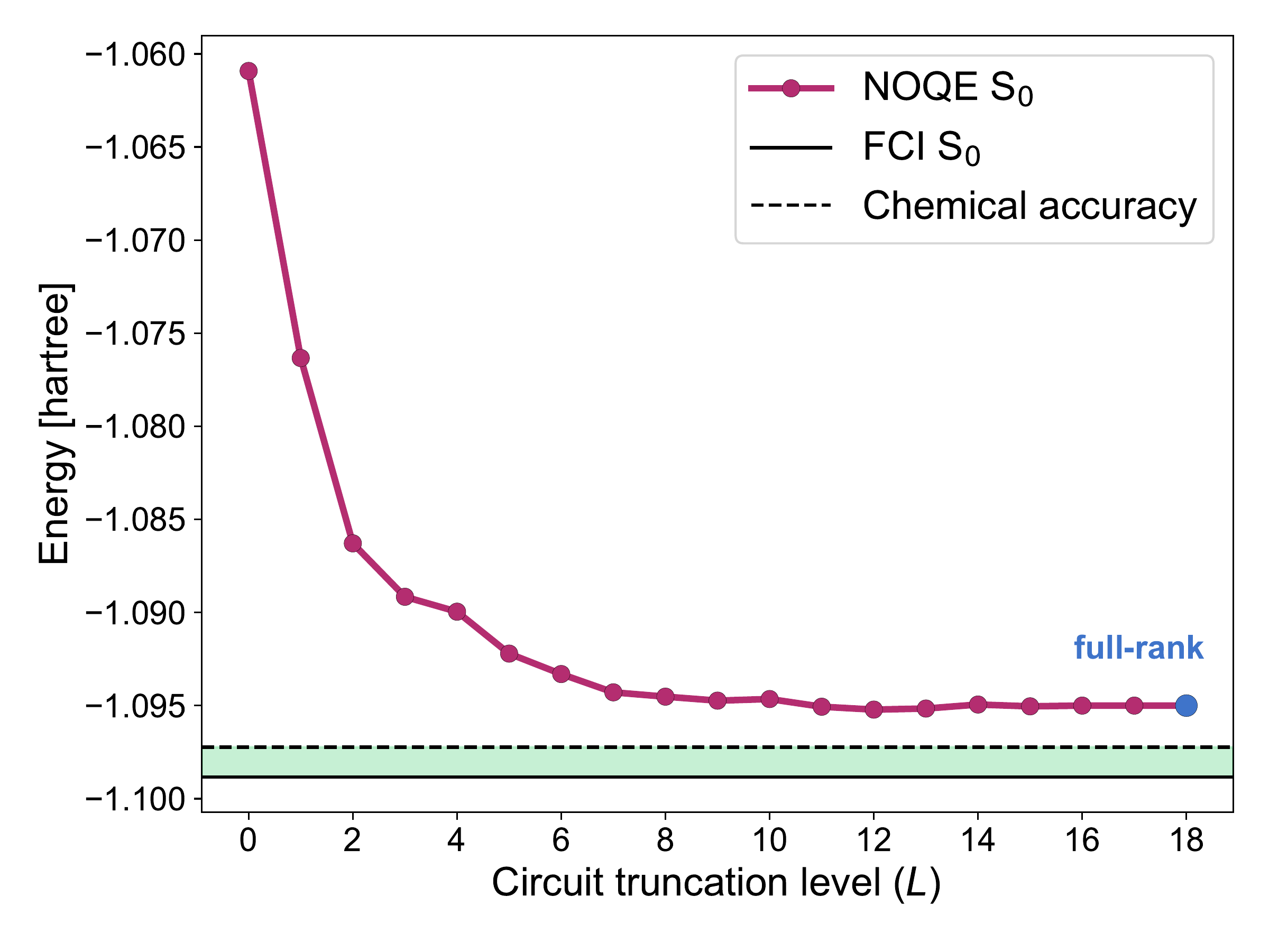}
       \subcaption{NOQE energy vs. $L$ at 1.2~\AA~(the Coulson-Fischer point of H$_2$).}
       \label{fig:h2-1p2-ranktruncationerrors} 
    \end{minipage}
    \caption{NOQE ground state (S$_0$) energy dependence on truncation level $L$ of the cluster tensor, representing the circuit truncation level, for H$_2$ with $s = 1.3$ in the 6-311G basis set at the equilibrium bond distance (panel \hyperref[fig:h2-0p75-ranktruncationerrors]{(a)}) and at the Coulson-Fischer point (panel \hyperref[fig:h2-1p2-ranktruncationerrors]{(b)}). The blue point labelled ``full-rank" shows the value $L_{\text{eff}}$. The green shaded region corresponds to the range of energy that is within chemical accuracy of the FCI result.
    }
    \label{fig:h2-ranktruncationerrors}
\end{figure}

\begin{figure*}[htb!]
    \begin{minipage}[t]{\columnwidth}
       \includegraphics[width=1\linewidth]{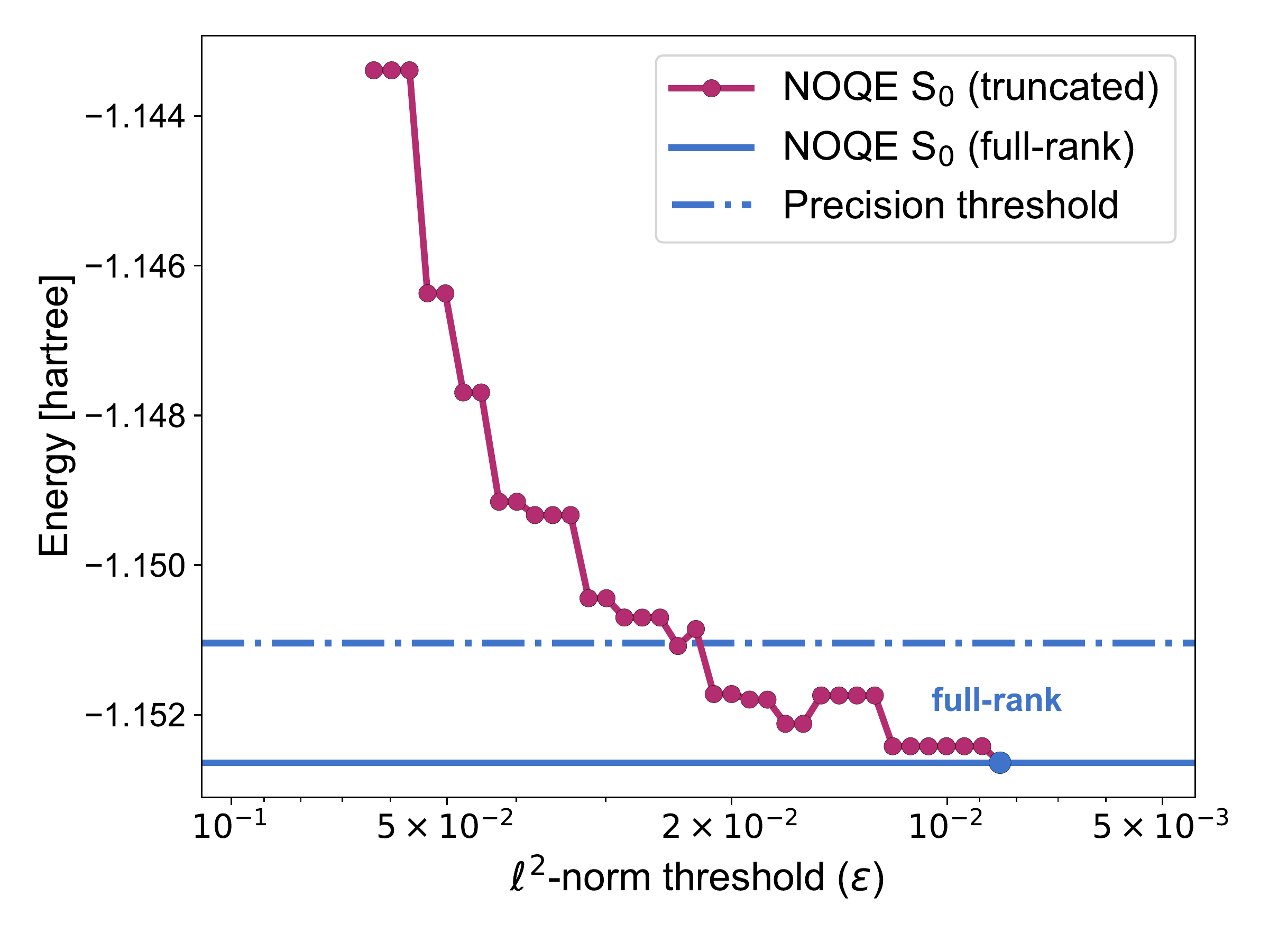}
       \subcaption{NOQE energy vs. $\varepsilon$ at 0.75~\AA~(the equilibrium bond distance of H$_2$).}
       \label{fig:h2-0p75-truncationerrors}
       \includegraphics[width=1\linewidth]{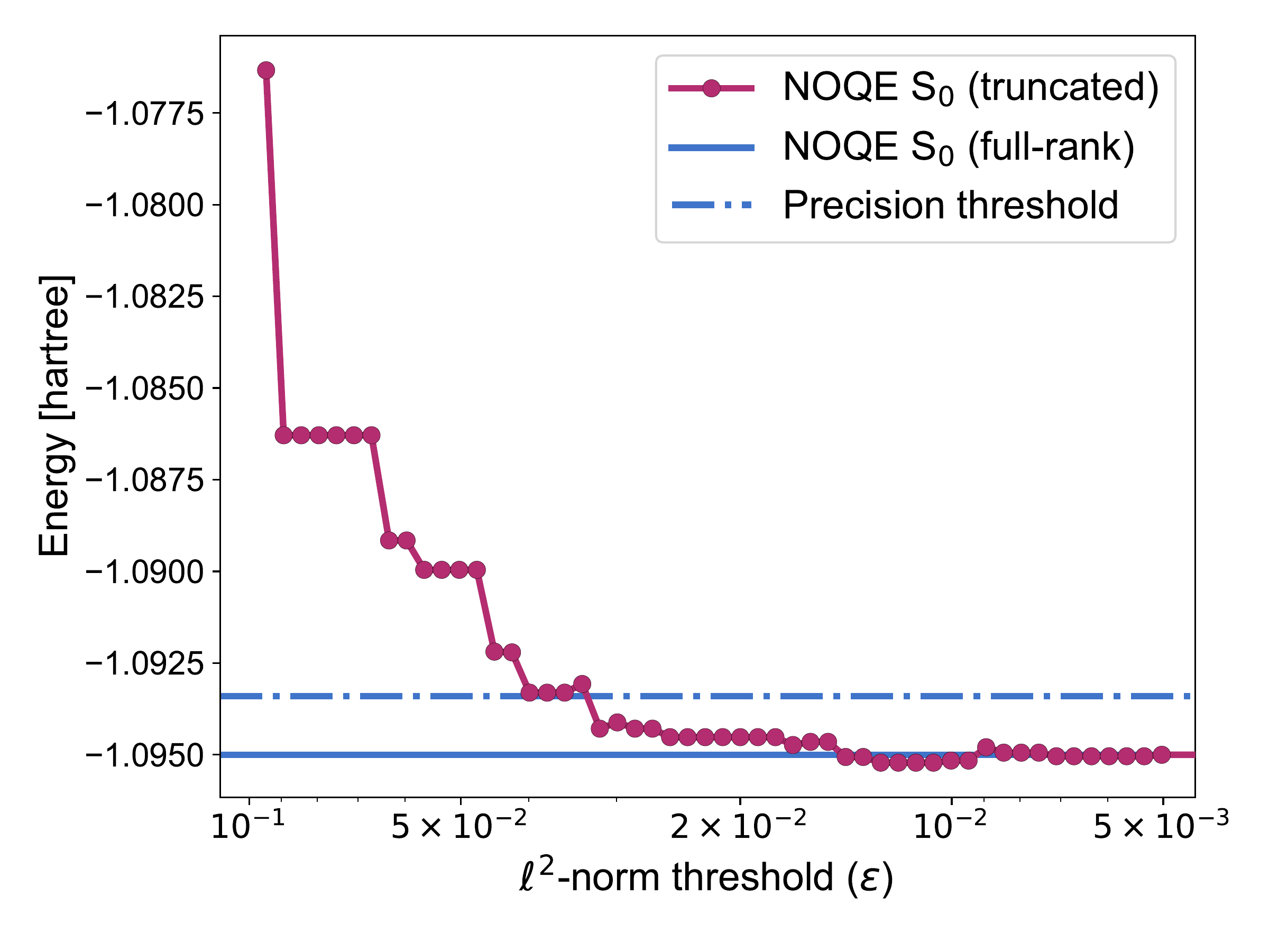}
       \subcaption{NOQE energy vs. $\varepsilon$ at 1.2~\AA~(the Coulson-Fischer point of H$_2$).  The effects of truncation of the two smallest singular values ($\sigma_i \leq10^{-4}$) give rise to energy changes less than $1 \times 10^{-6}$ Ha and are not shown in the plot.
       }
       \label{fig:h2-1p2-truncationerrors} 
    \end{minipage}
    \begin{minipage}[t]{\columnwidth}
       \includegraphics[width=1\linewidth]{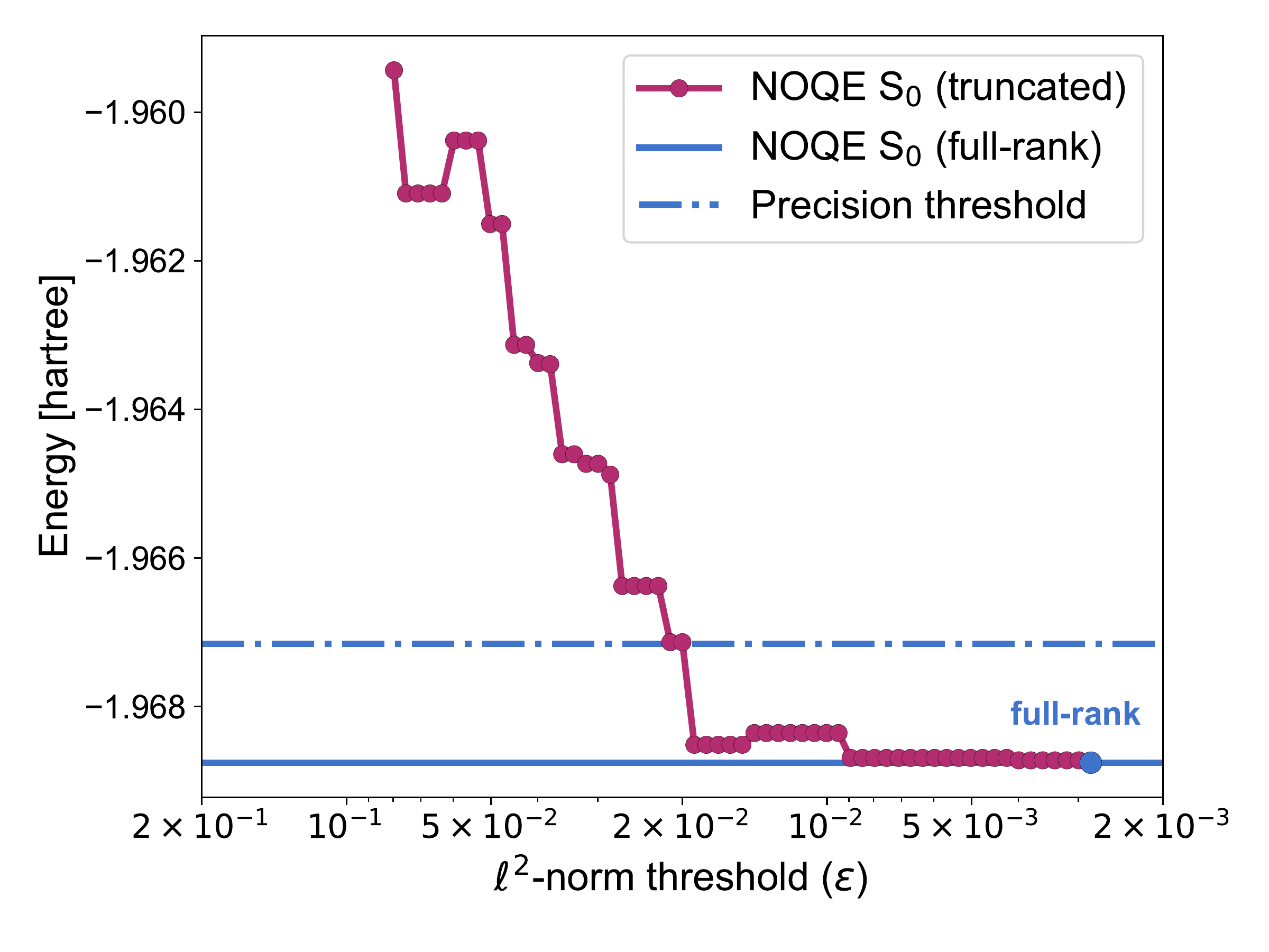}
       \subcaption{NOQE energy vs. $\varepsilon$ at 1.3~\AA~side length~(the equilibrium bond length for square H$_4$).}
       \label{fig:h4-1p3-truncationerrors}
       \includegraphics[width=1\linewidth]{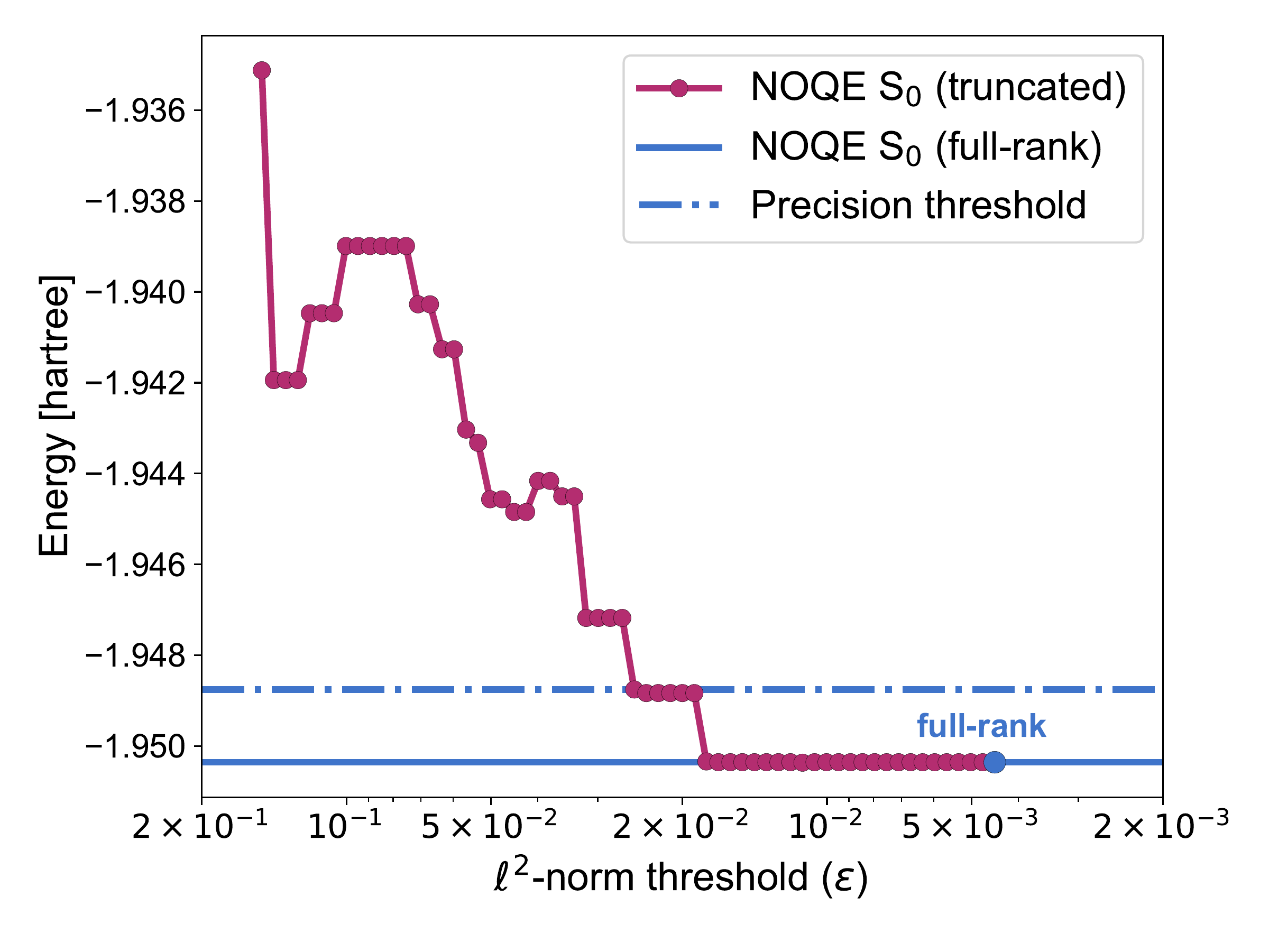}
       \subcaption{NOQE energy vs. $\varepsilon$ at 1.1~\AA~side length~(the Coulson-Fischer point for square H$_4$).}
       \label{fig:h4-1p1-truncationerrors} 
    \end{minipage}
    \caption{NOQE ground state (S$_0$) energy dependence on $\ell^2$-norm threshold level $\varepsilon$, for H$_2$/6-311G and H$_4$/STO-3G with $s = 1.3$ at the equilibrium geometry and at the Coulson-Fischer point. The blue points labelled ``full-rank" and the solid blue lines show the values with untruncated circuits. Singular values $\sigma_i$ of the cluster tensor that are removed by the $\ell^2$-norm truncation are set to zero and the corresponding circuit blocks are then omitted, leading to savings in the circuit depth. 
    }
    \label{fig:h2h4-truncationerrors}
\end{figure*}

The low-rank tensor decomposition of the cluster tensor Eqs.~(\ref{eq:ucc1_theory})-(\ref{eq:ucc2_theory}) that is given by \eq{sum_of_squares} (see \app{details-low-rank} for details) allows for a systematic reduction of the NOQE circuit. For single reference states, this can be done straightforwardly by truncation of the number of singular values $L$ of the cluster tensor $\matr{T}$ to reduce its effective rank (the analogous truncation of $\rho_L$ has been found to be less effective for the UCC operator~\cite{motta2021low}). We note that the error bounds presented in \app{error-estimation} may be used to guarantee a maximum NOQE energy error for a given choice of truncation. However, this is in practice a loose bound which overestimates the observed energy deviation when many singular values are discarded. While the $L$ truncation strategy is not directly applicable to general multi-reference states, because each state may have a different truncation value, for H$_2$ the cluster operators for the two non-orthogonal reference states are equivalent by spin symmetry. In this situation, a single value of $L$ can then be used for both reference states to analyze the effect of truncation on the NOQE energy.

\fig{h2-ranktruncationerrors} shows, for H$_2$ with NOUCC(2), the dependence of the NOQE ground state energy on the choice of truncation parameter $L$, using the Takagi decomposition of the cluster supermatrix $\matr{T}$, at the equilibrium internuclear distance (panel \hyperref[fig:h2-0p75-ranktruncationerrors]{(a)}) and at the CF point (panel \hyperref[fig:h2-1p2-ranktruncationerrors]{(b)}). The full-rank value is defined as $L_{\text{eff}} \leq N_{\text{occ}}N_{\text{virt}}= \eta(N - \eta)$, which is the residual number of singular values of $\matr{T}$ retained when all zero and near-zero singular values up to a certain fixed precision threshold (taken here as $10^{-12}$) are discarded, i.e., the effective full rank of $\matr{T}$. \fig{h2-0p75-ranktruncationerrors} demonstrates that, at the equilibrium bond length of 0.75~\AA, we can truncate the circuit by setting $L=L_\text{trunc}=11$ instead of $L=L_{\text{eff}}=16$, and thus reduce the circuit depth, while remaining within chemical accuracy of the FCI result. \fig{h2-1p2-ranktruncationerrors} shows the relatively more advantageous behavior of rank truncation when the bond is stretched to 1.2~\AA~(the CF point).  Although the full-rank NOQE energy lies slightly outside chemical accuracy, we find that the energy at truncated values of $L$ converges more quickly to the full-rank result.  This is because the longer bond distance enables the occupied and virtual orbitals to be more localized and reduces their mutual overlap, effectively lowering the rank of the $\matr{T}$ matrix.

For other multi-reference states, such as those of H$_4$, we must use an alternative truncation strategy to ensure consistency across the cluster tensor decompositions derived within UHF single-particle bases of different symmetry, which may have different $L_\text{eff}$. Since we are interested in the efficacy of the truncation procedure, we aim to find the minimum value of $L$, defined as $L_\text{trunc}$, for each single reference state, such that truncating all cluster operators to their individual $L_\text{trunc}$ values will yield an energy estimate within 1.6 mHa precision of the energy obtained using the full-rank cluster operators. Possible approaches include the use of a fixed tolerance threshold for the singular values of the cluster operator, or, alternatively, a norm-based criterion as in Ref.~\cite{motta2019efficient}. Here we employ a truncation strategy based on the vector $\ell^p$-norms.

We compute the vector of ordered singular values $\vec\sigma = \text{diag}(\matr{\Sigma})$, where $\matr{\Sigma}$ is the diagonal matrix of singular values that results from SVD or Takagi decomposition of $\matr{T}$ (see \app{details-low-rank}), and discard the largest subset of $L_\text{eff}-L$ singular values $\sigma_l$ satisfying 
\begin{equation} \label{eq:truncate_L}
\bigg(\sum_{l=L+1}^{L_\text{eff}}|\sigma_l|^p \bigg)^{1/p} \leq \varepsilon_p,
\end{equation}
with $\varepsilon_p$ a variable threshold. Here $\varepsilon_p$ is the independent variable and $L$ is the dependent variable to be determined. In this work, we use $p=2$ and define $\varepsilon \equiv \varepsilon_2$. We first establish the values of $L_{\text{eff}}$ for each of the reference states in the multi-reference ansatz, and then perform a sweep over $\varepsilon$, truncating the vectors of singular values of each reference state according to ~\eq{truncate_L}, until we find the largest value of $\varepsilon=\varepsilon_\text{max}$ that retains a desired level of accuracy in the multi-reference energy relative to the full-rank result. This sets the value of $L_\text{trunc}$ for each of the reference states, from which the fractional reduction in circuit depth can then be estimated (see \sec{circuit-resource-estimation}).

The left panels of \fig{h2h4-truncationerrors} show the NOQE ground state energy dependence on the $\ell^2$-norm threshold $\varepsilon$ for H$_2$/6-311G at the equilibrium bond distance (panel~\hyperref[fig:h2-0p75-truncationerrors]{(a)}) and the CF point (panel~\hyperref[fig:h2-1p2-truncationerrors]{(b)}). The right panels of \fig{h2h4-truncationerrors} show the corresponding plots for square H$_4$/STO-3G at the equilibrium geometry (side length 1.3~\AA, panel~\hyperref[fig:h4-1p3-truncationerrors]{(c)}) and the CF point (side length 1.1~\AA, panel~\hyperref[fig:h4-1p1-truncationerrors]{(d)}). For H$_4$ we can divide the reference states into two groups depending on the spatial distributions of the spins, as shown in \fig{num_refs}. There are four configurations ($J=1,2,4,5$) in which the two spins of the same sign are on the same edge of the square and two ($J=3,6$) where the spins of the same sign are diagonally opposite each other on the square. For each group we calculate the fractional circuit depth reductions after truncation, $L_\text{trunc}/L_\text{eff}$, which we report in \tab{h4-ltrunc-leff}. We also compute the total fractional reduction in the number of singular values over all references, $(\sum_{J=1}^M L_\text{trunc}^{(J)})/(\sum_{J=1}^M L_\text{eff}^{(J)})$, which we refer to here as the total cost reduction. We place emphasis on the single-reference circuit depth reductions since they are directly related to the requisite coherence times of qubits on the quantum processor. However, it should be noted that the total cost reduction estimates are not directly related to the coherence time, because the circuits for different matrix elements can be run independently or in parallel on different sets of qubits. These estimates can instead be related to the total gate count (\sec{circuit-resource-estimation}). The total cost reduction estimates are thus complementary to the single-reference circuit depths in discussing the overall effectiveness of rank truncation for a given NOQE system. 

We observe a total cost reduction relative to the full-rank calculation of 18.8\% and 15.6\% at the equilibrium and the CF point, respectively~(\tab{h2h4-ltrunc-leff}). We note that different truncation methods (e.g., using a different $\ell^p$-norm) result in different estimates of $L_{\text{trunc}}/L_{\text{eff}}$
and thus different percentages of total circuit cost reduction. 

\begin{center}
\begin{table}[h!]
\begin{subtable}{\columnwidth}
\begin{tabular}{ |c|c|c|c|c| }
\hline
Internuclear & $\ell^2$-norm & $L_{\text{trunc}}/L_{\text{eff}}$ & Total cost \\
distance & threshold & ($J=1,2$) & reduction \\
\hline
0.75~\AA & $2.4 \times 10^{-2}$ & 9/16 & 43.8\% \\ 
1.2~\AA  & $2.8 \times 10^{-2}$ & 7/18 & 61.1\% \\
\hline
\end{tabular}
\caption{
Calculations for H$_2$/6-311G ($s= 1.3$) at the equilibrium (0.75~\AA) and the Coulson-Fischer point (1.2~\AA).}
\label{tab:h2-ltrunc-leff}
\end{subtable}
\newline
\vspace*{\baselineskip}
\newline
\begin{subtable}{\columnwidth}
\begin{tabular}{ |c|c|c|c|c| } 
\hline
Side & $\ell^2$-norm & $L_{\text{trunc}}/L_{\text{eff}}$ & $L_{\text{trunc}}/L_{\text{eff}}$ & Total cost \\
length & threshold & ($J=1,2,4,5$) & ($J=3,6$) & reduction \\
\hline
1.3~\AA & $1.8 \times 10^{-2}$ & 10/12 & 6/8 & 18.8\% \\ 
1.1~\AA & $2.4 \times 10^{-2}$ & 11/12 & 5/8 & 15.6\% \\
\hline
\end{tabular}
\caption{ \label{tab:h4-ltrunc-leff}
Calculations for H$_4$/STO-3G ($s= 1.3$) at the equilibrium side length (1.3~\AA) and the Coulson-Fischer point (1.1~\AA). $J=1,2,4,5$ correspond to the `edge' spin configurations, $J=3,6$ to the `diagonal' spin configurations.
}
\end{subtable}
\caption{Calculations of total circuit cost reduction for H$_2$ and H$_4$. The $\ell^2$-norm threshold $\varepsilon_{\text{max}}$ is selected for each system, so that any $\varepsilon \leq \varepsilon_{\text{max}}$ will yield truncated energies that are within 1.6 mHa precision from the full-rank result.
}
\label{tab:h2h4-ltrunc-leff}
\end{table}
\end{center}

Comparing the overall circuit reduction estimates for H$_2$/6-311G ($N=12$) and H$_4$/STO-3G ($N=8)$, we see that the effectiveness of the low-rank truncation is larger for the greater basis set size,  in agreement with the conclusions of norm-based truncation in Ref.~\cite{motta2021low}. 
These results for H$_2$ and H$_4$ also suggest that the circuit reduction is more significant at larger distances, regardless of the relative location of the equilibrium and CF points. In general, the utility of low-rank decompositions is most apparent in cases where there are a large number of very small singular values. Thus, we could expect much greater reductions in circuit depth than those found here, when decomposing MP2 amplitudes corresponding to spatially localized occupied and virtual orbitals that are well-separated (or more generally, nearly orthogonal), as is the case in long, saturated hydrocarbons. 

\section{Circuit resource estimation}\label{sec:circuit-resource-estimation} 

In this section, we evaluate quantum resource estimates for the low-rank factorized NOUCC(2) algorithm as a function of (i) system size (taken as the number of spin orbitals $N$), and (ii) the number of radical sites $d$ that are directly involved in the strong static electron correlation. 
Specifically, we are analyzing the circuit in \fig{noqe-circuit}, with the small modification that the basis rotation is relative, i.e., NO reference $J$ is always rotated into $I$, $\opr{\mathcal{U}}_{J \rightarrow I}$, instead of a common global basis. 
Both approaches are equivalent, but the relative rotation reduces the gate complexity by a constant factor for each off-diagonal matrix element evaluated.

The number of NO reference states in this work is denoted as $M$. 
This scales binomially with the number of radical states $d$ and the number of electrons $\eta$ within the total $m_s=0$ subspace, according to \eq{NOQE_refs_paired} and \eq{NOQE_refs_general}. 
For example, for H$_2$ we have $M = \binom{2}{1} = 2$ reference states while for H$_4$ we have $M = \binom{4}{2} = 6$. 
Recall that it was found to be advantageous (in terms of maximizing error cancellation in spin gap quantities) when, for example, in the case of H$_4$, singlet, triplet, and quintet NOQE eigenstates are taken to be linear combinations of determinants in the $m_S = 0$ spin sector.  
The compute cost associated with classical diagonalization of an $M\times M$ matrix is therefore exponential with $d$ (as also for classical NOCI), but for many difficult molecular applications of interest, we can expect the number of radical sites $d$ to be less than $10$. 
There will be $M$ diagonal matrix elements of the NOQE Hamiltonian, $\matr{H}$, and $M(M-1)/2$ upper triangular matrix elements (not including the diagonals). 
One matrix element, $H_{IJ}$, requires two controlled $N$ by $N$ swap gates, two unitaries $e^{\hat\tau_I}$ and $e^{\hat\tau_J}$ to prepare the two NO reference states, and one additional unitary basis rotation $\opr{\mathcal{U}}_{J\rightarrow I}$ (we ignore the two Hadamard gates on the ancilla qubit).

When running quantum algorithms on NISQ devices, one critical quantity of interest is the number of two-qubit gates required, typically the CNOT gate count. 
We use the fact that a two-qubit Givens rotation, a paired number-operator rotation ($e^{-i\theta \opr{n}_p \opr{n}_q}$ with $p \neq q$), and a CSWAP gate require 4, 2, and 8 CNOT gates, respectively. 
We note that the number-operator pair rotations would incur additional CNOT gates on a linear architecture in the form of a SWAP network to connect the non-neighboring qubit pairs~\cite{kivlichan2018quantum,ogorman2019generalized}. 
However, in the current analysis, for simplicity, we shall assume full connectivity on the device.

Each $\opr{\mathcal{U}}_B$ basis rotation can be implemented with $2\binom{N/2}{2}$ Givens rotations, accounting for unrestricted orbitals. 
For the $I$-th UCCMP2 reference state, the number of basis rotation operators is $1 + m L$ (where $L \leq \text{rank}(\matr{T})$, and the number of $\opr{Y}^2$ terms per singular value is $m=4$ for SVD or $m=2$ for Takagi, for a total of $mL$ circuit blocks). 
Alternatively, with the UCJ approach in \eq{JastrowAnsatz}, only two basis rotations are needed (equivalent to $L=1$). 
In each circuit block there are $\binom{N}{2} = N(N-1)/2$ distinct number operator pair products (excluding diagonal terms). 
The controlled $N$ by $N$ qubit swap will require $N$ pairs of CSWAP gates (8 CNOTs each). 
Therefore the total number of CNOT gates can be decomposed into the following CSWAP, Givens, and number operator terms:
\begin{align}
    N_{\text{CNOT}}^{\text{Swap}} &= 8 \times 2 \times 2N = 32N \\
    N_{\text{CNOT}}^{\text{Givens}} &= 4\times k \left( 2 \times 2\binom{N/2}{2}\times(1+m L) + 2\binom{N/2}{2}\right) \label{eq:givens-cost} \\
    N_{\text{CNOT}}^{\opr{n}_p \opr{n}_q} &= 2 \times k\binom{N}{2}\times mL\times 2 = 4kmL\binom{N}{2},
\end{align}
Note that the $2\binom{N/2}{2}$ in \eq{givens-cost} corresponds to the $\opr{\mathcal{U}}_{J\rightarrow I}$ term, and $k$ is the $k$th order Trotter-Suzuki decomposition. 

The number of CNOT gates for one off-diagonal NOQE matrix element is then equal to
\begin{equation}
    N^{H_{IJ}}_{\text{CNOT}} =  N_{\text{CNOT}}^{\text{Swap}} + N_{\text{CNOT}}^{\text{Givens}} + N_{\text{CNOT}}^{\opr{n}_p \opr{n}_q},
\end{equation}
while the number of CNOT gates required for a diagonal term, $H_{II}$ is
\begin{equation}
    N^{H_{II}}_{\text{CNOT}} = 4 \times 2\binom{N/2}{2}\times(1+m L) +  N_{\text{CNOT}}^{n_p n_q}/2.
\end{equation}
Overall, the total number of CNOT gates, $N^{\text{Total}}_{\text{CNOT}}$, is then equal to
\begin{multline}
    N^{\text{Total}}_{\text{CNOT}} = \frac{M(M-1)}{2} N^{H_{IJ}}_{\text{CNOT}} + M N^{H_{II}}_{\text{CNOT}} \\
    \in \mathcal{O}(M^2 k N^2 L ).
\end{multline}

A second important resource count is the number of single-qubit non-Clifford $T$-gates, which is relevant to both NISQ and fault-tolerant quantum devices since these rotations enable universal quantum computation. 
Alternatively, a more general analogous quantity is the number of arbitrary single qubit rotation gates (e.g., $R_z$) which can be decomposed into a number of $T$-gates scaling as $1.15 \log_2 (1/\epsilon_{\text{syn}}) + 9.2$ with arbitrary synthesis error $\epsilon_{\text{syn}}$, using the result from Ref.~\cite{bocharov2015efficient}. 
A single Givens rotation, a single number operator product, and the controlled $N$ by $N$ qubit swap require 2, 3, and 7$N$ $R_z$ gates, respectively. 

We now consider all $N(N+1)/2$ relevant number operator products, including the diagonals, because the diagonal $\opr{n}_p \opr{n}_p$ terms contribute single-qubit rotations. 
In general, the number of $R_z$ gates, $N_R$, for generating each Takagi-factorized UCCMP2 reference ansatz is:
\begin{align}
    N_{R}^{\text{Swap}} &= 7 \times 2 \times 2N = 28N \\
    N_{R}^{\text{Givens}} &= 2\times k \left( 2 \times 2\binom{N/2}{2}\times(1+m L) + 2\binom{N/2}{2} \right) \label{givens-rz-cost} \\
    N_{R}^{\opr{n}_p \opr{n}_q} &= 3 \times k\frac{N(N+1)}{2} \times mL\times 2 = 3kmLN(N+1).
\end{align}
The total number of $R_z$ gates for evaluation of an off-diagonal matrix element of $H$ is given by
\begin{equation}
    N^{H_{IJ}}_{R} =  N_{R}^{\text{Swap}} + N_{R}^{\text{Givens}} + N_{R}^{\opr{n}_p \opr{n}_q},
\end{equation}
and the number of $R_z$ gates for evaluation of diagonal Hamiltonian matrix elements is
\begin{equation}
    N^{H_{II}}_{R} =  2 \times 2\binom{N/2}{2}\times(1+m L) +   N_{R}^{\opr{n}_p \opr{n}_q}/2.
\end{equation}
Overall, including the dependence on the number of NO basis states, $M$, the total number of $R_z$ gates can then be expressed as
\begin{equation}
    N^{\text{Total}}_{R} = \frac{M(M-1)}{2} N^{H_{IJ}}_{R} + MN^{H_{II}}_{R}
    \in \mathcal{O}(M^2 k N^2 L ),
\end{equation}
which scales the same as the CNOT gate complexity.

\begin{figure}[h!]
    \centering
        \includegraphics[width=\columnwidth]{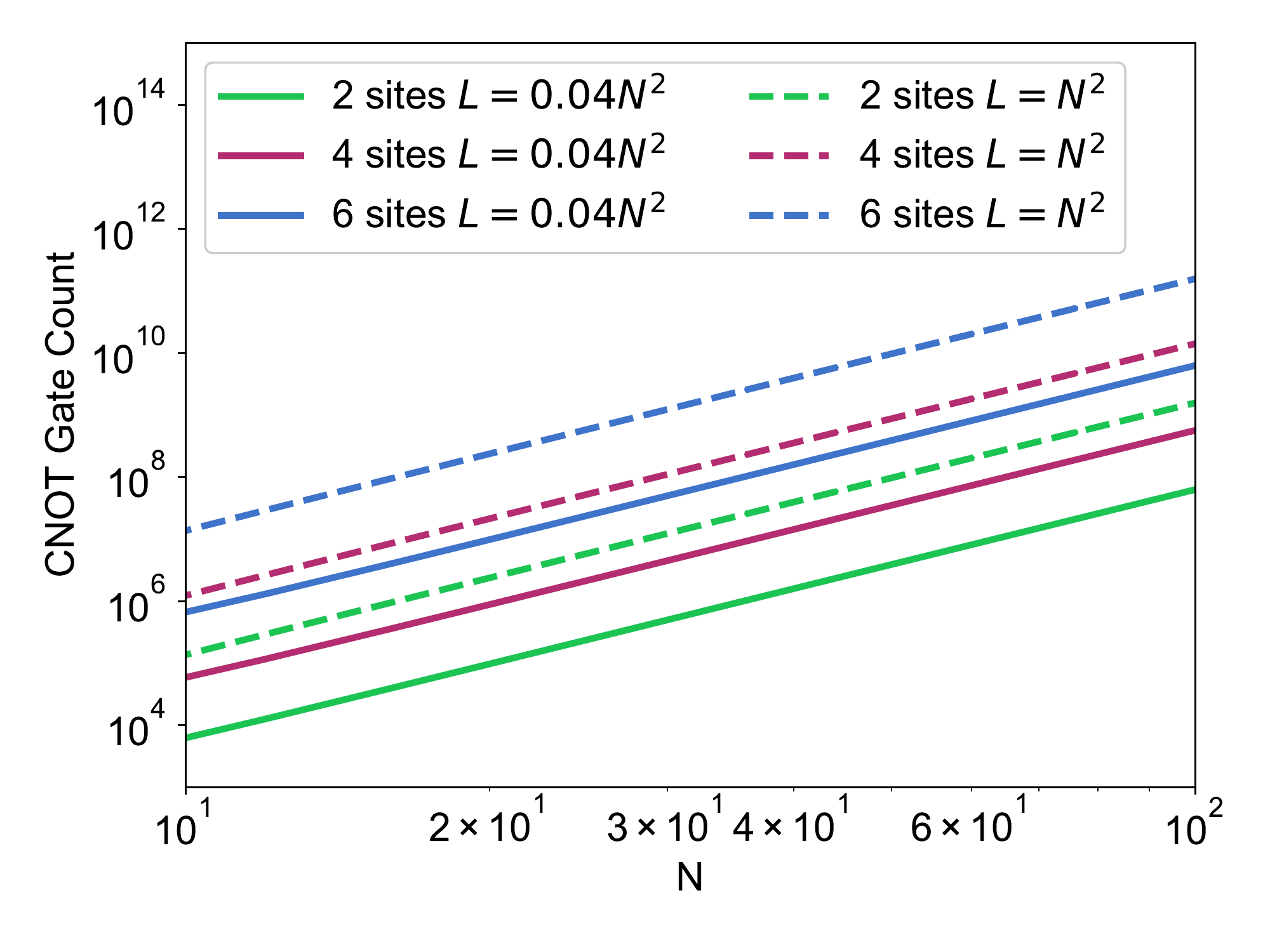}
    \caption{Scaling of the number of two-qubit gates with the number of spin-orbitals in the NOUCC(2) algorithm.  Data for 2, 4, and 6 radical sites are shown.  Without eigenvalue truncation, the $L$ index is equal to $N^2$ (dotted lines). With a truncation threshold that preserves chemical accuracy for alkane chains up to 8 C atoms, $L$ can be as low as 0.04$N^2$ (solid lines)~\cite{motta2021low}.}
    \label{fig:CNOTGateCount}
\end{figure}

We can use these resource counts to provide an empirical estimate for the overall gate counts required to achieve chemical accuracy for a given molecular system.
In the case of UCC, the asymptotic scaling of the two relevant truncation indices of \eq{lowrank_final} are  $\mathcal{O}(N^2)$ for $L$, and $\mathcal{O}(N)$ for $\rho_L$.  Using the truncation thresholds that were shown in Ref.~\cite{motta2021low} to preserve chemically accurate energies for alkane chains of increasing length (up to 8 carbon atoms), we estimate that the implied scaling prefactors for $L$ and $\rho_L$ are 0.04 and 1, respectively.  In other words, $L=0.04 N^2$ and $\rho_L = N$.  In NOUCC(2), UCC amplitudes are replaced with amplitudes obtained from MP2, decomposed in an analogous way, and we assume that similar levels of compression will be achieved.  The overall CNOT counts are plotted in \fig{CNOTGateCount} for 2 (H$_2$), 4 (square H$_4$), and 6 radical sites.  Sizable reductions in gate count result due to the eigenvalue truncation procedure described above, and we find that the addition of a single radical site increases the two-qubit gate count by approximately 1 order of magnitude.  We note that the overall scaling of the $R_z$ gate counts is very similar.  

The use of the UCJ ansatz in our NOQE framework is likely to be relatively advantageous, especially from a resource cost perspective.  
Recall that for H$_2$, NOUCJ($L=1$) was sufficient to produce comparable accuracy to NOUCC(2) with an untruncated factorization of the MP2 amplitudes.  
For a relatively more complicated system, Ref.~\cite{matsuzawa2020jastrow} demonstrated that the exact dissociation curve of the N$_2$ molecule, using an active space of six electrons in twelve spin-orbitals, was reproduced with satisfactory accuracy with the UCJ($L=2$) ansatz when variationally optimized (classically).  
While it remains to be verified, we are optimistic that, for general molecular systems, the $N^2$ scaling of $L$ can be reduced and in certain cases even excised.  
For illustrative purposes, the CNOT count for a given system size $N$ with NOUCJ($L=1$) is multiple orders of magnitude smaller than that from the Takagi-SVD decomposed NOUCC approach.  
For example, using two radical sites, 100 spin orbitals, and the NOUCC(2) routine assuming the 0.04 prefactor for $L$, it requires 6.3$\times 10^7$ CNOT gates while the $L=1$ NOUCJ approach requires only $1.3\times 10^5$ gates.
Clearly, this ansatz is a promising avenue forward and will be investigated in future work.  

In summary, our novel NOQE algorithm efficiently utilizes a quantum device to compute matrix elements of the Hamiltonian, and the overlap involving non-orthogonal reference states (where we have considered ans{\"a}tze of the UCC and UCJ form).  
For a fixed number of reference states involved in the Hamiltonian diagonalization, the total number of CNOT and $R_z$ gates required scales as $\mathcal{O}(N^2L)$ (assuming a first-order Trotter-Suzuki decomposition).  
The counts are summarized in \tab{countSummary}.  
With the UCJ($L=1$) ansatz, the number of these gates will scale quadratically with the number of spin orbitals, while for Tagaki/SVD-decomposed UCC amplitudes, the counts will scale quartically.

\begin{center}
\begin{table}[h!]
\begin{tabular}{ |c|c|c|} 
\hline
      & CNOT and $R_z$ gate complexity  \\
\hline
CSWAP & $\mathcal{O}(N)$  \\ 
$e^{\opr{\tau}}$ & $\mathcal{O}(kN^2L)$   \\ 
$\opr{\mathcal{U}}_{I \rightarrow 1}$ & $\mathcal{O}(N^2)$  \\ 
\hline
All $H_{IJ}$ ($I \neq J$) & $\mathcal{O}(M^2kN^2L)$  \\
\hline 
\end{tabular}
\caption{Asymptotic scaling of the number of CNOT and $R_z$ gates as a function of spin orbitals ($N$), NO reference states ($M$), Trotter-Suzuki decomposition order ($k$), and Tagaki decomposition threshold ($L$).}
\label{tab:countSummary}
\end{table}
\end{center}

\section{Discussion and outlook}\label{sec:discussion-and-outlook} 

In this work, we have presented a non-orthogonal quantum eigensolver, NOQE, that provides a novel electronic structure method for computing low-lying eigenstates of strongly correlated molecular systems. The NOQE method produces multireference wavefunctions in which classically-determined unitary coupled-cluster operators add dynamic correlation to each reference state.  The set of correlated reference states, in which the Hamiltonian is diagonalized, is not constrained to be orthogonal. Our analysis shows that correlating non-orthogonal reference states with a cluster operator captures a significant portion of the exact wavefunction in a highly compact manner, and furthermore, that evaluation of the resulting energies is possible to compute at polynomial cost with a quantum computer. This is in stark contrast to the exponentially scaling number of resources required to implement such an algorithm on a classical computer. Thus a non-orthogonal multireference eigensolver is possible to implement in a scalable manner only on a quantum computer. NOQE provides a flexible, compact, and rigorous description of both strong and dynamic electronic correlations, making it an attractive method for the calculation of electronic states of a wide range of molecular systems. 

A few comments are in order to clarify the advantage of our quantum NOQE methods vs. classical approaches.  We note that classical projected coupled-cluster methods with truncated cluster operators also show formal polynomial scaling in the number of spin-orbitals $N$. Indeed, the electronic ground states of the small H$_2$ and H$_4$ examples presented in this work can be exactly evaluated with polynomially scaling variants such as CCSD and CCSDTQ respectively, acting on a single reference Hartree-Fock determinant. However, excitations of order up to the number of radical sites $d$ must be included in the cluster operator in order to exactly model an arbitrary strongly correlated system with CC, resulting in a classical computational cost that scales as $\mathcal{O}(N^{d+4})$. This approach would therefore be infeasible for larger systems such as di-metal complexes, where $d$ is quite large. In practice, CC methods of higher order than CCSDTQ are seldom employed, and almost never beyond rather small basis sets~\cite{karton2011w4,hait2019levels}, restricting the practical utility of such classical methods to $d\le 4$. Even CCSDTQ requires a very large amount of resources for systems as small as benzene in the cc-pVDZ basis~\cite{eriksen2020ground}. Therefore, a multi-reference approach involving electron correlations between different radical sites is essential. The resource estimate analysis in this work suggests that implementation of the polynomial scaling NOQE algorithm on quantum computers could be feasible for strongly correlated dimetallic species within the longer term. 

Perhaps the closest classical analog to NOQE is the complete active space second-order perturbation theory (CASPT2~\cite{andersson1990second,andersson1992second}) and the closely related ``$N$-electron valence space second-order perturbation theory" (NEVPT2~\cite{angeli2001introduction}) approaches, where exact diagonalization is performed within a pre-defined ``active space" of a few electrons and orbitals assumed to be relevant for static correlation, followed by a perturbative treatment of dynamic correlation. This has obvious parallels with NOQE using exact diagonalization within a nonorthogonal basis for modeling strong correlation, with the unitary coupled-cluster operator, utilizing perturbative amplitudes in the NOUCC(2) implementation, accounting for dynamic correlation. While both CASPT2/NEVPT2 and NOQE require an exponentially growing number of states (with respect to the number of radical sites) in the exact diagonalization step,  the latter scales with a relatively softer exponential. Specifically, for the general case with $d$ radical sites, $\eta_{\alpha}$ up spins and $\eta_{\beta}$ down spins (where $\eta_\alpha+\eta_\beta<d$, $m_s=\eta_\alpha-\eta_\beta$), the subspace for exact diagonalization in CASPT2/NEVPT2 has a size $M_{\text{CAS}}$:
\begin{align}\label{eq:CAS_refs_general}
    M_{\text{CAS}}&= \dfrac{\left(d!\right)^2}{\eta_\alpha! \eta_\beta! (d-\eta_\alpha)!(d-\eta_\beta)!}
\end{align}
which, as mentioned above, is much larger than the corresponding NOQE diagonalization subspace given by \eq{NOQE_refs_general}. In particular, for the case of $\eta_{\alpha}=\eta_{\beta}=\dfrac{d}{2}$, $M_{\text{CAS}}=M_{\text{NOQE}}^2$. The reduction in subspace size for NOQE is due to the fact that the use of non-orthogonal reference states allows for a more compact representation of the multi-reference wavefunction required for describing strong correlation. Furthermore, CASPT2/NEVPT2 usually entails a self-consistent field (CASSCF~\cite{helgaker2014molecular}) stage in which the $M_{\text{CAS}}$ dimensional CAS wavefunction is iteratively optimized through repeated diagonalization and orbital rotations. In contrast, NOQE does not require any such optimization  and is ``one-shot" by construction. This leads to further computational efficiency compared to the classical CAS methods. 
We also note that the ground state NOQE method is bounded from below by the FCI ground state energy, which is not guaranteed to be the case for CASPT2/NEVPT2. Indeed, while dynamic correlation in NOQE is evaluated through perturbative amplitudes, the use of the UCC formalism and diagonalization of the Hamiltonian within the subspace spanned by NO states enables NOQE to be more robust against failures sometimes encountered in purely perturbative classical theories~\cite{roos1995multiconfigurational,ghigo2004modified,kurlancheek2009violations}. In particular, the diagonal elements of the Hamiltonian in NOQE are correct to the third order in perturbation theory, indicating that NOQE goes beyond classical second-order perturbation theories for dynamic correlation outside the exact diagonalization subspace.

In the context of previous quantum algorithms, NOQE possesses significant advantages over conventional variational quantum eigensolvers, for which a quantum device measures the energy while a classical device computes gradients and updates variational parameters for a given unitary ansatz. This results in a high measurement overhead from the input to the gradient and variational steps, both of which are absent in NOQE.
Instead, the NOQE approach benefits from using a quantum processor to compute both matrix elements of the Hamiltonian and the set of overlap matrix elements between dynamically correlated non-orthogonal states in parallel, at a low order polynomial cost for a fixed number of radical sites.

The NOQE algorithm also allows simultaneous calculation of both ground and excited electronic eigenstates of the molecular Hamiltonian within the Born-Oppenheimer description (i.e., for fixed nuclear positions). In particular, our analysis of H$_4$ with the NOUCC(2) ansatz shows that the algorithm can efficiently compute the relative ordering and energy gaps of a select number of low-lying eigenstates in a single calculation without incurring additional measurement overhead, whereas conventional VQE focuses on optimizing the ground state alone. We also note that NOQE amplitudes would provide a compact and high-quality guess for FCI eigenstates for use in fault-tolerant algorithms such as quantum phase estimation (QPE) for estimating eigenvalues.  The large improvement in fidelity over a single Slater Hartree-Fock reference is promising and efficient on a quantum computer, but needs to be investigated further to show the overall cost benefit analysis of this more involved state preparation technique for QPE.  It would be interesting here to compare the overall cost with that  using classically determined NOCI amplitudes, which when combined with efficient preparation of multi-determinant wavefunctions~\cite{tubman2018postponing} would similarly provide a polynomially scaling alternative but one with reduced prefactors compared to NOQE.

Another major computational advantage of NOQE over the conventional variational approach embodied by VQE is that our ``one-shot" method built on a multi-reference set of configuration interaction states, focuses on a chemistry-specific ansatz with significant input from classical quantum chemistry.  This results in a very high-quality set of non-orthogonal reference states that do not require iterative parameter optimization on a quantum device.  Instead, the expressivity of the output solution is provided by the multi-reference construction.  The result is a dramatic reduction in both the gate complexity and the total number of measurements required to extract molecular energies or energy differences from running the algorithm on quantum devices. Furthermore, NOQE completely avoids the issues associated with possibly encountering a large number of non-global minima in the energy landscape when performing variational optimization or getting trapped on barren plateaus with ultra-small gradients in all directions. These issues complicate the black-box use of VQE and are particularly problematic for strongly correlated systems such as square H$_4$. We note that the variational step of VQE does appear to provide some intrinsic error mitigation via the updating of circuit parameters in the classical optimization step, which is not present in NOQE.  Instead, for the NOQE algorithm as for other non-variational quantum algorithms, error mitigation on NISQ devices can be implemented by using the techniques of randomized compiling~\cite{wallman2016noise,hashim2020randomized}.

From the perspective of seeking energies and wavefunctions for strongly correlated systems, the use of classically inspired ans{\"a}tze, i.e., electronic wavefunction forms that can be motivated and justified by chemical or physical insights, is an advantageous feature. Furthermore, the maturity of wavefunction-based techniques for electronic structure calculations in classical quantum chemistry provides a well-paved roadmap for future improvements in NOQE. In this work, we have focused primarily on using amplitudes from second-order perturbation theory within a UCCD-like ansatz, which are then decomposed to readily prepare relatively low-depth quantum circuits. As shown here for the examples of H$_2$ and H$_4$, this choice already leads to encouraging levels of accuracy, although achieving the goal of chemical accuracy in one shot (i.e., within 1.6 mHa of the exact values) required amplitude scaling procedures. We expect that there are other \emph{ab initio} options for improving the ansatz states, beyond uniform or spin-component scaling, that do not dramatically increase the classical cost for preparing the ansatz states. Options include using amplitudes from CCSD or from energy-gap dependent regularized MP2~\cite{shee2021regularized}, as well as orbitals from methods other than Hartree-Fock~\cite{bertels2019third,rettig2020third}. Additionally, in this work, we took preliminary steps to explore the benefit of adding classically-optimized Jastrow correlators to the UHF states, building a NOUCJ ansatz for NOQE, which appears to enable a reduction in the cost of quantum circuits for comparable accuracy for the H$_2$ system.  In future work, we shall explore the performance of the classically variationally optimized NOUCJ ansatz for NOQE calculations of larger systems.     

We note that, as was also the case for classical NOCI calculations, the number of NO reference states in the NOQE approach still grows exponentially with the number of radical sites involved in the strong static correlation.  However, this is a formal scaling that in practice is neither problematic nor relevant to most systems of chemical interest, since the number of radical sites typically does not reach large values for molecular systems (especially naturally occurring ones). In fact, there is a myriad of molecular systems which are of great chemical interest that only require a relatively small number of strongly correlated sites ($d < 8$), which NOQE is well-suited to tackle, given appropriate quantum hardware. Such systems include di-copper subunits in metalloenzymes~\cite{gherman2009quantum,liakos2011interplay}, $n$-carbenes or long polyacenes which possess di- and poly-radical character~\cite{yang2016nature}, reduced states of metal complexes involving redox non-innocent ligands~\cite{NeesePorphyrin,derrick2020metal,shee2021revealing}, and systems ranging from small transition metal compounds to the OEC (4 transition metal sites), as mentioned in the introduction, and iron-sulfur clusters (2-8 Fe atoms)~\cite{sharma2014low,mejuto2022effect,li2019electronic}.

In summary, the NOQE method presented in this work is a promising quantum electronic structure algorithm that excels at the accurate computation of energy gaps between low-lying eigenstates of a strongly correlated molecular system over a wide range of internuclear distances, allowing the construction of potential energy surfaces for both ground and excited states. NOQE aims to systematically capture both static and dynamic correlation in a manner that is infeasible within a completely classical algorithm, whereas the quantum computer requires only a polynomial gate depth for a fixed number of radical sites. Initial results presented here for H$_2$ and H$_4$ are encouraging. Future work will focus on performing NOQE calculations on currently available noisy intermediate-scale quantum (NISQ) devices, investigating more sophisticated single-reference ansatz forms to employ in the NOQE multi-reference ansatz, and benchmarking the method for larger systems with more complex electronic structures.

\vspace{10pt}
\section*{Acknowledgments}

U.B., O.L., D.H., M.H.G., and K.B.W. were supported by the NSF QLCI program through grant number QMA-2016345. Early stages of this work were supported by the U.S. Department of Energy, Office of Science, Office of Advanced Scientific Computing Research, Quantum Algorithm Teams Program, under contract number DE-AC02-05CH11231 (U.B., W.H., and K.B.W.). J.S. acknowledges funding from the National Institute of General Medical Sciences of the National Institutes of Health under award number F32GM142231. T.F.S. acknowledges support as a Quantum Postdoctoral Fellow at the Simons Institute for the Theory of Computing, which is funded by NSF QLCI Grant No. 2016245.

\bibliography{refs}{}

\begin{thebibliography}{115}%
\makeatletter
\providecommand \@ifxundefined [1]{%
 \@ifx{#1\undefined}
}%
\providecommand \@ifnum [1]{%
 \ifnum #1\expandafter \@firstoftwo
 \else \expandafter \@secondoftwo
 \fi
}%
\providecommand \@ifx [1]{%
 \ifx #1\expandafter \@firstoftwo
 \else \expandafter \@secondoftwo
 \fi
}%
\providecommand \natexlab [1]{#1}%
\providecommand \enquote  [1]{``#1''}%
\providecommand \bibnamefont  [1]{#1}%
\providecommand \bibfnamefont [1]{#1}%
\providecommand \citenamefont [1]{#1}%
\providecommand \href@noop [0]{\@secondoftwo}%
\providecommand \href [0]{\begingroup \@sanitize@url \@href}%
\providecommand \@href[1]{\@@startlink{#1}\@@href}%
\providecommand \@@href[1]{\endgroup#1\@@endlink}%
\providecommand \@sanitize@url [0]{\catcode `\\12\catcode `\$12\catcode
  `\&12\catcode `\#12\catcode `\^12\catcode `\_12\catcode `\%12\relax}%
\providecommand \@@startlink[1]{}%
\providecommand \@@endlink[0]{}%
\providecommand \url  [0]{\begingroup\@sanitize@url \@url }%
\providecommand \@url [1]{\endgroup\@href {#1}{\urlprefix }}%
\providecommand \urlprefix  [0]{URL }%
\providecommand \Eprint [0]{\href }%
\providecommand \doibase [0]{http://dx.doi.org/}%
\providecommand \selectlanguage [0]{\@gobble}%
\providecommand \bibinfo  [0]{\@secondoftwo}%
\providecommand \bibfield  [0]{\@secondoftwo}%
\providecommand \translation [1]{[#1]}%
\providecommand \BibitemOpen [0]{}%
\providecommand \bibitemStop [0]{}%
\providecommand \bibitemNoStop [0]{.\EOS\space}%
\providecommand \EOS [0]{\spacefactor3000\relax}%
\providecommand \BibitemShut  [1]{\csname bibitem#1\endcsname}%
\let\auto@bib@innerbib\@empty
\bibitem [{\citenamefont {O'Gorman}\ \emph {et~al.}(2021)\citenamefont
  {O'Gorman}, \citenamefont {Irani}, \citenamefont {Whitfield},\ and\
  \citenamefont {Fefferman}}]{o2021electronic}%
  \BibitemOpen
  \bibfield  {author} {\bibinfo {author} {\bibfnamefont {Bryan}\ \bibnamefont
  {O'Gorman}}, \bibinfo {author} {\bibfnamefont {Sandy}\ \bibnamefont {Irani}},
  \bibinfo {author} {\bibfnamefont {James}\ \bibnamefont {Whitfield}}, \ and\
  \bibinfo {author} {\bibfnamefont {Bill}\ \bibnamefont {Fefferman}},\
  }\bibfield  {title} {\enquote {\bibinfo {title} {Electronic structure in a
  fixed basis is {QMA}-complete},}\ }\href@noop {} {\bibfield  {journal}
  {\bibinfo  {journal} {arXiv preprint arXiv:2103.08215}\ } (\bibinfo {year}
  {2021})}\BibitemShut {NoStop}%
\bibitem [{\citenamefont {Boys}\ and\ \citenamefont
  {Handy}(1969)}]{boys1969calculation}%
  \BibitemOpen
  \bibfield  {author} {\bibinfo {author} {\bibfnamefont {Samuel~Francis}\
  \bibnamefont {Boys}}\ and\ \bibinfo {author} {\bibfnamefont
  {Nicholas~Charles}\ \bibnamefont {Handy}},\ }\bibfield  {title} {\enquote
  {\bibinfo {title} {A calculation for the energies and wavefunctions for
  states of neon with full electronic correlation accuracy},}\ }\href@noop {}
  {\bibfield  {journal} {\bibinfo  {journal} {Proc. R. Soc. Lond. A}\ }\textbf
  {\bibinfo {volume} {310}},\ \bibinfo {pages} {63--78} (\bibinfo {year}
  {1969})}\BibitemShut {NoStop}%
\bibitem [{\citenamefont {Small}\ and\ \citenamefont
  {Head-Gordon}(2011)}]{small2011post}%
  \BibitemOpen
  \bibfield  {author} {\bibinfo {author} {\bibfnamefont {David~W}\ \bibnamefont
  {Small}}\ and\ \bibinfo {author} {\bibfnamefont {Martin}\ \bibnamefont
  {Head-Gordon}},\ }\bibfield  {title} {\enquote {\bibinfo {title} {Post-modern
  valence bond theory for strongly correlated electron spins},}\ }\href@noop {}
  {\bibfield  {journal} {\bibinfo  {journal} {Phys. Chem. Chem. Phys.}\
  }\textbf {\bibinfo {volume} {13}},\ \bibinfo {pages} {19285--19297} (\bibinfo
  {year} {2011})}\BibitemShut {NoStop}%
\bibitem [{\citenamefont {Shee}\ \emph
  {et~al.}(2021{\natexlab{a}})\citenamefont {Shee}, \citenamefont
  {Loipersberger}, \citenamefont {Hait}, \citenamefont {Lee},\ and\
  \citenamefont {Head-Gordon}}]{shee2021revealing}%
  \BibitemOpen
  \bibfield  {author} {\bibinfo {author} {\bibfnamefont {James}\ \bibnamefont
  {Shee}}, \bibinfo {author} {\bibfnamefont {Matthias}\ \bibnamefont
  {Loipersberger}}, \bibinfo {author} {\bibfnamefont {Diptarka}\ \bibnamefont
  {Hait}}, \bibinfo {author} {\bibfnamefont {Joonho}\ \bibnamefont {Lee}}, \
  and\ \bibinfo {author} {\bibfnamefont {Martin}\ \bibnamefont {Head-Gordon}},\
  }\bibfield  {title} {\enquote {\bibinfo {title} {Revealing the nature of
  electron correlation in transition metal complexes with symmetry breaking and
  chemical intuition},}\ }\href@noop {} {\bibfield  {journal} {\bibinfo
  {journal} {J. Chem. Phys.}\ }\textbf {\bibinfo {volume} {154}},\ \bibinfo
  {pages} {194109} (\bibinfo {year} {2021}{\natexlab{a}})}\BibitemShut
  {NoStop}%
\bibitem [{\citenamefont {Imada}\ \emph {et~al.}(1998)\citenamefont {Imada},
  \citenamefont {Fujimori},\ and\ \citenamefont {Tokura}}]{imada1998metal}%
  \BibitemOpen
  \bibfield  {author} {\bibinfo {author} {\bibfnamefont {Masatoshi}\
  \bibnamefont {Imada}}, \bibinfo {author} {\bibfnamefont {Atsushi}\
  \bibnamefont {Fujimori}}, \ and\ \bibinfo {author} {\bibfnamefont
  {Yoshinori}\ \bibnamefont {Tokura}},\ }\bibfield  {title} {\enquote {\bibinfo
  {title} {Metal-insulator transitions},}\ }\href@noop {} {\bibfield  {journal}
  {\bibinfo  {journal} {Rev. Mod. Phys.}\ }\textbf {\bibinfo {volume} {70}},\
  \bibinfo {pages} {1039} (\bibinfo {year} {1998})}\BibitemShut {NoStop}%
\bibitem [{\citenamefont {Witzke}\ \emph {et~al.}(2020)\citenamefont {Witzke},
  \citenamefont {Hait}, \citenamefont {Chakarawet}, \citenamefont
  {Head-Gordon},\ and\ \citenamefont {Tilley}}]{witzke2020bimetallic}%
  \BibitemOpen
  \bibfield  {author} {\bibinfo {author} {\bibfnamefont {Ryan~J}\ \bibnamefont
  {Witzke}}, \bibinfo {author} {\bibfnamefont {Diptarka}\ \bibnamefont {Hait}},
  \bibinfo {author} {\bibfnamefont {Khetpakorn}\ \bibnamefont {Chakarawet}},
  \bibinfo {author} {\bibfnamefont {Martin}\ \bibnamefont {Head-Gordon}}, \
  and\ \bibinfo {author} {\bibfnamefont {T~Don}\ \bibnamefont {Tilley}},\
  }\bibfield  {title} {\enquote {\bibinfo {title} {Bimetallic mechanism for
  alkyne cyclotrimerization with a two-coordinate {Fe} precatalyst},}\
  }\href@noop {} {\bibfield  {journal} {\bibinfo  {journal} {ACS Catal.}\
  }\textbf {\bibinfo {volume} {10}},\ \bibinfo {pages} {7800--7807} (\bibinfo
  {year} {2020})}\BibitemShut {NoStop}%
\bibitem [{\citenamefont {Gould}\ \emph {et~al.}(2022)\citenamefont {Gould},
  \citenamefont {McClain}, \citenamefont {Reta}, \citenamefont {Kragskow},
  \citenamefont {Marchiori}, \citenamefont {Lachman}, \citenamefont {Choi},
  \citenamefont {Analytis}, \citenamefont {Britt}, \citenamefont {Chilton}
  \emph {et~al.}}]{gould2022ultrahard}%
  \BibitemOpen
  \bibfield  {author} {\bibinfo {author} {\bibfnamefont {Colin~A}\ \bibnamefont
  {Gould}}, \bibinfo {author} {\bibfnamefont {K~Randall}\ \bibnamefont
  {McClain}}, \bibinfo {author} {\bibfnamefont {Daniel}\ \bibnamefont {Reta}},
  \bibinfo {author} {\bibfnamefont {Jon~GC}\ \bibnamefont {Kragskow}}, \bibinfo
  {author} {\bibfnamefont {David~A}\ \bibnamefont {Marchiori}}, \bibinfo
  {author} {\bibfnamefont {Ella}\ \bibnamefont {Lachman}}, \bibinfo {author}
  {\bibfnamefont {Eun-Sang}\ \bibnamefont {Choi}}, \bibinfo {author}
  {\bibfnamefont {James~G}\ \bibnamefont {Analytis}}, \bibinfo {author}
  {\bibfnamefont {R~David}\ \bibnamefont {Britt}}, \bibinfo {author}
  {\bibfnamefont {Nicholas~F}\ \bibnamefont {Chilton}},  \emph {et~al.},\
  }\bibfield  {title} {\enquote {\bibinfo {title} {Ultrahard magnetism from
  mixed-valence dilanthanide complexes with metal-metal bonding},}\ }\href@noop
  {} {\bibfield  {journal} {\bibinfo  {journal} {Science}\ }\textbf {\bibinfo
  {volume} {375}},\ \bibinfo {pages} {198--202} (\bibinfo {year}
  {2022})}\BibitemShut {NoStop}%
\bibitem [{\citenamefont {Askerka}\ \emph {et~al.}(2017)\citenamefont
  {Askerka}, \citenamefont {Brudvig},\ and\ \citenamefont
  {Batista}}]{askerka2017o2}%
  \BibitemOpen
  \bibfield  {author} {\bibinfo {author} {\bibfnamefont {Mikhail}\ \bibnamefont
  {Askerka}}, \bibinfo {author} {\bibfnamefont {Gary~W}\ \bibnamefont
  {Brudvig}}, \ and\ \bibinfo {author} {\bibfnamefont {Victor~S}\ \bibnamefont
  {Batista}},\ }\bibfield  {title} {\enquote {\bibinfo {title} {The
  {O}$_2$-evolving complex of photosystem {II: Recent} insights from quantum
  mechanics/molecular mechanics {(QM/MM)}, extended {X-ray} absorption fine
  structure ({EXAFS}), and femtosecond {X-ray} crystallography data},}\
  }\href@noop {} {\bibfield  {journal} {\bibinfo  {journal} {Acc. Chem. Res.}\
  }\textbf {\bibinfo {volume} {50}},\ \bibinfo {pages} {41--48} (\bibinfo
  {year} {2017})}\BibitemShut {NoStop}%
\bibitem [{\citenamefont {Umena}\ \emph {et~al.}(2011)\citenamefont {Umena},
  \citenamefont {Kawakami}, \citenamefont {Shen},\ and\ \citenamefont
  {Kamiya}}]{umena2011crystal}%
  \BibitemOpen
  \bibfield  {author} {\bibinfo {author} {\bibfnamefont {Yasufumi}\
  \bibnamefont {Umena}}, \bibinfo {author} {\bibfnamefont {Keisuke}\
  \bibnamefont {Kawakami}}, \bibinfo {author} {\bibfnamefont {Jian-Ren}\
  \bibnamefont {Shen}}, \ and\ \bibinfo {author} {\bibfnamefont {Nobuo}\
  \bibnamefont {Kamiya}},\ }\bibfield  {title} {\enquote {\bibinfo {title}
  {Crystal structure of oxygen-evolving photosystem {II} at a resolution of
  1.9~{{\AA}}},}\ }\href@noop {} {\bibfield  {journal} {\bibinfo  {journal}
  {Nature}\ }\textbf {\bibinfo {volume} {473}},\ \bibinfo {pages} {55--60}
  (\bibinfo {year} {2011})}\BibitemShut {NoStop}%
\bibitem [{\citenamefont {Raymond}\ and\ \citenamefont
  {Blankenship}(2008)}]{raymond2008origin}%
  \BibitemOpen
  \bibfield  {author} {\bibinfo {author} {\bibfnamefont {Jason}\ \bibnamefont
  {Raymond}}\ and\ \bibinfo {author} {\bibfnamefont {Robert~E}\ \bibnamefont
  {Blankenship}},\ }\bibfield  {title} {\enquote {\bibinfo {title} {The origin
  of the oxygen-evolving complex},}\ }\href@noop {} {\bibfield  {journal}
  {\bibinfo  {journal} {Coord. Chem. Rev.}\ }\textbf {\bibinfo {volume}
  {252}},\ \bibinfo {pages} {377--383} (\bibinfo {year} {2008})}\BibitemShut
  {NoStop}%
\bibitem [{\citenamefont {Aspuru-Guzik}\ \emph {et~al.}(2005)\citenamefont
  {Aspuru-Guzik}, \citenamefont {Dutoi}, \citenamefont {Love},\ and\
  \citenamefont {Head-Gordon}}]{aspuru2005simulated}%
  \BibitemOpen
  \bibfield  {author} {\bibinfo {author} {\bibfnamefont {Al{\'a}n}\
  \bibnamefont {Aspuru-Guzik}}, \bibinfo {author} {\bibfnamefont {Anthony~D}\
  \bibnamefont {Dutoi}}, \bibinfo {author} {\bibfnamefont {Peter~J}\
  \bibnamefont {Love}}, \ and\ \bibinfo {author} {\bibfnamefont {Martin}\
  \bibnamefont {Head-Gordon}},\ }\bibfield  {title} {\enquote {\bibinfo {title}
  {Simulated quantum computation of molecular energies},}\ }\href@noop {}
  {\bibfield  {journal} {\bibinfo  {journal} {Science}\ }\textbf {\bibinfo
  {volume} {309}},\ \bibinfo {pages} {1704--1707} (\bibinfo {year}
  {2005})}\BibitemShut {NoStop}%
\bibitem [{\citenamefont {Peruzzo}\ \emph {et~al.}(2014)\citenamefont
  {Peruzzo}, \citenamefont {McClean}, \citenamefont {Shadbolt}, \citenamefont
  {Yung}, \citenamefont {Zhou}, \citenamefont {Love}, \citenamefont
  {Aspuru-Guzik},\ and\ \citenamefont {O’brien}}]{peruzzo2014variational}%
  \BibitemOpen
  \bibfield  {author} {\bibinfo {author} {\bibfnamefont {Alberto}\ \bibnamefont
  {Peruzzo}}, \bibinfo {author} {\bibfnamefont {Jarrod}\ \bibnamefont
  {McClean}}, \bibinfo {author} {\bibfnamefont {Peter}\ \bibnamefont
  {Shadbolt}}, \bibinfo {author} {\bibfnamefont {Man-Hong}\ \bibnamefont
  {Yung}}, \bibinfo {author} {\bibfnamefont {Xiao-Qi}\ \bibnamefont {Zhou}},
  \bibinfo {author} {\bibfnamefont {Peter~J}\ \bibnamefont {Love}}, \bibinfo
  {author} {\bibfnamefont {Al{\'a}n}\ \bibnamefont {Aspuru-Guzik}}, \ and\
  \bibinfo {author} {\bibfnamefont {Jeremy~L}\ \bibnamefont {O’brien}},\
  }\bibfield  {title} {\enquote {\bibinfo {title} {A variational eigenvalue
  solver on a photonic quantum processor},}\ }\href@noop {} {\bibfield
  {journal} {\bibinfo  {journal} {Nat. Commun.}\ }\textbf {\bibinfo {volume}
  {5}},\ \bibinfo {pages} {1--7} (\bibinfo {year} {2014})}\BibitemShut
  {NoStop}%
\bibitem [{\citenamefont {McClean}\ \emph {et~al.}(2016)\citenamefont
  {McClean}, \citenamefont {Romero}, \citenamefont {Babbush},\ and\
  \citenamefont {Aspuru-Guzik}}]{mcclean2016theory}%
  \BibitemOpen
  \bibfield  {author} {\bibinfo {author} {\bibfnamefont {Jarrod~R}\
  \bibnamefont {McClean}}, \bibinfo {author} {\bibfnamefont {Jonathan}\
  \bibnamefont {Romero}}, \bibinfo {author} {\bibfnamefont {Ryan}\ \bibnamefont
  {Babbush}}, \ and\ \bibinfo {author} {\bibfnamefont {Al{\'a}n}\ \bibnamefont
  {Aspuru-Guzik}},\ }\bibfield  {title} {\enquote {\bibinfo {title} {The theory
  of variational hybrid quantum-classical algorithms},}\ }\href@noop {}
  {\bibfield  {journal} {\bibinfo  {journal} {New J. Phys}\ }\textbf {\bibinfo
  {volume} {18}},\ \bibinfo {pages} {023023} (\bibinfo {year}
  {2016})}\BibitemShut {NoStop}%
\bibitem [{\citenamefont {Taube}\ and\ \citenamefont
  {Bartlett}(2006)}]{taube2006new}%
  \BibitemOpen
  \bibfield  {author} {\bibinfo {author} {\bibfnamefont {Andrew~G}\
  \bibnamefont {Taube}}\ and\ \bibinfo {author} {\bibfnamefont {Rodney~J}\
  \bibnamefont {Bartlett}},\ }\bibfield  {title} {\enquote {\bibinfo {title}
  {New perspectives on unitary coupled-cluster theory},}\ }\href@noop {}
  {\bibfield  {journal} {\bibinfo  {journal} {Int. J. Quantum Chem.}\ }\textbf
  {\bibinfo {volume} {106}},\ \bibinfo {pages} {3393--3401} (\bibinfo {year}
  {2006})}\BibitemShut {NoStop}%
\bibitem [{\citenamefont {McClean}\ \emph {et~al.}(2018)\citenamefont
  {McClean}, \citenamefont {Boixo}, \citenamefont {Smelyanskiy}, \citenamefont
  {Babbush},\ and\ \citenamefont {Neven}}]{mcclean2018barren}%
  \BibitemOpen
  \bibfield  {author} {\bibinfo {author} {\bibfnamefont {Jarrod~R}\
  \bibnamefont {McClean}}, \bibinfo {author} {\bibfnamefont {Sergio}\
  \bibnamefont {Boixo}}, \bibinfo {author} {\bibfnamefont {Vadim~N}\
  \bibnamefont {Smelyanskiy}}, \bibinfo {author} {\bibfnamefont {Ryan}\
  \bibnamefont {Babbush}}, \ and\ \bibinfo {author} {\bibfnamefont {Hartmut}\
  \bibnamefont {Neven}},\ }\bibfield  {title} {\enquote {\bibinfo {title}
  {Barren plateaus in quantum neural network training landscapes},}\
  }\href@noop {} {\bibfield  {journal} {\bibinfo  {journal} {Nat. Commun.}\
  }\textbf {\bibinfo {volume} {9}},\ \bibinfo {pages} {1--6} (\bibinfo {year}
  {2018})}\BibitemShut {NoStop}%
\bibitem [{\citenamefont {Uvarov}\ \emph {et~al.}(2020)\citenamefont {Uvarov},
  \citenamefont {Biamonte},\ and\ \citenamefont
  {Yudin}}]{uvarov2020variational}%
  \BibitemOpen
  \bibfield  {author} {\bibinfo {author} {\bibfnamefont {Alexey}\ \bibnamefont
  {Uvarov}}, \bibinfo {author} {\bibfnamefont {Jacob~D}\ \bibnamefont
  {Biamonte}}, \ and\ \bibinfo {author} {\bibfnamefont {Dmitry}\ \bibnamefont
  {Yudin}},\ }\bibfield  {title} {\enquote {\bibinfo {title} {Variational
  quantum eigensolver for frustrated quantum systems},}\ }\href@noop {}
  {\bibfield  {journal} {\bibinfo  {journal} {Phys. Rev. B}\ }\textbf {\bibinfo
  {volume} {102}},\ \bibinfo {pages} {075104} (\bibinfo {year}
  {2020})}\BibitemShut {NoStop}%
\bibitem [{\citenamefont {Arrasmith}\ \emph {et~al.}(2021)\citenamefont
  {Arrasmith}, \citenamefont {Cerezo}, \citenamefont {Czarnik}, \citenamefont
  {Cincio},\ and\ \citenamefont {Coles}}]{arrasmith2021effect}%
  \BibitemOpen
  \bibfield  {author} {\bibinfo {author} {\bibfnamefont {Andrew}\ \bibnamefont
  {Arrasmith}}, \bibinfo {author} {\bibfnamefont {M}~\bibnamefont {Cerezo}},
  \bibinfo {author} {\bibfnamefont {Piotr}\ \bibnamefont {Czarnik}}, \bibinfo
  {author} {\bibfnamefont {Lukasz}\ \bibnamefont {Cincio}}, \ and\ \bibinfo
  {author} {\bibfnamefont {Patrick~J}\ \bibnamefont {Coles}},\ }\bibfield
  {title} {\enquote {\bibinfo {title} {Effect of barren plateaus on
  gradient-free optimization},}\ }\href@noop {} {\bibfield  {journal} {\bibinfo
   {journal} {Quantum}\ }\textbf {\bibinfo {volume} {5}},\ \bibinfo {pages}
  {558} (\bibinfo {year} {2021})}\BibitemShut {NoStop}%
\bibitem [{\citenamefont {Huggins}\ \emph {et~al.}(2020)\citenamefont
  {Huggins}, \citenamefont {Lee}, \citenamefont {Baek}, \citenamefont
  {O’Gorman},\ and\ \citenamefont {Whaley}}]{huggins2020non}%
  \BibitemOpen
  \bibfield  {author} {\bibinfo {author} {\bibfnamefont {William~J}\
  \bibnamefont {Huggins}}, \bibinfo {author} {\bibfnamefont {Joonho}\
  \bibnamefont {Lee}}, \bibinfo {author} {\bibfnamefont {Unpil}\ \bibnamefont
  {Baek}}, \bibinfo {author} {\bibfnamefont {Bryan}\ \bibnamefont
  {O’Gorman}}, \ and\ \bibinfo {author} {\bibfnamefont {K~Birgitta}\
  \bibnamefont {Whaley}},\ }\bibfield  {title} {\enquote {\bibinfo {title} {A
  non-orthogonal variational quantum eigensolver},}\ }\href@noop {} {\bibfield
  {journal} {\bibinfo  {journal} {New J. Phys.}\ }\textbf {\bibinfo {volume}
  {22}},\ \bibinfo {pages} {073009} (\bibinfo {year} {2020})}\BibitemShut
  {NoStop}%
\bibitem [{\citenamefont {Broer}\ and\ \citenamefont
  {Nieuwpoort}(1981)}]{broer1981broken}%
  \BibitemOpen
  \bibfield  {author} {\bibinfo {author} {\bibfnamefont {R}~\bibnamefont
  {Broer}}\ and\ \bibinfo {author} {\bibfnamefont {WC}~\bibnamefont
  {Nieuwpoort}},\ }\bibfield  {title} {\enquote {\bibinfo {title} {Broken
  orbital-symmetry and the description of hole states in the tetrahedral
  {[CrO4]}-anion. {I. Introductory} considerations and calculations on oxygen
  1s hole states},}\ }\href@noop {} {\bibfield  {journal} {\bibinfo  {journal}
  {Chem. Phys.}\ }\textbf {\bibinfo {volume} {54}},\ \bibinfo {pages}
  {291--303} (\bibinfo {year} {1981})}\BibitemShut {NoStop}%
\bibitem [{\citenamefont {Thom}\ and\ \citenamefont
  {Head-Gordon}(2009)}]{thom2009hartree}%
  \BibitemOpen
  \bibfield  {author} {\bibinfo {author} {\bibfnamefont {Alex~JW}\ \bibnamefont
  {Thom}}\ and\ \bibinfo {author} {\bibfnamefont {Martin}\ \bibnamefont
  {Head-Gordon}},\ }\bibfield  {title} {\enquote {\bibinfo {title}
  {{Hartree-Fock} solutions as a quasidiabatic basis for nonorthogonal
  configuration interaction},}\ }\href@noop {} {\bibfield  {journal} {\bibinfo
  {journal} {J. Chem. Phys.}\ }\textbf {\bibinfo {volume} {131}},\ \bibinfo
  {pages} {124113} (\bibinfo {year} {2009})}\BibitemShut {NoStop}%
\bibitem [{\citenamefont {Sundstrom}\ and\ \citenamefont
  {Head-Gordon}(2014)}]{sundstrom2014non}%
  \BibitemOpen
  \bibfield  {author} {\bibinfo {author} {\bibfnamefont {Eric~J}\ \bibnamefont
  {Sundstrom}}\ and\ \bibinfo {author} {\bibfnamefont {Martin}\ \bibnamefont
  {Head-Gordon}},\ }\bibfield  {title} {\enquote {\bibinfo {title}
  {Non-orthogonal configuration interaction for the calculation of
  multielectron excited states},}\ }\href@noop {} {\bibfield  {journal}
  {\bibinfo  {journal} {J. Chem. Phys.}\ }\textbf {\bibinfo {volume} {140}},\
  \bibinfo {pages} {114103} (\bibinfo {year} {2014})}\BibitemShut {NoStop}%
\bibitem [{\citenamefont {Yost}\ and\ \citenamefont
  {Head-Gordon}(2016)}]{yost2016size}%
  \BibitemOpen
  \bibfield  {author} {\bibinfo {author} {\bibfnamefont {Shane~R}\ \bibnamefont
  {Yost}}\ and\ \bibinfo {author} {\bibfnamefont {Martin}\ \bibnamefont
  {Head-Gordon}},\ }\bibfield  {title} {\enquote {\bibinfo {title} {Size
  consistent formulations of the perturb-then-diagonalize {M{\o}ller-Plesset}
  perturbation theory correction to non-orthogonal configuration
  interaction},}\ }\href@noop {} {\bibfield  {journal} {\bibinfo  {journal} {J.
  Chem. Phys.}\ }\textbf {\bibinfo {volume} {145}},\ \bibinfo {pages} {054105}
  (\bibinfo {year} {2016})}\BibitemShut {NoStop}%
\bibitem [{\citenamefont {Parrish}\ and\ \citenamefont
  {McMahon}(2019)}]{Parrish2019-dv}%
  \BibitemOpen
  \bibfield  {author} {\bibinfo {author} {\bibfnamefont {Robert~M}\
  \bibnamefont {Parrish}}\ and\ \bibinfo {author} {\bibfnamefont {Peter~L}\
  \bibnamefont {McMahon}},\ }\bibfield  {title} {\enquote {\bibinfo {title}
  {Quantum filter diagonalization: Quantum eigendecomposition without full
  quantum phase estimation},}\ }\href@noop {} {\bibfield  {journal} {\bibinfo
  {journal} {arXiv preprint arXiv:1909.08925}\ } (\bibinfo {year}
  {2019})}\BibitemShut {NoStop}%
\bibitem [{\citenamefont {Kyriienko}(2020)}]{Kyriienko2020-yc}%
  \BibitemOpen
  \bibfield  {author} {\bibinfo {author} {\bibfnamefont {Oleksandr}\
  \bibnamefont {Kyriienko}},\ }\bibfield  {title} {\enquote {\bibinfo {title}
  {Quantum inverse iteration algorithm for programmable quantum simulators},}\
  }\href@noop {} {\bibfield  {journal} {\bibinfo  {journal} {npj Quantum Inf.}\
  }\textbf {\bibinfo {volume} {6}},\ \bibinfo {pages} {1--8} (\bibinfo {year}
  {2020})}\BibitemShut {NoStop}%
\bibitem [{\citenamefont {Stair}\ \emph {et~al.}(2020)\citenamefont {Stair},
  \citenamefont {Huang},\ and\ \citenamefont {Evangelista}}]{Stair2020-co}%
  \BibitemOpen
  \bibfield  {author} {\bibinfo {author} {\bibfnamefont {Nicholas~H}\
  \bibnamefont {Stair}}, \bibinfo {author} {\bibfnamefont {Renke}\ \bibnamefont
  {Huang}}, \ and\ \bibinfo {author} {\bibfnamefont {Francesco~A}\ \bibnamefont
  {Evangelista}},\ }\bibfield  {title} {\enquote {\bibinfo {title} {A
  multireference quantum krylov algorithm for strongly correlated electrons},}\
  }\href@noop {} {\bibfield  {journal} {\bibinfo  {journal} {J. Chem. Theory.
  Comput.}\ }\textbf {\bibinfo {volume} {16}},\ \bibinfo {pages} {2236--2245}
  (\bibinfo {year} {2020})}\BibitemShut {NoStop}%
\bibitem [{\citenamefont {Klymko}\ \emph {et~al.}(2022)\citenamefont {Klymko},
  \citenamefont {Mejuto-Zaera}, \citenamefont {Cotton}, \citenamefont
  {Wudarski}, \citenamefont {Urbanek}, \citenamefont {Hait}, \citenamefont
  {Head-Gordon}, \citenamefont {Whaley}, \citenamefont {Moussa}, \citenamefont
  {Wiebe} \emph {et~al.}}]{klymko2022real}%
  \BibitemOpen
  \bibfield  {author} {\bibinfo {author} {\bibfnamefont {Katherine}\
  \bibnamefont {Klymko}}, \bibinfo {author} {\bibfnamefont {Carlos}\
  \bibnamefont {Mejuto-Zaera}}, \bibinfo {author} {\bibfnamefont {Stephen~J}\
  \bibnamefont {Cotton}}, \bibinfo {author} {\bibfnamefont {Filip}\
  \bibnamefont {Wudarski}}, \bibinfo {author} {\bibfnamefont {Miroslav}\
  \bibnamefont {Urbanek}}, \bibinfo {author} {\bibfnamefont {Diptarka}\
  \bibnamefont {Hait}}, \bibinfo {author} {\bibfnamefont {Martin}\ \bibnamefont
  {Head-Gordon}}, \bibinfo {author} {\bibfnamefont {K~Birgitta}\ \bibnamefont
  {Whaley}}, \bibinfo {author} {\bibfnamefont {Jonathan}\ \bibnamefont
  {Moussa}}, \bibinfo {author} {\bibfnamefont {Nathan}\ \bibnamefont {Wiebe}},
  \emph {et~al.},\ }\bibfield  {title} {\enquote {\bibinfo {title} {Real-time
  evolution for ultracompact hamiltonian eigenstates on quantum hardware},}\
  }\href@noop {} {\bibfield  {journal} {\bibinfo  {journal} {PRX Quantum}\
  }\textbf {\bibinfo {volume} {3}},\ \bibinfo {pages} {020323} (\bibinfo {year}
  {2022})}\BibitemShut {NoStop}%
\bibitem [{\citenamefont {Cohn}\ \emph {et~al.}(2021)\citenamefont {Cohn},
  \citenamefont {Motta},\ and\ \citenamefont {Parrish}}]{Cohn2021-mq}%
  \BibitemOpen
  \bibfield  {author} {\bibinfo {author} {\bibfnamefont {Jeffrey}\ \bibnamefont
  {Cohn}}, \bibinfo {author} {\bibfnamefont {Mario}\ \bibnamefont {Motta}}, \
  and\ \bibinfo {author} {\bibfnamefont {Robert~M}\ \bibnamefont {Parrish}},\
  }\bibfield  {title} {\enquote {\bibinfo {title} {Quantum filter
  diagonalization with double-factorized {Hamiltonians}},}\ }\href@noop {}
  {\bibfield  {journal} {\bibinfo  {journal} {arXiv preprint arXiv:2104.08957}\
  } (\bibinfo {year} {2021})}\BibitemShut {NoStop}%
\bibitem [{\citenamefont {Szabo}\ and\ \citenamefont
  {Ostlund}(1996)}]{szabo2012modern}%
  \BibitemOpen
  \bibfield  {author} {\bibinfo {author} {\bibfnamefont {Attila}\ \bibnamefont
  {Szabo}}\ and\ \bibinfo {author} {\bibfnamefont {Neil~S}\ \bibnamefont
  {Ostlund}},\ }\href@noop {} {\emph {\bibinfo {title} {{Modern quantum
  chemistry: {Introduction} to advanced electronic structure theory}}}}\
  (\bibinfo  {publisher} {Dover Publications, Inc.},\ \bibinfo {address}
  {Mineola, New York},\ \bibinfo {year} {1996})\ pp.\ \bibinfo {pages}
  {286--296}\BibitemShut {NoStop}%
\bibitem [{\citenamefont {Whitfield}\ \emph {et~al.}(2013)\citenamefont
  {Whitfield}, \citenamefont {Love},\ and\ \citenamefont
  {Aspuru-Guzik}}]{whitfield2013computational}%
  \BibitemOpen
  \bibfield  {author} {\bibinfo {author} {\bibfnamefont {James~Daniel}\
  \bibnamefont {Whitfield}}, \bibinfo {author} {\bibfnamefont {Peter~John}\
  \bibnamefont {Love}}, \ and\ \bibinfo {author} {\bibfnamefont {Al{\'a}n}\
  \bibnamefont {Aspuru-Guzik}},\ }\bibfield  {title} {\enquote {\bibinfo
  {title} {Computational complexity in electronic structure},}\ }\href@noop {}
  {\bibfield  {journal} {\bibinfo  {journal} {Phys. Chem. Chem. Phys.}\
  }\textbf {\bibinfo {volume} {15}},\ \bibinfo {pages} {397--411} (\bibinfo
  {year} {2013})}\BibitemShut {NoStop}%
\bibitem [{\citenamefont {Cremer}(2011)}]{cremer2011moller}%
  \BibitemOpen
  \bibfield  {author} {\bibinfo {author} {\bibfnamefont {Dieter}\ \bibnamefont
  {Cremer}},\ }\bibfield  {title} {\enquote {\bibinfo {title}
  {{M{\o}ller--Plesset} perturbation theory: from small molecule methods to
  methods for thousands of atoms},}\ }\href@noop {} {\bibfield  {journal}
  {\bibinfo  {journal} {Wiley Interdiscip. Rev. Comput. Mol. Sci.}\ }\textbf
  {\bibinfo {volume} {1}},\ \bibinfo {pages} {509--530} (\bibinfo {year}
  {2011})}\BibitemShut {NoStop}%
\bibitem [{\citenamefont {Bartlett}\ and\ \citenamefont
  {Musia{\l}}(2007)}]{bartlett2007coupled}%
  \BibitemOpen
  \bibfield  {author} {\bibinfo {author} {\bibfnamefont {Rodney~J}\
  \bibnamefont {Bartlett}}\ and\ \bibinfo {author} {\bibfnamefont {Monika}\
  \bibnamefont {Musia{\l}}},\ }\bibfield  {title} {\enquote {\bibinfo {title}
  {Coupled-cluster theory in quantum chemistry},}\ }\href@noop {} {\bibfield
  {journal} {\bibinfo  {journal} {Rev. Mod. Phys.}\ }\textbf {\bibinfo {volume}
  {79}},\ \bibinfo {pages} {291} (\bibinfo {year} {2007})}\BibitemShut
  {NoStop}%
\bibitem [{\citenamefont {Helgaker}\ \emph {et~al.}(2014)\citenamefont
  {Helgaker}, \citenamefont {Jorgensen},\ and\ \citenamefont
  {Olsen}}]{helgaker2014molecular}%
  \BibitemOpen
  \bibfield  {author} {\bibinfo {author} {\bibfnamefont {Trygve}\ \bibnamefont
  {Helgaker}}, \bibinfo {author} {\bibfnamefont {Poul}\ \bibnamefont
  {Jorgensen}}, \ and\ \bibinfo {author} {\bibfnamefont {Jeppe}\ \bibnamefont
  {Olsen}},\ }\href@noop {} {\emph {\bibinfo {title} {Molecular
  electronic-structure theory}}}\ (\bibinfo  {publisher} {John Wiley \& Sons},\
  \bibinfo {year} {2014})\BibitemShut {NoStop}%
\bibitem [{\citenamefont {Vogiatzis}\ \emph {et~al.}(2017)\citenamefont
  {Vogiatzis}, \citenamefont {Ma}, \citenamefont {Olsen}, \citenamefont
  {Gagliardi},\ and\ \citenamefont {De~Jong}}]{vogiatzis2017pushing}%
  \BibitemOpen
  \bibfield  {author} {\bibinfo {author} {\bibfnamefont {Konstantinos~D}\
  \bibnamefont {Vogiatzis}}, \bibinfo {author} {\bibfnamefont {Dongxia}\
  \bibnamefont {Ma}}, \bibinfo {author} {\bibfnamefont {Jeppe}\ \bibnamefont
  {Olsen}}, \bibinfo {author} {\bibfnamefont {Laura}\ \bibnamefont
  {Gagliardi}}, \ and\ \bibinfo {author} {\bibfnamefont {Wibe~A}\ \bibnamefont
  {De~Jong}},\ }\bibfield  {title} {\enquote {\bibinfo {title} {Pushing
  configuration-interaction to the limit: Towards massively parallel {MCSCF}
  calculations},}\ }\href@noop {} {\bibfield  {journal} {\bibinfo  {journal}
  {J. Chem. Phys.}\ }\textbf {\bibinfo {volume} {147}},\ \bibinfo {pages}
  {184111} (\bibinfo {year} {2017})}\BibitemShut {NoStop}%
\bibitem [{\citenamefont {Holmes}\ \emph {et~al.}(2016)\citenamefont {Holmes},
  \citenamefont {Tubman},\ and\ \citenamefont {Umrigar}}]{holmes2016heat}%
  \BibitemOpen
  \bibfield  {author} {\bibinfo {author} {\bibfnamefont {Adam~A}\ \bibnamefont
  {Holmes}}, \bibinfo {author} {\bibfnamefont {Norm~M}\ \bibnamefont {Tubman}},
  \ and\ \bibinfo {author} {\bibfnamefont {CJ}~\bibnamefont {Umrigar}},\
  }\bibfield  {title} {\enquote {\bibinfo {title} {Heat-bath configuration
  interaction: An efficient selected configuration interaction algorithm
  inspired by heat-bath sampling},}\ }\href@noop {} {\bibfield  {journal}
  {\bibinfo  {journal} {J. Chem. Theory Comput.}\ }\textbf {\bibinfo {volume}
  {12}},\ \bibinfo {pages} {3674--3680} (\bibinfo {year} {2016})}\BibitemShut
  {NoStop}%
\bibitem [{\citenamefont {Levine}\ \emph {et~al.}(2020)\citenamefont {Levine},
  \citenamefont {Hait}, \citenamefont {Tubman}, \citenamefont {Lehtola},
  \citenamefont {Whaley},\ and\ \citenamefont
  {Head-Gordon}}]{levine2020casscf}%
  \BibitemOpen
  \bibfield  {author} {\bibinfo {author} {\bibfnamefont {Daniel~S}\
  \bibnamefont {Levine}}, \bibinfo {author} {\bibfnamefont {Diptarka}\
  \bibnamefont {Hait}}, \bibinfo {author} {\bibfnamefont {Norm~M}\ \bibnamefont
  {Tubman}}, \bibinfo {author} {\bibfnamefont {Susi}\ \bibnamefont {Lehtola}},
  \bibinfo {author} {\bibfnamefont {K~Birgitta}\ \bibnamefont {Whaley}}, \ and\
  \bibinfo {author} {\bibfnamefont {Martin}\ \bibnamefont {Head-Gordon}},\
  }\bibfield  {title} {\enquote {\bibinfo {title} {Casscf with extremely large
  active spaces using the adaptive sampling configuration interaction
  method},}\ }\href@noop {} {\bibfield  {journal} {\bibinfo  {journal} {J.
  Chem. Theory Comput.}\ }\textbf {\bibinfo {volume} {16}},\ \bibinfo {pages}
  {2340--2354} (\bibinfo {year} {2020})}\BibitemShut {NoStop}%
\bibitem [{\citenamefont {Tubman}\ \emph {et~al.}(2020)\citenamefont {Tubman},
  \citenamefont {Freeman}, \citenamefont {Levine}, \citenamefont {Hait},
  \citenamefont {Head-Gordon},\ and\ \citenamefont
  {Whaley}}]{tubman2020modern}%
  \BibitemOpen
  \bibfield  {author} {\bibinfo {author} {\bibfnamefont {Norm~M}\ \bibnamefont
  {Tubman}}, \bibinfo {author} {\bibfnamefont {C~Daniel}\ \bibnamefont
  {Freeman}}, \bibinfo {author} {\bibfnamefont {Daniel~S}\ \bibnamefont
  {Levine}}, \bibinfo {author} {\bibfnamefont {Diptarka}\ \bibnamefont {Hait}},
  \bibinfo {author} {\bibfnamefont {Martin}\ \bibnamefont {Head-Gordon}}, \
  and\ \bibinfo {author} {\bibfnamefont {K~Birgitta}\ \bibnamefont {Whaley}},\
  }\bibfield  {title} {\enquote {\bibinfo {title} {Modern approaches to exact
  diagonalization and selected configuration interaction with the adaptive
  sampling {CI} method},}\ }\href@noop {} {\bibfield  {journal} {\bibinfo
  {journal} {J. Chem. Theory Comput}\ }\textbf {\bibinfo {volume} {16}},\
  \bibinfo {pages} {2139--2159} (\bibinfo {year} {2020})}\BibitemShut {NoStop}%
\bibitem [{\citenamefont {Andersson}\ \emph {et~al.}(1990)\citenamefont
  {Andersson}, \citenamefont {Malmqvist}, \citenamefont {Roos}, \citenamefont
  {Sadlej},\ and\ \citenamefont {Wolinski}}]{andersson1990second}%
  \BibitemOpen
  \bibfield  {author} {\bibinfo {author} {\bibfnamefont {Kerstin}\ \bibnamefont
  {Andersson}}, \bibinfo {author} {\bibfnamefont {Per~Aake}\ \bibnamefont
  {Malmqvist}}, \bibinfo {author} {\bibfnamefont {Bj{\"o}rn~O}\ \bibnamefont
  {Roos}}, \bibinfo {author} {\bibfnamefont {Andrzej~J}\ \bibnamefont
  {Sadlej}}, \ and\ \bibinfo {author} {\bibfnamefont {Krzysztof}\ \bibnamefont
  {Wolinski}},\ }\bibfield  {title} {\enquote {\bibinfo {title} {Second-order
  perturbation theory with a {CASSCF} reference function},}\ }\href@noop {}
  {\bibfield  {journal} {\bibinfo  {journal} {J. Phys. Chem.}\ }\textbf
  {\bibinfo {volume} {94}},\ \bibinfo {pages} {5483--5488} (\bibinfo {year}
  {1990})}\BibitemShut {NoStop}%
\bibitem [{\citenamefont {Andersson}\ \emph {et~al.}(1992)\citenamefont
  {Andersson}, \citenamefont {Malmqvist},\ and\ \citenamefont
  {Roos}}]{andersson1992second}%
  \BibitemOpen
  \bibfield  {author} {\bibinfo {author} {\bibfnamefont {Kerstin}\ \bibnamefont
  {Andersson}}, \bibinfo {author} {\bibfnamefont {Per-{\AA}ke}\ \bibnamefont
  {Malmqvist}}, \ and\ \bibinfo {author} {\bibfnamefont {Bj{\"o}rn~O}\
  \bibnamefont {Roos}},\ }\bibfield  {title} {\enquote {\bibinfo {title}
  {Second-order perturbation theory with a complete active space
  self-consistent field reference function},}\ }\href@noop {} {\bibfield
  {journal} {\bibinfo  {journal} {J. Chem. Phys.}\ }\textbf {\bibinfo {volume}
  {96}},\ \bibinfo {pages} {1218--1226} (\bibinfo {year} {1992})}\BibitemShut
  {NoStop}%
\bibitem [{\citenamefont {Angeli}\ \emph {et~al.}(2001)\citenamefont {Angeli},
  \citenamefont {Cimiraglia}, \citenamefont {Evangelisti}, \citenamefont
  {Leininger},\ and\ \citenamefont {Malrieu}}]{angeli2001introduction}%
  \BibitemOpen
  \bibfield  {author} {\bibinfo {author} {\bibfnamefont {Celestino}\
  \bibnamefont {Angeli}}, \bibinfo {author} {\bibfnamefont {Renzo}\
  \bibnamefont {Cimiraglia}}, \bibinfo {author} {\bibfnamefont {S}~\bibnamefont
  {Evangelisti}}, \bibinfo {author} {\bibfnamefont {T}~\bibnamefont
  {Leininger}}, \ and\ \bibinfo {author} {\bibfnamefont {J-P}\ \bibnamefont
  {Malrieu}},\ }\bibfield  {title} {\enquote {\bibinfo {title} {Introduction of
  $n$-electron valence states for multireference perturbation theory},}\
  }\href@noop {} {\bibfield  {journal} {\bibinfo  {journal} {J. Chem. Phys.}\
  }\textbf {\bibinfo {volume} {114}},\ \bibinfo {pages} {10252--10264}
  (\bibinfo {year} {2001})}\BibitemShut {NoStop}%
\bibitem [{\citenamefont {Gill}\ \emph {et~al.}(1988)\citenamefont {Gill},
  \citenamefont {Pople}, \citenamefont {Radom},\ and\ \citenamefont
  {Nobes}}]{gill1988does}%
  \BibitemOpen
  \bibfield  {author} {\bibinfo {author} {\bibfnamefont {Peter M~W}\
  \bibnamefont {Gill}}, \bibinfo {author} {\bibfnamefont {John~A}\ \bibnamefont
  {Pople}}, \bibinfo {author} {\bibfnamefont {Leo}\ \bibnamefont {Radom}}, \
  and\ \bibinfo {author} {\bibfnamefont {Ross~H}\ \bibnamefont {Nobes}},\
  }\bibfield  {title} {\enquote {\bibinfo {title} {Why does unrestricted
  {M{\o}ller--Plesset} perturbation theory converge so slowly for
  spin-contaminated wave functions?}}\ }\href@noop {} {\bibfield  {journal}
  {\bibinfo  {journal} {J. Chem. Phys.}\ }\textbf {\bibinfo {volume} {89}},\
  \bibinfo {pages} {7307--7314} (\bibinfo {year} {1988})}\BibitemShut {NoStop}%
\bibitem [{\citenamefont {Mayhall}\ \emph {et~al.}(2014)\citenamefont
  {Mayhall}, \citenamefont {Horn}, \citenamefont {Sundstrom},\ and\
  \citenamefont {Head-Gordon}}]{mayhall2014spin}%
  \BibitemOpen
  \bibfield  {author} {\bibinfo {author} {\bibfnamefont {Nicholas~J}\
  \bibnamefont {Mayhall}}, \bibinfo {author} {\bibfnamefont {Paul~R}\
  \bibnamefont {Horn}}, \bibinfo {author} {\bibfnamefont {Eric~J}\ \bibnamefont
  {Sundstrom}}, \ and\ \bibinfo {author} {\bibfnamefont {Martin}\ \bibnamefont
  {Head-Gordon}},\ }\bibfield  {title} {\enquote {\bibinfo {title} {Spin-flip
  non-orthogonal configuration interaction: a variational and almost black-box
  method for describing strongly correlated molecules},}\ }\href@noop {}
  {\bibfield  {journal} {\bibinfo  {journal} {Phys. Chem. Chem. Phys.}\
  }\textbf {\bibinfo {volume} {16}},\ \bibinfo {pages} {22694--22705} (\bibinfo
  {year} {2014})}\BibitemShut {NoStop}%
\bibitem [{\citenamefont {Shavitt}\ and\ \citenamefont
  {Bartlett}(2009)}]{shavitt2009many}%
  \BibitemOpen
  \bibfield  {author} {\bibinfo {author} {\bibfnamefont {Isaiah}\ \bibnamefont
  {Shavitt}}\ and\ \bibinfo {author} {\bibfnamefont {Rodney~J}\ \bibnamefont
  {Bartlett}},\ }\href@noop {} {\emph {\bibinfo {title} {Many-body methods in
  chemistry and physics: {MBPT} and coupled-cluster theory}}}\ (\bibinfo
  {publisher} {Cambridge university press},\ \bibinfo {year}
  {2009})\BibitemShut {NoStop}%
\bibitem [{\citenamefont {Sch{\"u}tz}\ \emph {et~al.}(1999)\citenamefont
  {Sch{\"u}tz}, \citenamefont {Hetzer},\ and\ \citenamefont
  {Werner}}]{schutz1999low}%
  \BibitemOpen
  \bibfield  {author} {\bibinfo {author} {\bibfnamefont {Martin}\ \bibnamefont
  {Sch{\"u}tz}}, \bibinfo {author} {\bibfnamefont {Georg}\ \bibnamefont
  {Hetzer}}, \ and\ \bibinfo {author} {\bibfnamefont {Hans-Joachim}\
  \bibnamefont {Werner}},\ }\bibfield  {title} {\enquote {\bibinfo {title}
  {Low-order scaling local electron correlation methods. i. linear scaling
  local mp2},}\ }\href@noop {} {\bibfield  {journal} {\bibinfo  {journal} {The
  Journal of chemical physics}\ }\textbf {\bibinfo {volume} {111}},\ \bibinfo
  {pages} {5691--5705} (\bibinfo {year} {1999})}\BibitemShut {NoStop}%
\bibitem [{\citenamefont {Riplinger}\ \emph {et~al.}(2016)\citenamefont
  {Riplinger}, \citenamefont {Pinski}, \citenamefont {Becker}, \citenamefont
  {Valeev},\ and\ \citenamefont {Neese}}]{riplinger2016sparse}%
  \BibitemOpen
  \bibfield  {author} {\bibinfo {author} {\bibfnamefont {Christoph}\
  \bibnamefont {Riplinger}}, \bibinfo {author} {\bibfnamefont {Peter}\
  \bibnamefont {Pinski}}, \bibinfo {author} {\bibfnamefont {Ute}\ \bibnamefont
  {Becker}}, \bibinfo {author} {\bibfnamefont {Edward~F}\ \bibnamefont
  {Valeev}}, \ and\ \bibinfo {author} {\bibfnamefont {Frank}\ \bibnamefont
  {Neese}},\ }\bibfield  {title} {\enquote {\bibinfo {title} {Sparse maps—a
  systematic infrastructure for reduced-scaling electronic structure methods.
  ii. linear scaling domain based pair natural orbital coupled cluster
  theory},}\ }\href@noop {} {\bibfield  {journal} {\bibinfo  {journal} {The
  Journal of chemical physics}\ }\textbf {\bibinfo {volume} {144}},\ \bibinfo
  {pages} {024109} (\bibinfo {year} {2016})}\BibitemShut {NoStop}%
\bibitem [{\citenamefont {Subotnik}\ \emph {et~al.}(2006)\citenamefont
  {Subotnik}, \citenamefont {Sodt},\ and\ \citenamefont
  {Head-Gordon}}]{subotnik2006near}%
  \BibitemOpen
  \bibfield  {author} {\bibinfo {author} {\bibfnamefont {Joseph~E}\
  \bibnamefont {Subotnik}}, \bibinfo {author} {\bibfnamefont {Alex}\
  \bibnamefont {Sodt}}, \ and\ \bibinfo {author} {\bibfnamefont {Martin}\
  \bibnamefont {Head-Gordon}},\ }\bibfield  {title} {\enquote {\bibinfo {title}
  {A near linear-scaling smooth local coupled cluster algorithm for electronic
  structure},}\ }\href@noop {} {\bibfield  {journal} {\bibinfo  {journal} {The
  Journal of chemical physics}\ }\textbf {\bibinfo {volume} {125}},\ \bibinfo
  {pages} {074116} (\bibinfo {year} {2006})}\BibitemShut {NoStop}%
\bibitem [{\citenamefont {Matsuzawa}\ and\ \citenamefont
  {Kurashige}(2020)}]{matsuzawa2020jastrow}%
  \BibitemOpen
  \bibfield  {author} {\bibinfo {author} {\bibfnamefont {Yuta}\ \bibnamefont
  {Matsuzawa}}\ and\ \bibinfo {author} {\bibfnamefont {Yuki}\ \bibnamefont
  {Kurashige}},\ }\bibfield  {title} {\enquote {\bibinfo {title}
  {{Jastrow-type} decomposition in quantum chemistry for low-depth quantum
  circuits},}\ }\href@noop {} {\bibfield  {journal} {\bibinfo  {journal} {J.
  Chem. Theory Comput.}\ }\textbf {\bibinfo {volume} {16}},\ \bibinfo {pages}
  {944--952} (\bibinfo {year} {2020})}\BibitemShut {NoStop}%
\bibitem [{\citenamefont {McArdle}\ \emph {et~al.}(2020)\citenamefont
  {McArdle}, \citenamefont {Endo}, \citenamefont {Aspuru-Guzik}, \citenamefont
  {Benjamin},\ and\ \citenamefont {Yuan}}]{mcardle2020quantum}%
  \BibitemOpen
  \bibfield  {author} {\bibinfo {author} {\bibfnamefont {Sam}\ \bibnamefont
  {McArdle}}, \bibinfo {author} {\bibfnamefont {Suguru}\ \bibnamefont {Endo}},
  \bibinfo {author} {\bibfnamefont {Al{\'a}n}\ \bibnamefont {Aspuru-Guzik}},
  \bibinfo {author} {\bibfnamefont {Simon~C}\ \bibnamefont {Benjamin}}, \ and\
  \bibinfo {author} {\bibfnamefont {Xiao}\ \bibnamefont {Yuan}},\ }\bibfield
  {title} {\enquote {\bibinfo {title} {Quantum computational chemistry},}\
  }\href@noop {} {\bibfield  {journal} {\bibinfo  {journal} {Rev. Mod. Phys.}\
  }\textbf {\bibinfo {volume} {92}},\ \bibinfo {pages} {015003} (\bibinfo
  {year} {2020})}\BibitemShut {NoStop}%
\bibitem [{\citenamefont {Epstein}(1926)}]{epstein1926}%
  \BibitemOpen
  \bibfield  {author} {\bibinfo {author} {\bibfnamefont {P.~S.}\ \bibnamefont
  {Epstein}},\ }\bibfield  {title} {\enquote {\bibinfo {title} {The stark
  effect from the point of view of {Schroedinger's} quantum theory},}\
  }\href@noop {} {\bibfield  {journal} {\bibinfo  {journal} {Phys. Rev.}\
  }\textbf {\bibinfo {volume} {28}} (\bibinfo {year} {1926})}\BibitemShut
  {NoStop}%
\bibitem [{\citenamefont {Nesbet}(1955)}]{nesbet1955}%
  \BibitemOpen
  \bibfield  {author} {\bibinfo {author} {\bibfnamefont {RK}~\bibnamefont
  {Nesbet}},\ }\bibfield  {title} {\enquote {\bibinfo {title} {Configuration
  interaction in orbital theories},}\ }\href@noop {} {\bibfield  {journal}
  {\bibinfo  {journal} {Proc. R. Soc. Lond. A}\ }\textbf {\bibinfo {volume}
  {230}},\ \bibinfo {pages} {312--321} (\bibinfo {year} {1955})}\BibitemShut
  {NoStop}%
\bibitem [{\citenamefont {Murray}\ and\ \citenamefont
  {Davidson}(1992)}]{murray1992different}%
  \BibitemOpen
  \bibfield  {author} {\bibinfo {author} {\bibfnamefont {Christopher}\
  \bibnamefont {Murray}}\ and\ \bibinfo {author} {\bibfnamefont {Ernest~R}\
  \bibnamefont {Davidson}},\ }\bibfield  {title} {\enquote {\bibinfo {title}
  {Different forms of perturbation theory for the calculation of the
  correlation energy},}\ }\href@noop {} {\bibfield  {journal} {\bibinfo
  {journal} {Int. J. Quantum Chem.}\ }\textbf {\bibinfo {volume} {43}},\
  \bibinfo {pages} {755--768} (\bibinfo {year} {1992})}\BibitemShut {NoStop}%
\bibitem [{\citenamefont {Grimme}(2003)}]{grimme2003improved}%
  \BibitemOpen
  \bibfield  {author} {\bibinfo {author} {\bibfnamefont {Stefan}\ \bibnamefont
  {Grimme}},\ }\bibfield  {title} {\enquote {\bibinfo {title} {Improved
  second-order {M{\o}ller--Plesset} perturbation theory by separate scaling of
  parallel-and antiparallel-spin pair correlation energies},}\ }\href@noop {}
  {\bibfield  {journal} {\bibinfo  {journal} {J. Chem. Phys.}\ }\textbf
  {\bibinfo {volume} {118}},\ \bibinfo {pages} {9095--9102} (\bibinfo {year}
  {2003})}\BibitemShut {NoStop}%
\bibitem [{\citenamefont {Jung}\ \emph {et~al.}(2004)\citenamefont {Jung},
  \citenamefont {Lochan}, \citenamefont {Dutoi},\ and\ \citenamefont
  {Head-Gordon}}]{jung2004scaled}%
  \BibitemOpen
  \bibfield  {author} {\bibinfo {author} {\bibfnamefont {Yousung}\ \bibnamefont
  {Jung}}, \bibinfo {author} {\bibfnamefont {Rohini~C}\ \bibnamefont {Lochan}},
  \bibinfo {author} {\bibfnamefont {Anthony~D}\ \bibnamefont {Dutoi}}, \ and\
  \bibinfo {author} {\bibfnamefont {Martin}\ \bibnamefont {Head-Gordon}},\
  }\bibfield  {title} {\enquote {\bibinfo {title} {Scaled opposite-spin second
  order {M{\o}ller--Plesset} correlation energy: {An} economical electronic
  structure method},}\ }\href@noop {} {\bibfield  {journal} {\bibinfo
  {journal} {J. Chem. Phys.}\ }\textbf {\bibinfo {volume} {121}},\ \bibinfo
  {pages} {9793--9802} (\bibinfo {year} {2004})}\BibitemShut {NoStop}%
\bibitem [{\citenamefont {Motta}\ \emph {et~al.}(2021)\citenamefont {Motta},
  \citenamefont {Ye}, \citenamefont {McClean}, \citenamefont {Li},
  \citenamefont {Minnich}, \citenamefont {Babbush},\ and\ \citenamefont
  {Chan}}]{motta2021low}%
  \BibitemOpen
  \bibfield  {author} {\bibinfo {author} {\bibfnamefont {Mario}\ \bibnamefont
  {Motta}}, \bibinfo {author} {\bibfnamefont {Erika}\ \bibnamefont {Ye}},
  \bibinfo {author} {\bibfnamefont {Jarrod~R}\ \bibnamefont {McClean}},
  \bibinfo {author} {\bibfnamefont {Zhendong}\ \bibnamefont {Li}}, \bibinfo
  {author} {\bibfnamefont {Austin~J}\ \bibnamefont {Minnich}}, \bibinfo
  {author} {\bibfnamefont {Ryan}\ \bibnamefont {Babbush}}, \ and\ \bibinfo
  {author} {\bibfnamefont {Garnet Kin-Lic}\ \bibnamefont {Chan}},\ }\bibfield
  {title} {\enquote {\bibinfo {title} {Low rank representations for quantum
  simulation of electronic structure},}\ }\href@noop {} {\bibfield  {journal}
  {\bibinfo  {journal} {npj Quantum Inf.}\ }\textbf {\bibinfo {volume} {7}},\
  \bibinfo {pages} {1--7} (\bibinfo {year} {2021})}\BibitemShut {NoStop}%
\bibitem [{\citenamefont {Rubin}\ \emph {et~al.}(2022)\citenamefont {Rubin},
  \citenamefont {Lee},\ and\ \citenamefont {Babbush}}]{rubin_compressing_2021}%
  \BibitemOpen
  \bibfield  {author} {\bibinfo {author} {\bibfnamefont {Nicholas~C}\
  \bibnamefont {Rubin}}, \bibinfo {author} {\bibfnamefont {Joonho}\
  \bibnamefont {Lee}}, \ and\ \bibinfo {author} {\bibfnamefont {Ryan}\
  \bibnamefont {Babbush}},\ }\bibfield  {title} {\enquote {\bibinfo {title}
  {Compressing many-body fermion operators under unitary constraints},}\
  }\href@noop {} {\bibfield  {journal} {\bibinfo  {journal} {J. Chem. Theory
  Comput.}\ }\textbf {\bibinfo {volume} {18}},\ \bibinfo {pages} {1480--1488}
  (\bibinfo {year} {2022})}\BibitemShut {NoStop}%
\bibitem [{\citenamefont {Peng}\ and\ \citenamefont
  {Kowalski}(2017)}]{peng2017highly}%
  \BibitemOpen
  \bibfield  {author} {\bibinfo {author} {\bibfnamefont {Bo}~\bibnamefont
  {Peng}}\ and\ \bibinfo {author} {\bibfnamefont {Karol}\ \bibnamefont
  {Kowalski}},\ }\bibfield  {title} {\enquote {\bibinfo {title} {Highly
  efficient and scalable compound decomposition of two-electron integral tensor
  and its application in coupled cluster calculations},}\ }\href@noop {}
  {\bibfield  {journal} {\bibinfo  {journal} {J. Chem. Theory Comput.}\
  }\textbf {\bibinfo {volume} {13}},\ \bibinfo {pages} {4179--4192} (\bibinfo
  {year} {2017})}\BibitemShut {NoStop}%
\bibitem [{\citenamefont {Motta}\ \emph {et~al.}(2019)\citenamefont {Motta},
  \citenamefont {Shee}, \citenamefont {Zhang},\ and\ \citenamefont
  {Chan}}]{motta2019efficient}%
  \BibitemOpen
  \bibfield  {author} {\bibinfo {author} {\bibfnamefont {Mario}\ \bibnamefont
  {Motta}}, \bibinfo {author} {\bibfnamefont {James}\ \bibnamefont {Shee}},
  \bibinfo {author} {\bibfnamefont {Shiwei}\ \bibnamefont {Zhang}}, \ and\
  \bibinfo {author} {\bibfnamefont {Garnet Kin-Lic}\ \bibnamefont {Chan}},\
  }\bibfield  {title} {\enquote {\bibinfo {title} {Efficient ab initio
  auxiliary-field quantum {Monte Carlo} calculations in {Gaussian} bases via
  low-rank tensor decomposition},}\ }\href@noop {} {\bibfield  {journal}
  {\bibinfo  {journal} {J. Chem. Theory Comput.}\ }\textbf {\bibinfo {volume}
  {15}},\ \bibinfo {pages} {3510--3521} (\bibinfo {year} {2019})}\BibitemShut
  {NoStop}%
\bibitem [{\citenamefont {Kivlichan}\ \emph {et~al.}(2018)\citenamefont
  {Kivlichan}, \citenamefont {McClean}, \citenamefont {Wiebe}, \citenamefont
  {Gidney}, \citenamefont {Aspuru-Guzik}, \citenamefont {Chan},\ and\
  \citenamefont {Babbush}}]{kivlichan2018quantum}%
  \BibitemOpen
  \bibfield  {author} {\bibinfo {author} {\bibfnamefont {Ian~D}\ \bibnamefont
  {Kivlichan}}, \bibinfo {author} {\bibfnamefont {Jarrod}\ \bibnamefont
  {McClean}}, \bibinfo {author} {\bibfnamefont {Nathan}\ \bibnamefont {Wiebe}},
  \bibinfo {author} {\bibfnamefont {Craig}\ \bibnamefont {Gidney}}, \bibinfo
  {author} {\bibfnamefont {Al{\'a}n}\ \bibnamefont {Aspuru-Guzik}}, \bibinfo
  {author} {\bibfnamefont {Garnet Kin-Lic}\ \bibnamefont {Chan}}, \ and\
  \bibinfo {author} {\bibfnamefont {Ryan}\ \bibnamefont {Babbush}},\ }\bibfield
   {title} {\enquote {\bibinfo {title} {Quantum simulation of electronic
  structure with linear depth and connectivity},}\ }\href@noop {} {\bibfield
  {journal} {\bibinfo  {journal} {Phys. Rev. Lett.}\ }\textbf {\bibinfo
  {volume} {120}},\ \bibinfo {pages} {110501} (\bibinfo {year}
  {2018})}\BibitemShut {NoStop}%
\bibitem [{\citenamefont {Whitfield}\ \emph {et~al.}(2011)\citenamefont
  {Whitfield}, \citenamefont {Biamonte},\ and\ \citenamefont
  {Aspuru-Guzik}}]{whitfield2011simulation}%
  \BibitemOpen
  \bibfield  {author} {\bibinfo {author} {\bibfnamefont {James~D}\ \bibnamefont
  {Whitfield}}, \bibinfo {author} {\bibfnamefont {Jacob}\ \bibnamefont
  {Biamonte}}, \ and\ \bibinfo {author} {\bibfnamefont {Al{\'a}n}\ \bibnamefont
  {Aspuru-Guzik}},\ }\bibfield  {title} {\enquote {\bibinfo {title} {Simulation
  of electronic structure {Hamiltonians} using quantum computers},}\
  }\href@noop {} {\bibfield  {journal} {\bibinfo  {journal} {Mol. Phys.}\
  }\textbf {\bibinfo {volume} {109}},\ \bibinfo {pages} {735--750} (\bibinfo
  {year} {2011})}\BibitemShut {NoStop}%
\bibitem [{\citenamefont {Hastings}\ \emph {et~al.}(2015)\citenamefont
  {Hastings}, \citenamefont {Wecker}, \citenamefont {Bauer},\ and\
  \citenamefont {Troyer}}]{hastings2015improving}%
  \BibitemOpen
  \bibfield  {author} {\bibinfo {author} {\bibfnamefont {Matthew~B}\
  \bibnamefont {Hastings}}, \bibinfo {author} {\bibfnamefont {Dave}\
  \bibnamefont {Wecker}}, \bibinfo {author} {\bibfnamefont {Bela}\ \bibnamefont
  {Bauer}}, \ and\ \bibinfo {author} {\bibfnamefont {Matthias}\ \bibnamefont
  {Troyer}},\ }\bibfield  {title} {\enquote {\bibinfo {title} {Improving
  quantum algorithms for quantum chemistry},}\ }\href@noop {} {\bibfield
  {journal} {\bibinfo  {journal} {Quantum Inf. \& Comput.}\ }\textbf {\bibinfo
  {volume} {15}},\ \bibinfo {pages} {1--21} (\bibinfo {year}
  {2015})}\BibitemShut {NoStop}%
\bibitem [{\citenamefont {Kato}(1957)}]{kato1957eigenfunctions}%
  \BibitemOpen
  \bibfield  {author} {\bibinfo {author} {\bibfnamefont {Tosio}\ \bibnamefont
  {Kato}},\ }\bibfield  {title} {\enquote {\bibinfo {title} {On the
  eigenfunctions of many-particle systems in quantum mechanics},}\ }\href@noop
  {} {\bibfield  {journal} {\bibinfo  {journal} {Commun. Pure Appl. Math.}\
  }\textbf {\bibinfo {volume} {10}},\ \bibinfo {pages} {151--177} (\bibinfo
  {year} {1957})}\BibitemShut {NoStop}%
\bibitem [{\citenamefont {Hammond}\ \emph {et~al.}(1994)\citenamefont
  {Hammond}, \citenamefont {Lester},\ and\ \citenamefont
  {Reynolds}}]{hammond1994monte}%
  \BibitemOpen
  \bibfield  {author} {\bibinfo {author} {\bibfnamefont {Brian~L}\ \bibnamefont
  {Hammond}}, \bibinfo {author} {\bibfnamefont {William~A}\ \bibnamefont
  {Lester}}, \ and\ \bibinfo {author} {\bibfnamefont {Peter~James}\
  \bibnamefont {Reynolds}},\ }\href@noop {} {\emph {\bibinfo {title} {{Monte
  Carlo} methods in ab initio quantum chemistry}}},\ Vol.~\bibinfo {volume}
  {1}\ (\bibinfo  {publisher} {World Scientific},\ \bibinfo {year}
  {1994})\BibitemShut {NoStop}%
\bibitem [{\citenamefont {Reynolds}\ \emph {et~al.}(1982)\citenamefont
  {Reynolds}, \citenamefont {Ceperley}, \citenamefont {Alder},\ and\
  \citenamefont {Lester~Jr}}]{reynolds1982fixed}%
  \BibitemOpen
  \bibfield  {author} {\bibinfo {author} {\bibfnamefont {Peter~J}\ \bibnamefont
  {Reynolds}}, \bibinfo {author} {\bibfnamefont {David~M}\ \bibnamefont
  {Ceperley}}, \bibinfo {author} {\bibfnamefont {Berni~J}\ \bibnamefont
  {Alder}}, \ and\ \bibinfo {author} {\bibfnamefont {William~A}\ \bibnamefont
  {Lester~Jr}},\ }\bibfield  {title} {\enquote {\bibinfo {title} {Fixed-node
  quantum {Monte Carlo} for molecules},}\ }\href@noop {} {\bibfield  {journal}
  {\bibinfo  {journal} {J. Chem. Phys.}\ }\textbf {\bibinfo {volume} {77}},\
  \bibinfo {pages} {5593--5603} (\bibinfo {year} {1982})}\BibitemShut {NoStop}%
\bibitem [{\citenamefont
  {Neuscamman}(2013{\natexlab{a}})}]{neuscamman2013communication}%
  \BibitemOpen
  \bibfield  {author} {\bibinfo {author} {\bibfnamefont {Eric}\ \bibnamefont
  {Neuscamman}},\ }\bibfield  {title} {\enquote {\bibinfo {title}
  {Communication: {A Jastrow} factor coupled cluster theory for weak and strong
  electron correlation},}\ }\href@noop {} {\bibfield  {journal} {\bibinfo
  {journal} {J. Chem. Phys.}\ }\textbf {\bibinfo {volume} {139}},\ \bibinfo
  {pages} {181101} (\bibinfo {year} {2013}{\natexlab{a}})}\BibitemShut
  {NoStop}%
\bibitem [{\citenamefont
  {Neuscamman}(2013{\natexlab{b}})}]{neuscamman2013jastrow}%
  \BibitemOpen
  \bibfield  {author} {\bibinfo {author} {\bibfnamefont {Eric}\ \bibnamefont
  {Neuscamman}},\ }\bibfield  {title} {\enquote {\bibinfo {title} {The
  {Jastrow} antisymmetric geminal power in {Hilbert} space: {Theory,}
  benchmarking, and application to a novel transition state},}\ }\href@noop {}
  {\bibfield  {journal} {\bibinfo  {journal} {J. Chem. Phys.}\ }\textbf
  {\bibinfo {volume} {139}},\ \bibinfo {pages} {194105} (\bibinfo {year}
  {2013}{\natexlab{b}})}\BibitemShut {NoStop}%
\bibitem [{\citenamefont {Neuscamman}(2016)}]{neuscamman2016improved}%
  \BibitemOpen
  \bibfield  {author} {\bibinfo {author} {\bibfnamefont {Eric}\ \bibnamefont
  {Neuscamman}},\ }\bibfield  {title} {\enquote {\bibinfo {title} {Improved
  optimization for the cluster {Jastrow} antisymmetric geminal power and tests
  on triple-bond dissociations},}\ }\href@noop {} {\bibfield  {journal}
  {\bibinfo  {journal} {J. Chem. Theory Comput.}\ }\textbf {\bibinfo {volume}
  {12}},\ \bibinfo {pages} {3149--3159} (\bibinfo {year} {2016})}\BibitemShut
  {NoStop}%
\bibitem [{\citenamefont {Epperly}\ \emph {et~al.}(2021)\citenamefont
  {Epperly}, \citenamefont {Lin},\ and\ \citenamefont
  {Nakatsukasa}}]{epperly_theory_2021}%
  \BibitemOpen
  \bibfield  {author} {\bibinfo {author} {\bibfnamefont {Ethan~N}\ \bibnamefont
  {Epperly}}, \bibinfo {author} {\bibfnamefont {Lin}\ \bibnamefont {Lin}}, \
  and\ \bibinfo {author} {\bibfnamefont {Yuji}\ \bibnamefont {Nakatsukasa}},\
  }\bibfield  {title} {\enquote {\bibinfo {title} {A theory of quantum subspace
  diagonalization},}\ }\href@noop {} {\bibfield  {journal} {\bibinfo  {journal}
  {arXiv preprint arXiv:2110.07492}\ } (\bibinfo {year} {2021})}\BibitemShut
  {NoStop}%
\bibitem [{\citenamefont {Huggins}\ \emph {et~al.}(2021)\citenamefont
  {Huggins}, \citenamefont {McClean}, \citenamefont {Rubin}, \citenamefont
  {Jiang}, \citenamefont {Wiebe}, \citenamefont {Whaley},\ and\ \citenamefont
  {Babbush}}]{Huggins2021-vu}%
  \BibitemOpen
  \bibfield  {author} {\bibinfo {author} {\bibfnamefont {William~J}\
  \bibnamefont {Huggins}}, \bibinfo {author} {\bibfnamefont {Jarrod~R}\
  \bibnamefont {McClean}}, \bibinfo {author} {\bibfnamefont {Nicholas~C}\
  \bibnamefont {Rubin}}, \bibinfo {author} {\bibfnamefont {Zhang}\ \bibnamefont
  {Jiang}}, \bibinfo {author} {\bibfnamefont {Nathan}\ \bibnamefont {Wiebe}},
  \bibinfo {author} {\bibfnamefont {K~Birgitta}\ \bibnamefont {Whaley}}, \ and\
  \bibinfo {author} {\bibfnamefont {Ryan}\ \bibnamefont {Babbush}},\ }\bibfield
   {title} {\enquote {\bibinfo {title} {Efficient and noise resilient
  measurements for quantum chemistry on near-term quantum computers},}\
  }\href@noop {} {\bibfield  {journal} {\bibinfo  {journal} {npj Quantum Inf.}\
  }\textbf {\bibinfo {volume} {7}},\ \bibinfo {pages} {1--9} (\bibinfo {year}
  {2021})}\BibitemShut {NoStop}%
\bibitem [{\citenamefont {McClean}\ \emph {et~al.}(2020)\citenamefont
  {McClean}, \citenamefont {Rubin}, \citenamefont {Sung}, \citenamefont
  {Kivlichan}, \citenamefont {Bonet-Monroig}, \citenamefont {Cao},
  \citenamefont {Dai}, \citenamefont {Fried}, \citenamefont {Gidney},
  \citenamefont {Gimby} \emph {et~al.}}]{mcclean2020openfermion}%
  \BibitemOpen
  \bibfield  {author} {\bibinfo {author} {\bibfnamefont {Jarrod~R}\
  \bibnamefont {McClean}}, \bibinfo {author} {\bibfnamefont {Nicholas~C}\
  \bibnamefont {Rubin}}, \bibinfo {author} {\bibfnamefont {Kevin~J}\
  \bibnamefont {Sung}}, \bibinfo {author} {\bibfnamefont {Ian~D}\ \bibnamefont
  {Kivlichan}}, \bibinfo {author} {\bibfnamefont {Xavier}\ \bibnamefont
  {Bonet-Monroig}}, \bibinfo {author} {\bibfnamefont {Yudong}\ \bibnamefont
  {Cao}}, \bibinfo {author} {\bibfnamefont {Chengyu}\ \bibnamefont {Dai}},
  \bibinfo {author} {\bibfnamefont {E~Schuyler}\ \bibnamefont {Fried}},
  \bibinfo {author} {\bibfnamefont {Craig}\ \bibnamefont {Gidney}}, \bibinfo
  {author} {\bibfnamefont {Brendan}\ \bibnamefont {Gimby}},  \emph {et~al.},\
  }\bibfield  {title} {\enquote {\bibinfo {title} {{OpenFermion:} the
  electronic structure package for quantum computers},}\ }\href@noop {}
  {\bibfield  {journal} {\bibinfo  {journal} {Quantum Sci. Technol.}\ }\textbf
  {\bibinfo {volume} {5}},\ \bibinfo {pages} {034014} (\bibinfo {year}
  {2020})}\BibitemShut {NoStop}%
\bibitem [{\citenamefont {Sun}\ \emph {et~al.}(2020)\citenamefont {Sun},
  \citenamefont {Zhang}, \citenamefont {Banerjee}, \citenamefont {Bao},
  \citenamefont {Barbry}, \citenamefont {Blunt}, \citenamefont {Bogdanov},
  \citenamefont {Booth}, \citenamefont {Chen}, \citenamefont {Cui} \emph
  {et~al.}}]{sun2020recent}%
  \BibitemOpen
  \bibfield  {author} {\bibinfo {author} {\bibfnamefont {Qiming}\ \bibnamefont
  {Sun}}, \bibinfo {author} {\bibfnamefont {Xing}\ \bibnamefont {Zhang}},
  \bibinfo {author} {\bibfnamefont {Samragni}\ \bibnamefont {Banerjee}},
  \bibinfo {author} {\bibfnamefont {Peng}\ \bibnamefont {Bao}}, \bibinfo
  {author} {\bibfnamefont {Marc}\ \bibnamefont {Barbry}}, \bibinfo {author}
  {\bibfnamefont {Nick~S}\ \bibnamefont {Blunt}}, \bibinfo {author}
  {\bibfnamefont {Nikolay~A}\ \bibnamefont {Bogdanov}}, \bibinfo {author}
  {\bibfnamefont {George~H}\ \bibnamefont {Booth}}, \bibinfo {author}
  {\bibfnamefont {Jia}\ \bibnamefont {Chen}}, \bibinfo {author} {\bibfnamefont
  {Zhi-Hao}\ \bibnamefont {Cui}},  \emph {et~al.},\ }\bibfield  {title}
  {\enquote {\bibinfo {title} {Recent developments in the {PySCF} program
  package},}\ }\href@noop {} {\bibfield  {journal} {\bibinfo  {journal} {J.
  Chem. Phys.}\ }\textbf {\bibinfo {volume} {153}},\ \bibinfo {pages} {024109}
  (\bibinfo {year} {2020})}\BibitemShut {NoStop}%
\bibitem [{\citenamefont {Epifanovsky}\ \emph {et~al.}(2021)\citenamefont
  {Epifanovsky} \emph {et~al.}}]{epifanovsky2021software}%
  \BibitemOpen
  \bibfield  {author} {\bibinfo {author} {\bibfnamefont {Evgeny}\ \bibnamefont
  {Epifanovsky}} \emph {et~al.},\ }\bibfield  {title} {\enquote {\bibinfo
  {title} {Software for the frontiers of quantum chemistry: An overview of
  developments in the {Q-Chem} 5 package},}\ }\href@noop {} {\bibfield
  {journal} {\bibinfo  {journal} {J. Chem. Phys.}\ }\textbf {\bibinfo {volume}
  {155}},\ \bibinfo {pages} {084801} (\bibinfo {year} {2021})}\BibitemShut
  {NoStop}%
\bibitem [{\citenamefont {Boys}(1960)}]{boys1960construction}%
  \BibitemOpen
  \bibfield  {author} {\bibinfo {author} {\bibfnamefont {S~Francis}\
  \bibnamefont {Boys}},\ }\bibfield  {title} {\enquote {\bibinfo {title}
  {Construction of some molecular orbitals to be approximately invariant for
  changes from one molecule to another},}\ }\href@noop {} {\bibfield  {journal}
  {\bibinfo  {journal} {Rev. Mod. Phys.}\ }\textbf {\bibinfo {volume} {32}},\
  \bibinfo {pages} {296} (\bibinfo {year} {1960})}\BibitemShut {NoStop}%
\bibitem [{\citenamefont {Edmiston}\ and\ \citenamefont
  {Ruedenberg}(1963)}]{edmiston1963localized}%
  \BibitemOpen
  \bibfield  {author} {\bibinfo {author} {\bibfnamefont {Clyde}\ \bibnamefont
  {Edmiston}}\ and\ \bibinfo {author} {\bibfnamefont {Klaus}\ \bibnamefont
  {Ruedenberg}},\ }\bibfield  {title} {\enquote {\bibinfo {title} {Localized
  atomic and molecular orbitals},}\ }\href@noop {} {\bibfield  {journal}
  {\bibinfo  {journal} {Rev. Mod. Phys.}\ }\textbf {\bibinfo {volume} {35}},\
  \bibinfo {pages} {457} (\bibinfo {year} {1963})}\BibitemShut {NoStop}%
\bibitem [{\citenamefont {Burton}\ and\ \citenamefont
  {Wales}(2021)}]{burton2020energy}%
  \BibitemOpen
  \bibfield  {author} {\bibinfo {author} {\bibfnamefont {Hugh~GA}\ \bibnamefont
  {Burton}}\ and\ \bibinfo {author} {\bibfnamefont {David~J}\ \bibnamefont
  {Wales}},\ }\bibfield  {title} {\enquote {\bibinfo {title} {Energy landscapes
  for electronic structure},}\ }\href@noop {} {\bibfield  {journal} {\bibinfo
  {journal} {J. Chem. Theory Comput.}\ }\textbf {\bibinfo {volume} {17}},\
  \bibinfo {pages} {151–--169} (\bibinfo {year} {2021})}\BibitemShut
  {NoStop}%
\bibitem [{\citenamefont {Hait}\ and\ \citenamefont
  {Head-Gordon}(2020)}]{hait2020excited}%
  \BibitemOpen
  \bibfield  {author} {\bibinfo {author} {\bibfnamefont {Diptarka}\
  \bibnamefont {Hait}}\ and\ \bibinfo {author} {\bibfnamefont {Martin}\
  \bibnamefont {Head-Gordon}},\ }\bibfield  {title} {\enquote {\bibinfo {title}
  {Excited state orbital optimization via minimizing the square of the
  gradient: {General} approach and application to singly and doubly excited
  states via density functional theory},}\ }\href@noop {} {\bibfield  {journal}
  {\bibinfo  {journal} {J. Chem. Theory Comput.}\ }\textbf {\bibinfo {volume}
  {16}},\ \bibinfo {pages} {1699--1710} (\bibinfo {year} {2020})}\BibitemShut
  {NoStop}%
\bibitem [{\citenamefont {Hait}\ and\ \citenamefont
  {Head-Gordon}(2021)}]{hait2021orbital}%
  \BibitemOpen
  \bibfield  {author} {\bibinfo {author} {\bibfnamefont {Diptarka}\
  \bibnamefont {Hait}}\ and\ \bibinfo {author} {\bibfnamefont {Martin}\
  \bibnamefont {Head-Gordon}},\ }\bibfield  {title} {\enquote {\bibinfo {title}
  {Orbital optimized density functional theory for electronic excited
  states},}\ }\href@noop {} {\bibfield  {journal} {\bibinfo  {journal} {J.
  Chem. Phys. Lett.}\ }\textbf {\bibinfo {volume} {12}},\ \bibinfo {pages}
  {4517--4529} (\bibinfo {year} {2021})}\BibitemShut {NoStop}%
\bibitem [{\citenamefont {{The Cirq
  Developers}}(2019)}]{The_Cirq_Developers2019-xo}%
  \BibitemOpen
  \bibfield  {author} {\bibinfo {author} {\bibnamefont {{The Cirq
  Developers}}},\ }\href {https://github.com/quantumlib/Cirq} {\enquote
  {\bibinfo {title} {Cirq},}\ } (\bibinfo {year} {2019})\BibitemShut {NoStop}%
\bibitem [{\citenamefont {Coulson}\ and\ \citenamefont
  {Fischer}(1949)}]{coulson1949xxxiv}%
  \BibitemOpen
  \bibfield  {author} {\bibinfo {author} {\bibfnamefont {Charles~Alfred}\
  \bibnamefont {Coulson}}\ and\ \bibinfo {author} {\bibfnamefont {Inga}\
  \bibnamefont {Fischer}},\ }\bibfield  {title} {\enquote {\bibinfo {title}
  {{XXXIV. Notes on the molecular orbital treatment of the hydrogen
  molecule}},}\ }\href@noop {} {\bibfield  {journal} {\bibinfo  {journal}
  {Philos. Mag.}\ }\textbf {\bibinfo {volume} {40}},\ \bibinfo {pages}
  {386--393} (\bibinfo {year} {1949})}\BibitemShut {NoStop}%
\bibitem [{\citenamefont {Hait}\ \emph
  {et~al.}(2019{\natexlab{a}})\citenamefont {Hait}, \citenamefont {Rettig},\
  and\ \citenamefont {Head-Gordon}}]{hait2019wellbehaved}%
  \BibitemOpen
  \bibfield  {author} {\bibinfo {author} {\bibfnamefont {Diptarka}\
  \bibnamefont {Hait}}, \bibinfo {author} {\bibfnamefont {Adam}\ \bibnamefont
  {Rettig}}, \ and\ \bibinfo {author} {\bibfnamefont {Martin}\ \bibnamefont
  {Head-Gordon}},\ }\bibfield  {title} {\enquote {\bibinfo {title}
  {Well-behaved versus ill-behaved density functionals for single bond
  dissociation: Separating success from disaster functional by functional for
  stretched {H$_2$}},}\ }\href@noop {} {\bibfield  {journal} {\bibinfo
  {journal} {J. Chem. Phys.}\ }\textbf {\bibinfo {volume} {150}},\ \bibinfo
  {pages} {094115} (\bibinfo {year} {2019}{\natexlab{a}})}\BibitemShut
  {NoStop}%
\bibitem [{\citenamefont {Mardirossian}\ and\ \citenamefont
  {Head-Gordon}(2017)}]{mardirossian2017thirty}%
  \BibitemOpen
  \bibfield  {author} {\bibinfo {author} {\bibfnamefont {Narbe}\ \bibnamefont
  {Mardirossian}}\ and\ \bibinfo {author} {\bibfnamefont {Martin}\ \bibnamefont
  {Head-Gordon}},\ }\bibfield  {title} {\enquote {\bibinfo {title} {Thirty
  years of density functional theory in computational chemistry: {An} overview
  and extensive assessment of 200 density functionals},}\ }\href@noop {}
  {\bibfield  {journal} {\bibinfo  {journal} {Mol. Phys.}\ }\textbf {\bibinfo
  {volume} {115}},\ \bibinfo {pages} {2315--2372} (\bibinfo {year}
  {2017})}\BibitemShut {NoStop}%
\bibitem [{\citenamefont {Hait}\ \emph
  {et~al.}(2019{\natexlab{b}})\citenamefont {Hait}, \citenamefont {Rettig},\
  and\ \citenamefont {Head-Gordon}}]{hait2019beyond}%
  \BibitemOpen
  \bibfield  {author} {\bibinfo {author} {\bibfnamefont {Diptarka}\
  \bibnamefont {Hait}}, \bibinfo {author} {\bibfnamefont {Adam}\ \bibnamefont
  {Rettig}}, \ and\ \bibinfo {author} {\bibfnamefont {Martin}\ \bibnamefont
  {Head-Gordon}},\ }\bibfield  {title} {\enquote {\bibinfo {title} {Beyond the
  {Coulson-Fischer} point: {Characterizing single excitation CI and TDDFT for
  excited states in single bond dissociations}},}\ }\href@noop {} {\bibfield
  {journal} {\bibinfo  {journal} {Phys. Chem. Chem. Phys.}\ }\textbf {\bibinfo
  {volume} {21}},\ \bibinfo {pages} {21761--21775} (\bibinfo {year}
  {2019}{\natexlab{b}})}\BibitemShut {NoStop}%
\bibitem [{\citenamefont {Hehre}\ \emph {et~al.}(1969)\citenamefont {Hehre},
  \citenamefont {Stewart},\ and\ \citenamefont {Pople}}]{hehre1969self}%
  \BibitemOpen
  \bibfield  {author} {\bibinfo {author} {\bibfnamefont {Warren~J}\
  \bibnamefont {Hehre}}, \bibinfo {author} {\bibfnamefont {Robert~F}\
  \bibnamefont {Stewart}}, \ and\ \bibinfo {author} {\bibfnamefont {John~A}\
  \bibnamefont {Pople}},\ }\bibfield  {title} {\enquote {\bibinfo {title}
  {Self-consistent molecular-orbital methods. {I. Use of Gaussian expansions of
  Slater-type atomic orbitals}},}\ }\href@noop {} {\bibfield  {journal}
  {\bibinfo  {journal} {J. Chem. Phys.}\ }\textbf {\bibinfo {volume} {51}},\
  \bibinfo {pages} {2657--2664} (\bibinfo {year} {1969})}\BibitemShut {NoStop}%
\bibitem [{\citenamefont {Ditchfield}\ \emph {et~al.}(1971)\citenamefont
  {Ditchfield}, \citenamefont {Hehre},\ and\ \citenamefont
  {Pople}}]{ditchfield1971self}%
  \BibitemOpen
  \bibfield  {author} {\bibinfo {author} {\bibfnamefont {R}~\bibnamefont
  {Ditchfield}}, \bibinfo {author} {\bibfnamefont {W~J}\ \bibnamefont {Hehre}},
  \ and\ \bibinfo {author} {\bibfnamefont {John~A}\ \bibnamefont {Pople}},\
  }\bibfield  {title} {\enquote {\bibinfo {title} {Self-consistent
  molecular-orbital methods. {IX. An extended Gaussian-type basis for
  molecular-orbital studies of organic molecules}},}\ }\href@noop {} {\bibfield
   {journal} {\bibinfo  {journal} {J. Chem. Phys.}\ }\textbf {\bibinfo {volume}
  {54}},\ \bibinfo {pages} {724--728} (\bibinfo {year} {1971})}\BibitemShut
  {NoStop}%
\bibitem [{\citenamefont {Krishnan}\ \emph {et~al.}(1980)\citenamefont
  {Krishnan}, \citenamefont {Binkley}, \citenamefont {Seeger},\ and\
  \citenamefont {Pople}}]{krishnan1980self}%
  \BibitemOpen
  \bibfield  {author} {\bibinfo {author} {\bibfnamefont {R}~\bibnamefont
  {Krishnan}}, \bibinfo {author} {\bibfnamefont {J~Stephen}\ \bibnamefont
  {Binkley}}, \bibinfo {author} {\bibfnamefont {Rolf}\ \bibnamefont {Seeger}},
  \ and\ \bibinfo {author} {\bibfnamefont {John~A}\ \bibnamefont {Pople}},\
  }\bibfield  {title} {\enquote {\bibinfo {title} {Self-consistent molecular
  orbital methods. {XX. A basis set for correlated wave functions}},}\
  }\href@noop {} {\bibfield  {journal} {\bibinfo  {journal} {J. Chem. Phys.}\
  }\textbf {\bibinfo {volume} {72}},\ \bibinfo {pages} {650--654} (\bibinfo
  {year} {1980})}\BibitemShut {NoStop}%
\bibitem [{\citenamefont {O'Gorman}\ \emph {et~al.}(2019)\citenamefont
  {O'Gorman}, \citenamefont {Huggins}, \citenamefont {Rieffel},\ and\
  \citenamefont {Whaley}}]{ogorman2019generalized}%
  \BibitemOpen
  \bibfield  {author} {\bibinfo {author} {\bibfnamefont {Bryan}\ \bibnamefont
  {O'Gorman}}, \bibinfo {author} {\bibfnamefont {William~J}\ \bibnamefont
  {Huggins}}, \bibinfo {author} {\bibfnamefont {Eleanor~G}\ \bibnamefont
  {Rieffel}}, \ and\ \bibinfo {author} {\bibfnamefont {K~Birgitta}\
  \bibnamefont {Whaley}},\ }\bibfield  {title} {\enquote {\bibinfo {title}
  {Generalized swap networks for near-term quantum computing},}\ }\href@noop {}
  {\bibfield  {journal} {\bibinfo  {journal} {arXiv preprint arXiv:1905.05118}\
  } (\bibinfo {year} {2019})}\BibitemShut {NoStop}%
\bibitem [{\citenamefont {Bocharov}\ \emph {et~al.}(2015)\citenamefont
  {Bocharov}, \citenamefont {Roetteler},\ and\ \citenamefont
  {Svore}}]{bocharov2015efficient}%
  \BibitemOpen
  \bibfield  {author} {\bibinfo {author} {\bibfnamefont {Alex}\ \bibnamefont
  {Bocharov}}, \bibinfo {author} {\bibfnamefont {Martin}\ \bibnamefont
  {Roetteler}}, \ and\ \bibinfo {author} {\bibfnamefont {Krysta~M}\
  \bibnamefont {Svore}},\ }\bibfield  {title} {\enquote {\bibinfo {title}
  {Efficient synthesis of universal repeat-until-success quantum circuits},}\
  }\href@noop {} {\bibfield  {journal} {\bibinfo  {journal} {Phys. Rev. Lett.}\
  }\textbf {\bibinfo {volume} {114}},\ \bibinfo {pages} {080502} (\bibinfo
  {year} {2015})}\BibitemShut {NoStop}%
\bibitem [{\citenamefont {Karton}\ \emph {et~al.}(2011)\citenamefont {Karton},
  \citenamefont {Daon},\ and\ \citenamefont {Martin}}]{karton2011w4}%
  \BibitemOpen
  \bibfield  {author} {\bibinfo {author} {\bibfnamefont {Amir}\ \bibnamefont
  {Karton}}, \bibinfo {author} {\bibfnamefont {Shauli}\ \bibnamefont {Daon}}, \
  and\ \bibinfo {author} {\bibfnamefont {Jan~ML}\ \bibnamefont {Martin}},\
  }\bibfield  {title} {\enquote {\bibinfo {title} {{W4-11}: A high-confidence
  benchmark dataset for computational thermochemistry derived from
  first-principles {W4} data},}\ }\href@noop {} {\bibfield  {journal} {\bibinfo
   {journal} {Chem. Phys. Lett.}\ }\textbf {\bibinfo {volume} {510}},\ \bibinfo
  {pages} {165--178} (\bibinfo {year} {2011})}\BibitemShut {NoStop}%
\bibitem [{\citenamefont {Hait}\ \emph
  {et~al.}(2019{\natexlab{c}})\citenamefont {Hait}, \citenamefont {Tubman},
  \citenamefont {Levine}, \citenamefont {Whaley},\ and\ \citenamefont
  {Head-Gordon}}]{hait2019levels}%
  \BibitemOpen
  \bibfield  {author} {\bibinfo {author} {\bibfnamefont {Diptarka}\
  \bibnamefont {Hait}}, \bibinfo {author} {\bibfnamefont {Norman~M}\
  \bibnamefont {Tubman}}, \bibinfo {author} {\bibfnamefont {Daniel~S}\
  \bibnamefont {Levine}}, \bibinfo {author} {\bibfnamefont {K~Birgitta}\
  \bibnamefont {Whaley}}, \ and\ \bibinfo {author} {\bibfnamefont {Martin}\
  \bibnamefont {Head-Gordon}},\ }\bibfield  {title} {\enquote {\bibinfo {title}
  {What levels of coupled cluster theory are appropriate for transition metal
  systems? {A} study using near-exact quantum chemical values for 3d transition
  metal binary compounds},}\ }\href@noop {} {\bibfield  {journal} {\bibinfo
  {journal} {J. Chem Theory Comput.}\ }\textbf {\bibinfo {volume} {15}},\
  \bibinfo {pages} {5370--5385} (\bibinfo {year}
  {2019}{\natexlab{c}})}\BibitemShut {NoStop}%
\bibitem [{\citenamefont {Eriksen}\ \emph {et~al.}(2020)\citenamefont
  {Eriksen}, \citenamefont {Anderson}, \citenamefont {Deustua}, \citenamefont
  {Ghanem}, \citenamefont {Hait}, \citenamefont {Hoffmann}, \citenamefont
  {Lee}, \citenamefont {Levine}, \citenamefont {Magoulas}, \citenamefont {Shen}
  \emph {et~al.}}]{eriksen2020ground}%
  \BibitemOpen
  \bibfield  {author} {\bibinfo {author} {\bibfnamefont {Janus~J}\ \bibnamefont
  {Eriksen}}, \bibinfo {author} {\bibfnamefont {Tyler~A}\ \bibnamefont
  {Anderson}}, \bibinfo {author} {\bibfnamefont {J~Emiliano}\ \bibnamefont
  {Deustua}}, \bibinfo {author} {\bibfnamefont {Khaldoon}\ \bibnamefont
  {Ghanem}}, \bibinfo {author} {\bibfnamefont {Diptarka}\ \bibnamefont {Hait}},
  \bibinfo {author} {\bibfnamefont {Mark~R}\ \bibnamefont {Hoffmann}}, \bibinfo
  {author} {\bibfnamefont {Seunghoon}\ \bibnamefont {Lee}}, \bibinfo {author}
  {\bibfnamefont {Daniel~S}\ \bibnamefont {Levine}}, \bibinfo {author}
  {\bibfnamefont {Ilias}\ \bibnamefont {Magoulas}}, \bibinfo {author}
  {\bibfnamefont {Jun}\ \bibnamefont {Shen}},  \emph {et~al.},\ }\bibfield
  {title} {\enquote {\bibinfo {title} {The ground state electronic energy of
  benzene},}\ }\href@noop {} {\bibfield  {journal} {\bibinfo  {journal} {J.
  Phys. Chem. Lett.}\ }\textbf {\bibinfo {volume} {11}},\ \bibinfo {pages}
  {8922--8929} (\bibinfo {year} {2020})}\BibitemShut {NoStop}%
\bibitem [{\citenamefont {Roos}\ and\ \citenamefont
  {Andersson}(1995)}]{roos1995multiconfigurational}%
  \BibitemOpen
  \bibfield  {author} {\bibinfo {author} {\bibfnamefont {Bj{\"o}rn~O}\
  \bibnamefont {Roos}}\ and\ \bibinfo {author} {\bibfnamefont {Kerstin}\
  \bibnamefont {Andersson}},\ }\bibfield  {title} {\enquote {\bibinfo {title}
  {Multiconfigurational perturbation theory with level shift — the {Cr$_2$}
  potential revisited},}\ }\href@noop {} {\bibfield  {journal} {\bibinfo
  {journal} {Chem. Phys. Lett.}\ }\textbf {\bibinfo {volume} {245}},\ \bibinfo
  {pages} {215--223} (\bibinfo {year} {1995})}\BibitemShut {NoStop}%
\bibitem [{\citenamefont {Ghigo}\ \emph {et~al.}(2004)\citenamefont {Ghigo},
  \citenamefont {Roos},\ and\ \citenamefont {Malmqvist}}]{ghigo2004modified}%
  \BibitemOpen
  \bibfield  {author} {\bibinfo {author} {\bibfnamefont {Giovanni}\
  \bibnamefont {Ghigo}}, \bibinfo {author} {\bibfnamefont {Bj{\"o}rn~O}\
  \bibnamefont {Roos}}, \ and\ \bibinfo {author} {\bibfnamefont {Per-{\AA}ke}\
  \bibnamefont {Malmqvist}},\ }\bibfield  {title} {\enquote {\bibinfo {title}
  {A modified definition of the {zeroth-order Hamiltonian in
  multiconfigurational perturbation theory (CASPT2)}},}\ }\href@noop {}
  {\bibfield  {journal} {\bibinfo  {journal} {Chem. Phys. Lett.}\ }\textbf
  {\bibinfo {volume} {396}},\ \bibinfo {pages} {142--149} (\bibinfo {year}
  {2004})}\BibitemShut {NoStop}%
\bibitem [{\citenamefont {Kurlancheek}\ and\ \citenamefont
  {Head-Gordon}(2009)}]{kurlancheek2009violations}%
  \BibitemOpen
  \bibfield  {author} {\bibinfo {author} {\bibfnamefont {Westin}\ \bibnamefont
  {Kurlancheek}}\ and\ \bibinfo {author} {\bibfnamefont {Martin}\ \bibnamefont
  {Head-Gordon}},\ }\bibfield  {title} {\enquote {\bibinfo {title} {Violations
  of {$N$}-representability from spin-unrestricted orbitals in
  {M{\o}ller--Plesset} perturbation theory and related double-hybrid density
  functional theory},}\ }\href@noop {} {\bibfield  {journal} {\bibinfo
  {journal} {Mol. Phys.}\ }\textbf {\bibinfo {volume} {107}},\ \bibinfo {pages}
  {1223--1232} (\bibinfo {year} {2009})}\BibitemShut {NoStop}%
\bibitem [{\citenamefont {Tubman}\ \emph {et~al.}(2018)\citenamefont {Tubman},
  \citenamefont {Mejuto-Zaera}, \citenamefont {Epstein}, \citenamefont {Hait},
  \citenamefont {Levine}, \citenamefont {Huggins}, \citenamefont {Jiang},
  \citenamefont {McClean}, \citenamefont {Babbush}, \citenamefont {Head-Gordon}
  \emph {et~al.}}]{tubman2018postponing}%
  \BibitemOpen
  \bibfield  {author} {\bibinfo {author} {\bibfnamefont {Norm~M}\ \bibnamefont
  {Tubman}}, \bibinfo {author} {\bibfnamefont {Carlos}\ \bibnamefont
  {Mejuto-Zaera}}, \bibinfo {author} {\bibfnamefont {Jeffrey~M}\ \bibnamefont
  {Epstein}}, \bibinfo {author} {\bibfnamefont {Diptarka}\ \bibnamefont
  {Hait}}, \bibinfo {author} {\bibfnamefont {Daniel~S}\ \bibnamefont {Levine}},
  \bibinfo {author} {\bibfnamefont {William}\ \bibnamefont {Huggins}}, \bibinfo
  {author} {\bibfnamefont {Zhang}\ \bibnamefont {Jiang}}, \bibinfo {author}
  {\bibfnamefont {Jarrod~R}\ \bibnamefont {McClean}}, \bibinfo {author}
  {\bibfnamefont {Ryan}\ \bibnamefont {Babbush}}, \bibinfo {author}
  {\bibfnamefont {Martin}\ \bibnamefont {Head-Gordon}},  \emph {et~al.},\
  }\bibfield  {title} {\enquote {\bibinfo {title} {Postponing the orthogonality
  catastrophe: efficient state preparation for electronic structure simulations
  on quantum devices},}\ }\href@noop {} {\bibfield  {journal} {\bibinfo
  {journal} {arXiv preprint arXiv:1809.05523}\ } (\bibinfo {year}
  {2018})}\BibitemShut {NoStop}%
\bibitem [{\citenamefont {Wallman}\ and\ \citenamefont
  {Emerson}(2016)}]{wallman2016noise}%
  \BibitemOpen
  \bibfield  {author} {\bibinfo {author} {\bibfnamefont {Joel~J}\ \bibnamefont
  {Wallman}}\ and\ \bibinfo {author} {\bibfnamefont {Joseph}\ \bibnamefont
  {Emerson}},\ }\bibfield  {title} {\enquote {\bibinfo {title} {Noise tailoring
  for scalable quantum computation via randomized compiling},}\ }\href@noop {}
  {\bibfield  {journal} {\bibinfo  {journal} {Phys. Rev. A}\ }\textbf {\bibinfo
  {volume} {94}},\ \bibinfo {pages} {052325} (\bibinfo {year}
  {2016})}\BibitemShut {NoStop}%
\bibitem [{\citenamefont {Hashim}\ \emph {et~al.}(2021)\citenamefont {Hashim},
  \citenamefont {Naik}, \citenamefont {Morvan}, \citenamefont {Ville},
  \citenamefont {Mitchell}, \citenamefont {Kreikebaum}, \citenamefont {Davis},
  \citenamefont {Smith}, \citenamefont {Iancu}, \citenamefont {O’Brien} \emph
  {et~al.}}]{hashim2020randomized}%
  \BibitemOpen
  \bibfield  {author} {\bibinfo {author} {\bibfnamefont {Akel}\ \bibnamefont
  {Hashim}}, \bibinfo {author} {\bibfnamefont {Ravi~K}\ \bibnamefont {Naik}},
  \bibinfo {author} {\bibfnamefont {Alexis}\ \bibnamefont {Morvan}}, \bibinfo
  {author} {\bibfnamefont {Jean-Loup}\ \bibnamefont {Ville}}, \bibinfo {author}
  {\bibfnamefont {Bradley}\ \bibnamefont {Mitchell}}, \bibinfo {author}
  {\bibfnamefont {John~Mark}\ \bibnamefont {Kreikebaum}}, \bibinfo {author}
  {\bibfnamefont {Marc}\ \bibnamefont {Davis}}, \bibinfo {author}
  {\bibfnamefont {Ethan}\ \bibnamefont {Smith}}, \bibinfo {author}
  {\bibfnamefont {Costin}\ \bibnamefont {Iancu}}, \bibinfo {author}
  {\bibfnamefont {Kevin~P}\ \bibnamefont {O’Brien}},  \emph {et~al.},\
  }\bibfield  {title} {\enquote {\bibinfo {title} {Randomized compiling for
  scalable quantum computing on a noisy superconducting quantum processor},}\
  }\href@noop {} {\bibfield  {journal} {\bibinfo  {journal} {Phys. Rev. X}\
  }\textbf {\bibinfo {volume} {11}},\ \bibinfo {pages} {041039} (\bibinfo
  {year} {2021})}\BibitemShut {NoStop}%
\bibitem [{\citenamefont {Shee}\ \emph
  {et~al.}(2021{\natexlab{b}})\citenamefont {Shee}, \citenamefont
  {Loipersberger}, \citenamefont {Rettig}, \citenamefont {Lee},\ and\
  \citenamefont {Head-Gordon}}]{shee2021regularized}%
  \BibitemOpen
  \bibfield  {author} {\bibinfo {author} {\bibfnamefont {James}\ \bibnamefont
  {Shee}}, \bibinfo {author} {\bibfnamefont {Matthias}\ \bibnamefont
  {Loipersberger}}, \bibinfo {author} {\bibfnamefont {Adam}\ \bibnamefont
  {Rettig}}, \bibinfo {author} {\bibfnamefont {Joonho}\ \bibnamefont {Lee}}, \
  and\ \bibinfo {author} {\bibfnamefont {Martin}\ \bibnamefont {Head-Gordon}},\
  }\bibfield  {title} {\enquote {\bibinfo {title} {Regularized second-order
  {M{\o}ller--Plesset} theory: A more accurate alternative to conventional
  {MP2} for noncovalent interactions and transition metal thermochemistry for
  the same computational cost},}\ }\href@noop {} {\bibfield  {journal}
  {\bibinfo  {journal} {J. Phys. Chem. Lett.}\ }\textbf {\bibinfo {volume}
  {12}},\ \bibinfo {pages} {12084--12097} (\bibinfo {year}
  {2021}{\natexlab{b}})}\BibitemShut {NoStop}%
\bibitem [{\citenamefont {Bertels}\ \emph {et~al.}(2019)\citenamefont
  {Bertels}, \citenamefont {Lee},\ and\ \citenamefont
  {Head-Gordon}}]{bertels2019third}%
  \BibitemOpen
  \bibfield  {author} {\bibinfo {author} {\bibfnamefont {Luke~W}\ \bibnamefont
  {Bertels}}, \bibinfo {author} {\bibfnamefont {Joonho}\ \bibnamefont {Lee}}, \
  and\ \bibinfo {author} {\bibfnamefont {Martin}\ \bibnamefont {Head-Gordon}},\
  }\bibfield  {title} {\enquote {\bibinfo {title} {Third-order
  m{\o}ller--plesset perturbation theory made useful? choice of orbitals and
  scaling greatly improves accuracy for thermochemistry, kinetics, and
  intermolecular interactions},}\ }\href@noop {} {\bibfield  {journal}
  {\bibinfo  {journal} {J. Phys. Chem. Lett.}\ }\textbf {\bibinfo {volume}
  {10}},\ \bibinfo {pages} {4170--4176} (\bibinfo {year} {2019})}\BibitemShut
  {NoStop}%
\bibitem [{\citenamefont {Rettig}\ \emph {et~al.}(2020)\citenamefont {Rettig},
  \citenamefont {Hait}, \citenamefont {Bertels},\ and\ \citenamefont
  {Head-Gordon}}]{rettig2020third}%
  \BibitemOpen
  \bibfield  {author} {\bibinfo {author} {\bibfnamefont {Adam}\ \bibnamefont
  {Rettig}}, \bibinfo {author} {\bibfnamefont {Diptarka}\ \bibnamefont {Hait}},
  \bibinfo {author} {\bibfnamefont {Luke~W}\ \bibnamefont {Bertels}}, \ and\
  \bibinfo {author} {\bibfnamefont {Martin}\ \bibnamefont {Head-Gordon}},\
  }\bibfield  {title} {\enquote {\bibinfo {title} {Third-order
  {M{\o}ller--Plesset} theory made more useful? {The} role of density
  functional theory orbitals},}\ }\href@noop {} {\bibfield  {journal} {\bibinfo
   {journal} {J. Chem. Theory Comput.}\ }\textbf {\bibinfo {volume} {16}},\
  \bibinfo {pages} {7473--7489} (\bibinfo {year} {2020})}\BibitemShut {NoStop}%
\bibitem [{\citenamefont {Gherman}\ and\ \citenamefont
  {Cramer}(2009)}]{gherman2009quantum}%
  \BibitemOpen
  \bibfield  {author} {\bibinfo {author} {\bibfnamefont {Benjamin~F}\
  \bibnamefont {Gherman}}\ and\ \bibinfo {author} {\bibfnamefont
  {Christopher~J}\ \bibnamefont {Cramer}},\ }\bibfield  {title} {\enquote
  {\bibinfo {title} {Quantum chemical studies of molecules incorporating a
  {Cu$_2$O$_2^{2+}$} core},}\ }\href@noop {} {\bibfield  {journal} {\bibinfo
  {journal} {Coord. Chem. Rev.}\ }\textbf {\bibinfo {volume} {253}},\ \bibinfo
  {pages} {723--753} (\bibinfo {year} {2009})}\BibitemShut {NoStop}%
\bibitem [{\citenamefont {Liakos}\ and\ \citenamefont
  {Neese}(2011)}]{liakos2011interplay}%
  \BibitemOpen
  \bibfield  {author} {\bibinfo {author} {\bibfnamefont {Dimitrios~G}\
  \bibnamefont {Liakos}}\ and\ \bibinfo {author} {\bibfnamefont {Frank}\
  \bibnamefont {Neese}},\ }\bibfield  {title} {\enquote {\bibinfo {title}
  {Interplay of correlation and relativistic effects in correlated calculations
  on transition-metal complexes: The {(Cu$_2$O$_2$)$^{2+}$} core revisited},}\
  }\href@noop {} {\bibfield  {journal} {\bibinfo  {journal} {J. Chem. Theory
  Comput.}\ }\textbf {\bibinfo {volume} {7}},\ \bibinfo {pages} {1511--1523}
  (\bibinfo {year} {2011})}\BibitemShut {NoStop}%
\bibitem [{\citenamefont {Yang}\ \emph {et~al.}(2016)\citenamefont {Yang},
  \citenamefont {Davidson},\ and\ \citenamefont {Yang}}]{yang2016nature}%
  \BibitemOpen
  \bibfield  {author} {\bibinfo {author} {\bibfnamefont {Yang}\ \bibnamefont
  {Yang}}, \bibinfo {author} {\bibfnamefont {Ernest~R}\ \bibnamefont
  {Davidson}}, \ and\ \bibinfo {author} {\bibfnamefont {Weitao}\ \bibnamefont
  {Yang}},\ }\bibfield  {title} {\enquote {\bibinfo {title} {Nature of ground
  and electronic excited states of higher acenes},}\ }\href@noop {} {\bibfield
  {journal} {\bibinfo  {journal} {Proc. Natl. Acad. Sci.}\ }\textbf {\bibinfo
  {volume} {113}},\ \bibinfo {pages} {E5098--E5107} (\bibinfo {year}
  {2016})}\BibitemShut {NoStop}%
\bibitem [{\citenamefont {Römelt}\ \emph {et~al.}(2017)\citenamefont
  {Römelt}, \citenamefont {Song}, \citenamefont {Tarrago}, \citenamefont
  {Rees}, \citenamefont {van Gastel}, \citenamefont {Weyhermüller},
  \citenamefont {DeBeer}, \citenamefont {Bill}, \citenamefont {Neese},\ and\
  \citenamefont {Ye}}]{NeesePorphyrin}%
  \BibitemOpen
  \bibfield  {author} {\bibinfo {author} {\bibfnamefont {Christina}\
  \bibnamefont {Römelt}}, \bibinfo {author} {\bibfnamefont {Jinshuai}\
  \bibnamefont {Song}}, \bibinfo {author} {\bibfnamefont {Maxime}\ \bibnamefont
  {Tarrago}}, \bibinfo {author} {\bibfnamefont {Julian~A}\ \bibnamefont
  {Rees}}, \bibinfo {author} {\bibfnamefont {Maurice}\ \bibnamefont {van
  Gastel}}, \bibinfo {author} {\bibfnamefont {Thomas}\ \bibnamefont
  {Weyhermüller}}, \bibinfo {author} {\bibfnamefont {Serena}\ \bibnamefont
  {DeBeer}}, \bibinfo {author} {\bibfnamefont {Eckhard}\ \bibnamefont {Bill}},
  \bibinfo {author} {\bibfnamefont {Frank}\ \bibnamefont {Neese}}, \ and\
  \bibinfo {author} {\bibfnamefont {Shengfa}\ \bibnamefont {Ye}},\ }\bibfield
  {title} {\enquote {\bibinfo {title} {Electronic structure of a formal
  {Iron(0)} porphyrin complex relevant to {CO}$_2$ reduction},}\ }\href@noop {}
  {\bibfield  {journal} {\bibinfo  {journal} {Inorg. Chem.}\ }\textbf {\bibinfo
  {volume} {56}},\ \bibinfo {pages} {4745--4750} (\bibinfo {year}
  {2017})}\BibitemShut {NoStop}%
\bibitem [{\citenamefont {Derrick}\ \emph {et~al.}(2020)\citenamefont
  {Derrick}, \citenamefont {Loipersberger}, \citenamefont {Chatterjee},
  \citenamefont {Iovan}, \citenamefont {Smith}, \citenamefont {Chakarawet},
  \citenamefont {Yano}, \citenamefont {Long}, \citenamefont {Head-Gordon},\
  and\ \citenamefont {Chang}}]{derrick2020metal}%
  \BibitemOpen
  \bibfield  {author} {\bibinfo {author} {\bibfnamefont {Jeffrey~S}\
  \bibnamefont {Derrick}}, \bibinfo {author} {\bibfnamefont {Matthias}\
  \bibnamefont {Loipersberger}}, \bibinfo {author} {\bibfnamefont {Ruchira}\
  \bibnamefont {Chatterjee}}, \bibinfo {author} {\bibfnamefont {Diana~A}\
  \bibnamefont {Iovan}}, \bibinfo {author} {\bibfnamefont {Peter~T}\
  \bibnamefont {Smith}}, \bibinfo {author} {\bibfnamefont {Khetpakorn}\
  \bibnamefont {Chakarawet}}, \bibinfo {author} {\bibfnamefont {Junko}\
  \bibnamefont {Yano}}, \bibinfo {author} {\bibfnamefont {Jeffrey~R}\
  \bibnamefont {Long}}, \bibinfo {author} {\bibfnamefont {Martin}\ \bibnamefont
  {Head-Gordon}}, \ and\ \bibinfo {author} {\bibfnamefont {Christopher~J}\
  \bibnamefont {Chang}},\ }\bibfield  {title} {\enquote {\bibinfo {title}
  {Metal--ligand cooperativity via exchange coupling promotes iron-catalyzed
  electrochemical {CO}$_2$ reduction at low overpotentials},}\ }\href@noop {}
  {\bibfield  {journal} {\bibinfo  {journal} {J. Am. Chem. Soc.}\ }\textbf
  {\bibinfo {volume} {142}},\ \bibinfo {pages} {20489--20501} (\bibinfo {year}
  {2020})}\BibitemShut {NoStop}%
\bibitem [{\citenamefont {Sharma}\ \emph {et~al.}(2014)\citenamefont {Sharma},
  \citenamefont {Sivalingam}, \citenamefont {Neese},\ and\ \citenamefont
  {Chan}}]{sharma2014low}%
  \BibitemOpen
  \bibfield  {author} {\bibinfo {author} {\bibfnamefont {Sandeep}\ \bibnamefont
  {Sharma}}, \bibinfo {author} {\bibfnamefont {Kantharuban}\ \bibnamefont
  {Sivalingam}}, \bibinfo {author} {\bibfnamefont {Frank}\ \bibnamefont
  {Neese}}, \ and\ \bibinfo {author} {\bibfnamefont {Garnet~Kin}\ \bibnamefont
  {Chan}},\ }\bibfield  {title} {\enquote {\bibinfo {title} {Low-energy
  spectrum of iron--sulfur clusters directly from many-particle quantum
  mechanics},}\ }\href@noop {} {\bibfield  {journal} {\bibinfo  {journal}
  {Nature chemistry}\ }\textbf {\bibinfo {volume} {6}},\ \bibinfo {pages}
  {927--933} (\bibinfo {year} {2014})}\BibitemShut {NoStop}%
\bibitem [{\citenamefont {Mejuto-Zaera}\ \emph {et~al.}(2022)\citenamefont
  {Mejuto-Zaera}, \citenamefont {Tzeli}, \citenamefont {Williams-Young},
  \citenamefont {Tubman}, \citenamefont {Matoušek}, \citenamefont {Brabec},
  \citenamefont {Veis}, \citenamefont {Xantheas},\ and\ \citenamefont
  {de~Jong}}]{mejuto2022effect}%
  \BibitemOpen
  \bibfield  {author} {\bibinfo {author} {\bibfnamefont {Carlos}\ \bibnamefont
  {Mejuto-Zaera}}, \bibinfo {author} {\bibfnamefont {Demeter}\ \bibnamefont
  {Tzeli}}, \bibinfo {author} {\bibfnamefont {David}\ \bibnamefont
  {Williams-Young}}, \bibinfo {author} {\bibfnamefont {Norm~M}\ \bibnamefont
  {Tubman}}, \bibinfo {author} {\bibfnamefont {Mikuláš}\ \bibnamefont
  {Matoušek}}, \bibinfo {author} {\bibfnamefont {Jiri}\ \bibnamefont
  {Brabec}}, \bibinfo {author} {\bibfnamefont {Libor}\ \bibnamefont {Veis}},
  \bibinfo {author} {\bibfnamefont {Sotiris~S}\ \bibnamefont {Xantheas}}, \
  and\ \bibinfo {author} {\bibfnamefont {Wibe~A}\ \bibnamefont {de~Jong}},\
  }\bibfield  {title} {\enquote {\bibinfo {title} {The effect of geometry,
  spin, and orbital optimization in achieving accurate, correlated results for
  iron--sulfur cubanes},}\ }\href@noop {} {\bibfield  {journal} {\bibinfo
  {journal} {J. Chem. Theory Comput.}\ } (\bibinfo {year} {2022})}\BibitemShut
  {NoStop}%
\bibitem [{\citenamefont {Li}\ \emph {et~al.}(2019)\citenamefont {Li},
  \citenamefont {Li}, \citenamefont {Dattani}, \citenamefont {Umrigar},\ and\
  \citenamefont {Chan}}]{li2019electronic}%
  \BibitemOpen
  \bibfield  {author} {\bibinfo {author} {\bibfnamefont {Zhendong}\
  \bibnamefont {Li}}, \bibinfo {author} {\bibfnamefont {Junhao}\ \bibnamefont
  {Li}}, \bibinfo {author} {\bibfnamefont {Nikesh~S}\ \bibnamefont {Dattani}},
  \bibinfo {author} {\bibfnamefont {CJ}~\bibnamefont {Umrigar}}, \ and\
  \bibinfo {author} {\bibfnamefont {Garnet Kin-Lic}\ \bibnamefont {Chan}},\
  }\bibfield  {title} {\enquote {\bibinfo {title} {The electronic complexity of
  the ground-state of the {FeMo} cofactor of nitrogenase as relevant to quantum
  simulations},}\ }\href@noop {} {\bibfield  {journal} {\bibinfo  {journal} {J.
  Chem. Phys.}\ }\textbf {\bibinfo {volume} {150}},\ \bibinfo {pages} {024302}
  (\bibinfo {year} {2019})}\BibitemShut {NoStop}%
\bibitem [{\citenamefont {Bagus}\ and\ \citenamefont
  {Bennett}(1975)}]{bagus1975singlet}%
  \BibitemOpen
  \bibfield  {author} {\bibinfo {author} {\bibfnamefont {PS}~\bibnamefont
  {Bagus}}\ and\ \bibinfo {author} {\bibfnamefont {BI}~\bibnamefont
  {Bennett}},\ }\bibfield  {title} {\enquote {\bibinfo {title}
  {Singlet--triplet splittings as obtained from the {X}$\alpha$-scattered wave
  method: A theoretical analysis},}\ }\href@noop {} {\bibfield  {journal}
  {\bibinfo  {journal} {Int. J. Quantum Chem}\ }\textbf {\bibinfo {volume}
  {9}},\ \bibinfo {pages} {143--148} (\bibinfo {year} {1975})}\BibitemShut
  {NoStop}%
\bibitem [{\citenamefont {Jimenez-Hoyos}\ \emph {et~al.}(2011)\citenamefont
  {Jimenez-Hoyos}, \citenamefont {Henderson},\ and\ \citenamefont
  {Scuseria}}]{jimenez2011generalized}%
  \BibitemOpen
  \bibfield  {author} {\bibinfo {author} {\bibfnamefont {Carlos~A}\
  \bibnamefont {Jimenez-Hoyos}}, \bibinfo {author} {\bibfnamefont {Thomas~M}\
  \bibnamefont {Henderson}}, \ and\ \bibinfo {author} {\bibfnamefont
  {Gustavo~E}\ \bibnamefont {Scuseria}},\ }\bibfield  {title} {\enquote
  {\bibinfo {title} {Generalized {Hartree-Fock} description of molecular
  dissociation},}\ }\href@noop {} {\bibfield  {journal} {\bibinfo  {journal}
  {J. Chem. Theory Comput.}\ }\textbf {\bibinfo {volume} {7}},\ \bibinfo
  {pages} {2667--2674} (\bibinfo {year} {2011})}\BibitemShut {NoStop}%
\bibitem [{\citenamefont {Small}\ \emph {et~al.}(2015)\citenamefont {Small},
  \citenamefont {Sundstrom},\ and\ \citenamefont
  {Head-Gordon}}]{small2015simple}%
  \BibitemOpen
  \bibfield  {author} {\bibinfo {author} {\bibfnamefont {David~W}\ \bibnamefont
  {Small}}, \bibinfo {author} {\bibfnamefont {Eric~J}\ \bibnamefont
  {Sundstrom}}, \ and\ \bibinfo {author} {\bibfnamefont {Martin}\ \bibnamefont
  {Head-Gordon}},\ }\bibfield  {title} {\enquote {\bibinfo {title} {A simple
  way to test for collinearity in spin symmetry broken wave functions: {General
  theory and application to generalized Hartree-Fock}},}\ }\href@noop {}
  {\bibfield  {journal} {\bibinfo  {journal} {J. Chem. Phys.}\ }\textbf
  {\bibinfo {volume} {142}},\ \bibinfo {pages} {094112} (\bibinfo {year}
  {2015})}\BibitemShut {NoStop}%
\bibitem [{\citenamefont {Pritchard}\ \emph {et~al.}(2019)\citenamefont
  {Pritchard}, \citenamefont {Altarawy}, \citenamefont {Didier}, \citenamefont
  {Gibson},\ and\ \citenamefont {Windus}}]{pritchard2019new}%
  \BibitemOpen
  \bibfield  {author} {\bibinfo {author} {\bibfnamefont {Benjamin~P}\
  \bibnamefont {Pritchard}}, \bibinfo {author} {\bibfnamefont {Doaa}\
  \bibnamefont {Altarawy}}, \bibinfo {author} {\bibfnamefont {Brett}\
  \bibnamefont {Didier}}, \bibinfo {author} {\bibfnamefont {Tara~D}\
  \bibnamefont {Gibson}}, \ and\ \bibinfo {author} {\bibfnamefont {Theresa~L}\
  \bibnamefont {Windus}},\ }\bibfield  {title} {\enquote {\bibinfo {title} {New
  basis set exchange: {An} open, up-to-date resource for the molecular sciences
  community},}\ }\href@noop {} {\bibfield  {journal} {\bibinfo  {journal} {J.
  Chem. Inf. Model.}\ }\textbf {\bibinfo {volume} {59}},\ \bibinfo {pages}
  {4814--4820} (\bibinfo {year} {2019})}\BibitemShut {NoStop}%
\bibitem [{\citenamefont {{\v{C}}{\'\i}{\v{z}}ek}\ and\ \citenamefont
  {Paldus}(1967)}]{vcivzek1967stability}%
  \BibitemOpen
  \bibfield  {author} {\bibinfo {author} {\bibfnamefont {J}~\bibnamefont
  {{\v{C}}{\'\i}{\v{z}}ek}}\ and\ \bibinfo {author} {\bibfnamefont
  {J}~\bibnamefont {Paldus}},\ }\bibfield  {title} {\enquote {\bibinfo {title}
  {Stability conditions for the solutions of the {Hartree-Fock} equations for
  atomic and molecular systems. application to the pi-electron model of cyclic
  polyenes},}\ }\href@noop {} {\bibfield  {journal} {\bibinfo  {journal} {J.
  Chem. Phys.}\ }\textbf {\bibinfo {volume} {47}},\ \bibinfo {pages}
  {3976--3985} (\bibinfo {year} {1967})}\BibitemShut {NoStop}%
\bibitem [{\citenamefont {Seeger}\ and\ \citenamefont
  {Pople}(1977)}]{seeger1977self}%
  \BibitemOpen
  \bibfield  {author} {\bibinfo {author} {\bibfnamefont {Rolf}\ \bibnamefont
  {Seeger}}\ and\ \bibinfo {author} {\bibfnamefont {John~A}\ \bibnamefont
  {Pople}},\ }\bibfield  {title} {\enquote {\bibinfo {title} {Self-consistent
  molecular orbital methods. {XVIII. Constraints} and stability in
  {Hartree--Fock} theory},}\ }\href@noop {} {\bibfield  {journal} {\bibinfo
  {journal} {J. Chem. Phys.}\ }\textbf {\bibinfo {volume} {66}},\ \bibinfo
  {pages} {3045--3050} (\bibinfo {year} {1977})}\BibitemShut {NoStop}%
\bibitem [{\citenamefont {Childs}\ \emph {et~al.}(2021)\citenamefont {Childs},
  \citenamefont {Su}, \citenamefont {Tran}, \citenamefont {Wiebe},\ and\
  \citenamefont {Zhu}}]{childs2021}%
  \BibitemOpen
  \bibfield  {author} {\bibinfo {author} {\bibfnamefont {Andrew~M.}\
  \bibnamefont {Childs}}, \bibinfo {author} {\bibfnamefont {Yuan}\ \bibnamefont
  {Su}}, \bibinfo {author} {\bibfnamefont {Minh~C.}\ \bibnamefont {Tran}},
  \bibinfo {author} {\bibfnamefont {Nathan}\ \bibnamefont {Wiebe}}, \ and\
  \bibinfo {author} {\bibfnamefont {Shuchen}\ \bibnamefont {Zhu}},\ }\bibfield
  {title} {\enquote {\bibinfo {title} {Theory of {Trotter} error with
  commutator scaling},}\ }\href@noop {} {\bibfield  {journal} {\bibinfo
  {journal} {Phys. Rev. X}\ }\textbf {\bibinfo {volume} {11}},\ \bibinfo
  {pages} {011020} (\bibinfo {year} {2021})}\BibitemShut {NoStop}%
\bibitem [{\citenamefont {Magnus}(1954)}]{magnus1954exponential}%
  \BibitemOpen
  \bibfield  {author} {\bibinfo {author} {\bibfnamefont {Wilhelm}\ \bibnamefont
  {Magnus}},\ }\bibfield  {title} {\enquote {\bibinfo {title} {{On the
  exponential solution of differential equations for a linear operator}},}\
  }\href {\doibase 10.1002/cpa.3160070404} {\bibfield  {journal} {\bibinfo
  {journal} {Commun. Pure Appl. Math}\ }\textbf {\bibinfo {volume} {7}},\
  \bibinfo {pages} {649--673} (\bibinfo {year} {1954})}\BibitemShut {NoStop}%
\bibitem [{\citenamefont {Mathias}\ and\ \citenamefont
  {Li}(2004)}]{Mathias2004TheDG}%
  \BibitemOpen
  \bibfield  {author} {\bibinfo {author} {\bibfnamefont {Roy}\ \bibnamefont
  {Mathias}}\ and\ \bibinfo {author} {\bibfnamefont {Chi-Kwong}\ \bibnamefont
  {Li}},\ }\bibfield  {title} {\enquote {\bibinfo {title} {The definite
  generalized eigenvalue problem: A new perturbation theory},}\ \ }(\bibinfo
  {year} {2004})\BibitemShut {NoStop}%
\bibitem [{\citenamefont {Stewart}(1979)}]{stewart1979pertubation}%
  \BibitemOpen
  \bibfield  {author} {\bibinfo {author} {\bibfnamefont {Gilbert~W}\
  \bibnamefont {Stewart}},\ }\bibfield  {title} {\enquote {\bibinfo {title}
  {Pertubation bounds for the definite generalized eigenvalue problem},}\
  }\href@noop {} {\bibfield  {journal} {\bibinfo  {journal} {Linear algebra and
  its applications}\ }\textbf {\bibinfo {volume} {23}},\ \bibinfo {pages}
  {69--85} (\bibinfo {year} {1979})}\BibitemShut {NoStop}%
\end{thebibliography}%
\onecolumngrid
\newpage
\appendix
\section{Orbitals in quantum chemistry}\label{app:orbitals}

Classical quantum chemistry techniques are often described in terms of `orbitals' and `basis sets', nomenclature which might be somewhat inaccessible to those who are not practicing electronic structure theorists. Therefore, we provide a brief overview of some of the terms used by quantum chemists that are relevant to this work.

\subsection{Restricted, Unrestricted and General Many-electron Wavefunctions}

The simplest definition of an orbital is that it is a single particle/electron spatial wavefunction, like those of the hydrogen atom. Adding the spin degree of freedom gives  a `spin-orbital' $\phi(\vec{r},s)$, which may be viewed as a wavefunction that describes the continuous position distribution $\vec{r}$, and the discrete distribution of the spin projection $s$. Since an electron only has two possible eigenstates of spin projected onto a spatial axis, $\alpha$ and $\beta$ (corresponding to up and down spins, respectively), the spin projection distribution has just two discrete values and we can express the general spin-orbital as:
\begin{align}
    \phi(\vec{r},s)&=c_1\omega^\alpha(\vec{r})\alpha(s)+c_2\omega^\beta(\vec{r})\beta(s).\label{eq:ghfdef}
\end{align}
Here $\omega^\alpha(\vec{r}), \omega^\beta(\vec{r})$ are (complex-valued) functions of the spatial coordinates $\vec{r}$ which are referred to as the `spatial orbitals', and the constants $c_1$ and $c_2$ are complex numbers that control the relative weights of the two spin components $\alpha$ and $\beta$. 

This formulation of spin-orbitals is, however, very rarely used in practice. More common is the use of a formalism based on primitive spin-orbitals that are taken to be a product of separate spatial and spin components and are assumed real (as we did in the main paper). For example, we can write:  
\begin{align}
   \phi^\alpha(\vec{r},s)&=\omega^\alpha(\vec{r})\alpha(s)\label{eq:uhfdef1}\\
   \phi^\beta(\vec{r},s)&=\omega^\beta(\vec{r})\beta(s)\label{eq:uhfdef2}
\end{align}
In general, the spatial distribution of the up and down spins, given by $\omega^\alpha(\vec{r})$ and $\omega^\beta(\vec{r})$ respectively, can be very different. For the special case of systems with equal numbers of up and down spin, it is often assumed that the spatial distributions of the up and down spins are identical (`spin-symmetry'). Electrons therefore `pair up' in each occupied spatial orbital. Then, the $p$th alpha spin-orbital $\{\phi_p^\alpha(\vec{r},s)\}$ and the $p$th beta spin-orbital $\{\phi_p^\beta(\vec{r},s)\}$ have the same spatial component $\omega_p(\vec{r})$, leading to:
\begin{align}
    \phi_p^\alpha(\vec{r},s)&=\omega_p(\vec{r})\alpha(s)\label{eq:rhfdef1}\\
    \phi_p^\beta(\vec{r},s)&=\omega_p(\vec{r})\beta(s)\label{eq:rhfdef2}
\end{align}
Each of the above three cases is carried over to Hartree-Fock calculations. Generalized Hartree-Fock (GHF) uses the most general form of orbitals shown in \eq{ghfdef}. Unrestricted Hartree-Fock (UHF) utilizes the factorizable form shown in Eqs.~(\ref{eq:uhfdef1}) and (\ref{eq:uhfdef2}), while restricted Hartree-Fock (RHF) uses the more constrained form given by Eqs.~(\ref{eq:rhfdef1}) and (\ref{eq:rhfdef2}). The `spin-restricted' formalism employed in RHF calculations is perhaps the most familiar form of quantum chemistry because it is extensively used for ground-state calculations with closed-shell molecular systems. In this work, however, we utilize the more general UHF formalism, which is better suited for describing systems with unpaired electrons such as strongly correlated systems or excited states. 

It is worth noting, however, that the greater flexibility of the GHF and UHF approaches over RHF permits them to violate symmetries satisfied by the exact wavefunction. RHF wavefunctions are eigenstates of the $\opr{S}^2$ and $\opr{S}_z$ operators. UHF wavefunctions are only guaranteed to be eigenstates of $\opr{S}_z$, and need not be eigenstates of $\opr{S}^2$. In fact, UHF wavefunctions with $\langle \opr{S}^2\rangle \sim 1$ (i.e. approximately halfway between singlet and triplet) are quite often utilized to describe bond dissociations~\cite{szabo2012modern,hait2019beyond} or excited states~\cite{bagus1975singlet,hait2021orbital}. GHF wavefunctions are not required to be eigenstates of either $\opr{S}^2$ or $\opr{S}_z$ and are used to describe multiple bond dissociations~\cite{jimenez2011generalized,small2015simple}. Ref. ~\cite{jimenez2011generalized} provides a detailed discussion about the symmetries preserved by various variants of Hartree-Fock within non-relativistic quantum mechanics. 

\subsection{Basis Functions and Atomic Orbitals}

In quantum chemistry, the notions of `atomic orbitals' and `basis functions' or `basis sets' are also very important. An atomic orbital $\chi$ refers to a hydrogen atom-like spin-orbital, centered around a particular point in space (corresponding to the nucleus and assumed here to be the origin), that asymptotically decays to zero with increasing radial distance $r$ away from this central point. According to the exact solution for the electronic states of the H atom, the decay of $\chi$ should be rigorously exponential in $r$, but such `Slater type orbitals' are computationally challenging to work with. It is thus far more common to represent the spatial part with Gaussian functions of the form $x^ay^bz^c\displaystyle\sum\limits_k d_k e^{-A_k r^2}$ (where $l=a+b+c$ is the angular momentum quantum number for that orbital).  An individual function of this form is referred to as a Gaussian, atom-centered `basis function'. An `atomic orbital' is then a spin-orbital with the spatial component given by a single basis function, together with a single spin state $\alpha$ or $\beta$. We note that the name `atomic orbital' is actually somewhat of a misnomer, as these functions are seldom actual solutions to an atomic Hamiltonian. A more appropriate term is `atom-centered orbital', which is the term employed in this work.

For any atom, there are several `shells' (corresponding to the principal quantum number $n$ and angular momentum $l$ of the electronic states) that need to be represented with such basis functions. Each shell of every chemically distinct atom has its own finite set of coefficients $d_k$ and exponents $A_k$, whose optimized values are reported in the literature or can be accessed from the basis set exchange~\cite{pritchard2019new}. A collection of basis functions employed for specific molecular calculations is referred to as a `basis set'. Thus, basis sets define the spatial basis functions for all the different atoms and different shells within each atom involved in the calculation, by specifying the coefficients and exponents of the constituent Gaussians.  For example, the STO-3G basis set represents each spatial basis function as a sum of 3 Gaussians (hence the `3G'), with the Gaussian exponents and coefficients being unique for each (atom, shell) combination, e.g., the 1s orbital of H, or the 3p orbital of Cl. STO-3G is in fact a `minimal basis' in that each (atom, shell) combination has only one associated spatial basis function. As an example, when only the 1s orbital of H is included in a calculation, there is only 1 spatial basis function per atom in the STO-3G basis. There are then two atom-centered spin-orbitals - one corresponding to the up spin ($\alpha$) and the other to the down spin ($\beta$). 

For greater flexibility, it is common to assign multiple spatial basis functions to a given (atom, shell) combination, in order to permit orbital expansion, contraction, or polarization along a given direction within a molecular structure. This leads to larger basis sets.  An example is the 6-31G basis used in this work. In this case, each core (i.e., not valence) shell for any atom is assigned a single spatial basis function, made of a linear combination of 6 Gaussians. The valence shells however are assigned two basis functions, one constructed from 3 Gaussians and the other consisting of just 1 Gaussian. Similarly, the 6-311G basis used in this work has a single 6 Gaussian basis function for each core level of a given atom, while there are three functions per valence shell, one being a linear combination of 3 Gaussians and the other two consisting of just 1 Gaussian each.  Since 6-31G has two basis functions per valence shell, and 6-311G has three, they are referred to as `double' and `triple' zeta basis sets respectively (zeta referring to the number of basis functions assigned per valence shell).  Applying these basis sets to the example of the hydrogen atom, we note that the lowest level (1s) is the valence shell and there are no core levels. Therefore, for the H atom, the 6-31G basis set uses two basis functions per atom, while the 6-311G basis set uses three basis functions per atom, leading to 4 and 6 atom-centered spin-orbitals respectively. For calculations with the H$_2$ molecule, these basis sets will then yield 8 and 12 spin-orbitals (see \tab{num-qubits-h2}).

It is important to remember that the basis functions are all predefined by the pre-optimized coefficients and exponents. The basis functions are therefore best thought of as some predefined set of functions, that are not altered by the quantum chemistry protocol and are fixed parameters of the problem once a specific basis set has been chosen.  

\subsection{Molecular Orbitals and Coefficient Matrices}\label{app:orbs-coeffs}

Interesting chemical systems are generally larger than a single atom and are thus not well described by the bare atom-centered orbitals described above. The optimal orbitals for multi-atomic (i.e., molecular) chemical systems are linear combinations of atom-centered orbitals that are called 'molecular orbitals', often abbreviated as MOs. Consider some molecular species with a number of constituent atoms spread out in space. We denote the set of spin-orbitals centered on these atoms by $\{\chi_{\mu}\}$ and the molecular system MOs by $\{\phi_p\}$. Each spin-orbital $\{\chi_{\mu}\}$ consists of a spatial part given by a basis function, and a spin component $\alpha$ or $\beta$. Then the superposition principle motivates the construction of a linear combination of atom-centered spin-orbitals 
\begin{equation}
 \phi_p=\displaystyle\sum\limits C_{\mu p}\chi_\mu
 \label{eq:LCAO}
\end{equation}
where the coefficients $C_{\mu p}$ preceding the atomic orbitals parameterize the MO $\phi_p$ in terms of the specific basis set $\{\chi_{\mu}\}$. These coefficients  $C_{\mu p}$ form a matrix $\mathbf{C}$, often referred to as the `coefficient' or `C' matrix in the literature. $N$ linearly independent atom-centered orbitals can lead to $N$ orthonormal MOs (with the linearly dependent case leading to fewer MOs~\cite{szabo2012modern}).

In the Hartree-Fock optimization procedure, $\matr{C}$ constitutes the only degree of freedom. Specifically, given some molecular system and a spin-orbital basis set, $\{\chi_{\mu}\}$, there are now inflexible, predefined, linear combinations of Gaussians centered on individual atoms of the system. The Hartree-Fock optimization, which is a self-consistent iterative optimization procedure~\cite{szabo2012modern}, therefore optimizes the set of coefficients $\matr{C}$ such that the resulting MOs minimize the total system energy. It is important to note, however, that only MOs that are occupied by electrons contribute to the total energy, and so the Hartree-Fock protocol is perhaps best described as optimizing  $\{C_{\mu i}\}$ for defining the occupied MOs $\{\phi_i\}$ that minimize the energy. In other words, Hartree-Fock optimization separates the space spanned by the atom-centered orbitals into an occupied subspace and an empty or virtual subspace, as specified by $\{C_{\mu i}\}$ and $\{C_{\mu a}\}$, respectively.

It should be noted that there are alternative routes to performing Hartree-Fock optimization than with atom-centered orbitals. These include using plane waves, finite elements, multi-wavelets, and other representations. However, the atom-centered Gaussian orbital paradigm is nearly universal in molecular quantum chemistry, so we do not discuss alternative approaches here. 

\subsection{Coulson-Fischer Point}\label{app:cfpt}

The wavefunction for most closed-shell molecules at their equilibrium geometry is well described by a single RHF configuration. However, stretching a covalent bond in the gas phase to the dissociation limit generally results in the formation of open-shell ``radicals" with unpaired electrons. Some examples of this are stretching \ce{H2} to two H atoms, stretching one \ce{O-H} bond in \ce{H2O} to form the OH radical and an H atom, or stretching the \ce{C-C} bond in \ce{H3C-CH3} to form two \ce{CH3} radicals. The dissociation limit of open-shell species thus cannot be described by RHF, which assumes all electrons are perfectly paired. They however can be described by UHF solutions where this spin-symmetry is broken (i.e., the spatial density of electrons with spin up is not constrained to be identical to that of electrons with spin down). 

There must be a transition along the bond stretching coordinate between the RHF closed-shell behavior at equilibrium and the dissociation limit spin-symmetry broken UHF case. The point on the bond stretching coordinate where the RHF closed-shell solution can undergo spin-symmetry breaking to a lower energy solution is called the Coulson-Fischer (CF) point. At this point, the closed-shell RHF solution ceases to be a minimum of energy against perturbations that break the spatial symmetry between up and down spins, and the minimum energy UHF solutions now have distinct up and down spin spatial distributions. It is however important to remember that the true eigenstates do not have spin-symmetry breaking, and this feature is an artifact of the Hartree-Fock model that reflects its lack of electron correlation.

The mathematics underlying the CF point can be best understood by considering \ce{H2} in a minimal basis-set where only one 1s orbital (constructed from a sum of Gaussians) is available per atom. Let the atoms be labeled H$_A$ and H$_B$, with the basis functions centered therein being $\textrm{1s}_{\textrm{A}}$ and $\textrm{1s}_{\textrm{B}}$. From symmetry between the two atoms, we can define symmetry-adapted linear combinations $\ket{\sigma}$ and $\ket{\sigma^*}$:
\begin{align}
    \ket{\sigma}&=\dfrac{\ket{\textrm{1s}_{\textrm{A}}}+\ket{\textrm{1s}_{\textrm{B}}}}{\sqrt{2(1+\braket{\textrm{1s}_{\textrm{A}}}{\textrm{1s}_{\textrm{B}}}})}\\
    \ket{\sigma^*}&=\dfrac{\ket{\textrm{1s}_{\textrm{A}}}-\ket{\textrm{1s}_{\textrm{B}}}}{\sqrt{2(1-\braket{\textrm{1s}_{\textrm{A}}}{\textrm{1s}_{\textrm{B}}}})}
\end{align}
These two orthonormal spatial orbitals encompass all the available spatial degrees of freedom, and all MOs must then be linear combinations of $\ket{\sigma}$ and $\ket{\sigma^*}$. Let us define the spatial orbitals $\ket{\phi_\alpha}$ and $\ket{\phi_\beta}$ for the up and down spins as:
\begin{align}
    \ket{\phi_\alpha}&=\cos\theta\ket{\sigma}+\sin\theta\ket{\sigma^*}  \\
    \ket{\phi_\beta}&=\cos\theta\ket{\sigma}-\sin\theta\ket{\sigma^*}
\end{align}
At equilibrium, the ground state solution to the Hartree-Fock equations has $\theta=0$, corresponding to $\ket{\phi_\alpha}=\ket{\phi_\beta}=\ket{\sigma}$. Both electrons are therefore present in the $\ket{\sigma}$ level, leading to this being called the `bonding orbital'. At the dissociation limit, however, the stable solutions are:
\begin{align}
    \ket{\phi_\alpha}&=\textrm{1s}_{\textrm{A}} \\
    \ket{\phi_\beta}&=\textrm{1s}_{\textrm{B}}
\end{align}
or 
\begin{align}
    \ket{\phi_\alpha}&=\textrm{1s}_{\textrm{B}} \\
    \ket{\phi_\beta}&=\textrm{1s}_{\textrm{A}}
\end{align}
corresponding to $\theta=\pm \dfrac{\pi}{4}$ (since the overlap $\braket{\textrm{1s}_{\textrm{A}}}{\textrm{1s}_{\textrm{B}}}=0$ at dissociation). 

The $\theta=0$ solution corresponding to the RHF solution is a stationary point of energy at all points, in that $\left(\dfrac{dE}{d\theta}\right)_{\theta=0}=0$ for all internuclear separations. However, it is only a minimum up to the CF point. At the CF point $\left(\dfrac{d^2E}{d\theta^2}\right)_{\theta=0}=0$, while $\left(\dfrac{d^2E}{d\theta^2}\right)_{\theta=0} <0$ at longer distances. Minima of energy correspond instead to $\theta=\pm \Delta$, where the closed-form expression for $\Delta$ is quite complicated (See Ref.~\cite{szabo2012modern} for details), with $0\le \Delta \le \dfrac{\pi}{4}$. The variable $\theta$ can thus be viewed as an order parameter for this problem. 

Since the CF point involves (spin) symmetry breaking, some discontinuities in the energy derivatives are expected on crossing the CF point. The energy is continuous through the CF point, as are the first derivatives of the energy. However, second derivatives of the energy usually have a discontinuity at the CF point~\cite{hait2019wellbehaved}, as they are connected to the inverse of $\left(\dfrac{d^2E}{d\theta^2}\right)$, which goes to zero at the CF point. Molecular properties that are second derivatives of the energy (such as the force constant, which is the second derivative of the energy against the bond stretch) thus are discontinuous at this point, making the CF point transition analogous to a second-order phase transition. 

We note that there are other forms of instabilities (and associated symmetry) breaking in Hartree-Fock, aside from the aforementioned transition from RHF to spin-symmetry broken UHF at the CF point. The interested reader is directed to Refs.~\cite{vcivzek1967stability,seeger1977self} for further details.

\subsection{Summary of terms}

\begin{itemize}
    \item Spin-orbital: Orbital that accounts for both position and spin degrees of freedom, often abbreviated simply as `orbital'.
    \item Spatial orbital: Orbital that only accounts for position degrees of freedom. 
    \item Atom-centered orbital: Orbital centered at one point in space (corresponding to the atomic nucleus) and asymptotically decaying away from it. Commonly called an atomic orbital.
    \item Basis function: The spatial part of an atom-centered orbital. Generally (but not always) defined as a linear combination of the form $x^ay^bz^c\displaystyle\sum\limits_k d_k e^{-A_k r^2}$, where the coefficients $\{d_k, A_k\}$ are preoptimized and fixed, i.e., they are not degrees of freedom available to standard quantum chemistry calculations. 
    \item Basis set: A collection of basis functions, for each shell of each atom (for a given set of atoms). These are predefined in the literature (see, e.g., Ref. \cite{pritchard2019new}). They essentially define a way to discretize space centered around a given atom. 
    \item Molecular orbital: Orbitals for a general, multi-atom system.  In quantum chemistry, these are generally represented as a linear combination of atom-centered orbitals. In general, given $N$ atom-centered orbitals that are linearly independent, $N$ orthonormal molecular orbitals can be constructed from these. 
    \item Coefficient matrix: The set of coefficients that maps atom-centered orbitals to molecular orbitals. 
    \item Spin-general orbitals: Orbitals that can include both spin up and spin down components. Rare in quantum chemistry, but vital for relativistic effects using the Dirac equation.
    \item Spin-unrestricted orbitals: Orbitals that are factorizable into spatial and spin components. This is relatively common in quantum chemistry.
    \item Restricted orbitals: Special case where two spin orbitals are constrained to have the same spatial function, but with different spin components.
    \item Coulson-Fischer point: Point along the bond-stretching coordinate when it is energetically favorable to break spin-symmetry between up and down spins in mean-field models like Hartree-Fock or DFT. 
\end{itemize}

\section{Details of the low-rank factorization}\label{app:details-low-rank}

In the following we have taken the outline from Appendix A of Ref.~\cite{motta2021low}, with some clarifications and changes of notation. The unitary coupled-cluster doubles ansatz is 
\begin{equation}
    e^{\hat\tau} |\Phi_0\rangle
\end{equation}
where 
\begin{align}
    \hat\tau &\equiv \hat{T}-\hat{T}^\dag, \label{eq:ucc1}\\
    \hat{T} &= \sum_{pqrs=1}^N t_{ps,qr} a_p^\dag a_s a_q^\dag a_r,
    \label{eq:ucc2}
\end{align}
where we define the $N^2\times N^2$ supermatrix $\matr{T}$ as
\begin{equation}
    t_{ps,qr} \equiv \begin{cases} 
    t^\text{MP2}_{pqrs} & p<q;s<r;r,s\in\text{occ};p,q\in\text{virt} \\
    0 & \text{otherwise}.
    \end{cases}
\end{equation}
For the Takagi factorization we require $\matr{T}$ to be a symmetric matrix, so we define the symmetrized version of $\matr{T}$ as 
\begin{equation}
\matr{T}^\text{S} = \frac{1}{2}(\matr{T}+\matr{T}^\text{T}).
\end{equation}
This symmetrization may increase the number of nonzero singular values of the highly sparse $\matr{T}$ matrix by up to a factor of two, depending on the column filling. Performing the SVD gives
\begin{align}
    \matr{T} &= \matr{U}\matr{\Sigma} \matr{V}^\dag, \\
    t_{ps,qr} &= \sum_{l=1}^L u_{ps, l}\sigma_{l}v_{qr,l}^*,
\end{align}
where $\matr{U}$ and $\matr{V}^\dag$ are unitary and $\sigma_{l}$ are the singular values. We then define the operators
\begin{align}
    \hat{M}_l &\equiv \sqrt{\sigma_l}\sum_{pq=1}^N(u_{pq,l}+v_{pq,l}^*)\opr{a}_p^\dag \opr{a}_q, \\
    \hat{N}_l &\equiv \sqrt{\sigma_l}\sum_{pq=1}^N(u_{pq,l}-v_{pq,l}^*)\opr{a}_p^\dag \opr{a}_q.
\end{align}
Following this definition we arrive at
\begin{equation}
    \hat\tau = -i\sum_{l=1}^{L}\sum_{\mu=1}^4 \hat{Y}_{l\mu}^2
\end{equation}
where
\begin{align}
    \hat{Y}_{l1} &= \dfrac{1+i}{4}(\hat{M}_l - i\hat{M}_l^{\dag}) \\
    \hat{Y}_{l2} &= \dfrac{1+i}{4}(\hat{M}_l + i\hat{M}_l^{\dag}) \\
    \hat{Y}_{l3} &= \dfrac{1-i}{4}(\hat{N}_l - i\hat{N}_l^{\dag}) \\
    \hat{Y}_{l4} &= \dfrac{1-i}{4}(\hat{N}_l + i\hat{N}_l^{\dag})
\end{align}
are a set of four normal operators. Note that for the Takagi factorization we have $\matr{V}^\dag=\matr{U}^{\text{T}}$, so that $\hat{N}_l=0$, thus reducing the number of $\hat{Y}_{l\mu}$ operators per singular value from $m=4$ to $m=2$. Since these operators are normal, then the $N\times N$ matrix of the coefficients $\matr{Y}$ (representing $\hat{Y}_{l\mu}$, so all the following quantities have $l,\mu$ labels implied) has the eigendecomposition
\begin{align}
    \matr{Y} &= \matr{B}\matr{\Lambda} \matr{B}^\dag, \\
    y_{rs} &= \sum_{p=1}^\rho b_{rp}\lambda_p b_{sp}^*,
\end{align}
where $\matr{B}$ is a unitary matrix and $\lambda_p$ are the eigenvalues. Then we may write
\begin{equation}
    \hat{Y} = \sum_{rs=1}^N\sum_{p=1}^\rho b_{rp} \lambda_p b_{sp}^* \opr{a}_r^\dag \opr{a}_s =\sum_{p=1}^\rho\lambda_p\tilde a^\dag_p\tilde a_p = \sum_{p=1}^\rho\lambda_p\tilde n_p,
\end{equation} 
where the tilde'd operators are expressed in the rotated basis,
\begin{equation}
    \tilde a_p^\dag = \sum_{r=1}^N b_{rp} \opr{a}_r^\dag, \qquad \tilde a_p = \sum_{s=1}^N b_{sp}^* \opr{a}_s.
\end{equation}
Therefore we may express the $\hat{Y}^2_{l\mu}$ operators in terms of fermionic number operators by first rotating the entire qubit register into the appropriate basis. Then with the appropriate basis rotation operators determined according to Ref.~\cite{kivlichan2018quantum}, we arrive at
\begin{align}
    e^{\hat\tau} \approx \prod_{l=1}^L\prod_{\mu=1}^m\opr{\mathcal{U}}_B^{(l,\mu)\dag} \exp\left( -i\sum_{pq=1}^\rho\lambda_p^{(l,\mu)}\lambda_q^{(l,\mu)} \opr{n}_p\opr{n}_q\right) \opr{\mathcal{U}}_B^{(l,\mu)}.
    \label{finalAnsatz}
\end{align}
We may group neighboring basis rotation matrices into a single unitary, $\tilde{\mathcal{U}}_B$, to reduce the accumulated circuit depth incurred by the basis changes by roughly a factor of 2, yielding the final form written in the main text: 
\begin{align}
    e^{\hat{\tau}} \approx \opr{\mathcal{U}}_B^{(1,1)\dag}\prod_{l=1}^L\prod_{\mu=1}^m \exp\left(-i \sum_{pq}^{\rho_l}\lambda_p^{(l,\mu)}\lambda_q^{(l,\mu)} \opr{n}_p\opr{n}_q\right) \tilde{\mathcal{U}}_B^{(l,\mu)}
\end{align}

\section{Details of the NOQE circuit}\label{app:noqe-circuit}

The NOQE circuit of \fig{noqe-circuit} for the real component of $H_{IJ}$, following the initial Hadamard gate to prepare the ancilla register, starts with the $2N+1$ qubit input state $\ket{\Phi_\text{UHF}}\ket{\text{vac}}\ket{+}$, distributed over three registers. Register 1 contains $\ket{\Phi_\text{UHF}}$, an $N$-qubit UHF ansatz state, register 2 contains $\ket{\text{vac}}$, the vacuum (zero electron) state of an $N$-qubit register, and register 3 is a 1-qubit ancilla containing $\ket{+}$, the $+X$ eigenstate.  The inclusion of the additional phase gate $P_{\pi/2}$ for calculation of the imaginary component $H_{IJ}$ merely replaces this initial ancilla state by the $+Y$ eigenstate. Below we follow the steps in the circuit for the real part of $H_{IJ}$. An $N$-qubit controlled-SWAP operation is then performed, with the ancilla qubit as control, yielding the state 
\begin{equation}
\frac{1}{\sqrt 2} \bigg( \ket{\Phi_\text{UHF}}\ket{\text{vac}}\ket{0} + \ket{\text{vac}}\ket{\Phi_\text{UHF}}\ket{1} \bigg).
\end{equation}
The low rank cluster operators $e^{\hat{\tau}_I}$ and $e^{\hat{\tau}_J}$ are then applied to registers 1 and 2, followed by the basis rotations $\opr{\mathcal{U}}_{I\rightarrow1}$ and $\opr{\mathcal{U}}_{J\rightarrow1}$ to these same registers, generating the state
\begin{equation}
\frac{1}{\sqrt 2} \bigg( \ket{\phi_{I}}\ket{\text{vac}}\ket{0} + \ket{\text{vac}}\ket{\phi_{J}}\ket{1} \bigg),
\end{equation}
in which there is now a different NOUCC(2) ansatz in registers 1 and 2, but both are expressed in the same basis. We then perform a second controlled-SWAP conditioned on the ancilla qubit as before, after which register 2 is again in the vacuum state and can be ignored, yielding the entangled $N+1$ qubit state 
\begin{equation}
\frac{1}{\sqrt 2} \bigg( \ket{\phi_{I}}\ket{0} + \ket{\phi_{J}}\ket{1} \bigg).
\end{equation}
A second Hadamard gate on the ancilla is then performed and the ancilla (register 3) is measured in the $Z$ basis. It is straightforward to verify that, as in the simple Hadamard test, this results in the expectation value $\langle \opr{Z}_{\text{anc}} \rangle = \text{Re} \langle \phi_I | \phi_J \rangle$. The Hamiltonian is then measured on the state of the $N$ qubits in register 1, resulting in an expectation value that is conditioned on the outcome of the ancilla measurement. As shown in Ref.~\cite{huggins2020non}, the conditional outcome may be combined shotwise with the corresponding ancilla outcome to generate the expectation value $\text{Re} \langle \phi_I | \opr{H} | \phi_J \rangle $. Repeating the circuit with the ancilla prepared as the $+Y$ eigenstate, i.e., $\frac{1}{\sqrt{2}} (\ket{0}+i\ket{1})$, yields the imaginary components of the matrix elements, $\text{Im} \langle \phi_I | \phi_J \rangle$ and $\text{Im} \langle \phi_I | \opr{H} | \phi_J \rangle $.

\section{Trotter error bounds}\label{app:error-estimation}

Since we propose the use of non-variational parameters, we present some cursory analysis of the Trotter error propagated through to the calculated eigenvalues. We expect that the derived error bound is a significant overestimate of the true error in the eigenvalues, and we include it here as a starting point for further investigation. We also extend the analysis to include truncation errors due to the discarding of small singular values during the low-rank decomposition. We observe that both sources of error are incurred at the first factorization step (it is easy to see that Trotter error is only incurred at the level of the first factorization in the low-rank procedure, since the number operator terms after the second factorization commute exactly). After the first SVD or Takagi factorization, the ansatz is of the form
\begin{align}
    \opr{U} = e^{-i\sum_{k=1}^{mL}\hat{Y}^2_k},
\end{align}
where the $\hat{Y}^2_k$ are one-body-squared operators, with the index $k$ combining both the $l$ and $\mu$ indices. To bound the norm of these operators, we use the fact that the fermionic ladder operators have spectral norm bounded by $\|\opr{a}_p^\dag \opr{a}_q\|\leq1$, so that we may bound 
\begin{align}
\|\hat Y_k\|\leq\frac{\sqrt{\sigma_k}}{2\sqrt{2}}\sum_{pq}(|u_{pq,k}|+|v_{pq,k}|)=\frac{\sqrt{\sigma_k}}{2\sqrt{2}}(\|\vec{u}^{(k)}\|_1+\|\vec{v}^{(k)}\|_1),
\end{align}
where $\|\cdot\|_1$ here denotes the vector 1-norm. The usual expression to relate this to the vector 2-norm would then give $\|\vec{u}^{(k)}\|_1\leq\sqrt{N^2}\|\vec{u}^{(k)}\|_2=N$, but we may exploit the sparsity in the singular vectors to instead write 
\begin{align}
\|\hat{Y}_{k}\| \leq \frac{\sqrt{\sigma_k}}{2\sqrt{2}}\left(\sqrt{\text{nnz}(\vec{u}^{(k)})}+\sqrt{\text{nnz}(\vec{v}^{(k)})}\right),
\end{align}
where $\text{nnz}(\cdot)$ counts the number of nonzero elements in a vector. For the Takagi factorization we have $\vec{u}^{(k)}=\vec{v}^{(k)}$, so that
\begin{align}
    \|\hat Y_k^2\| \leq \frac{\sigma_k\text{nnz}(\vec{u}^{(k)})}{2}.
\end{align}

\subsection{Trotter error}

Let $\opr{U}_I$ denote the exact unitary $e^{\hat\tau_I}$, $\tilde U_I$ refer to the approximated form on the RHS of \eq{lowrank_final}, and $\Delta \opr{U}_I\equiv \opr{U}_I- \tilde U_I$ denote the difference in the Taylor expansions of the exact and approximated unitaries.  We can estimate the $\|\Delta \opr{U}_I\|$ by the largest non-cancelling terms in the expansion of $\opr{U}_I- \tilde U_I$ at finite Trotter order~\cite{childs2021}. After the first SVD or Takagi factorization, with $p$-th order Trotterization, and with the abbreviations $\opr{U}= \opr{U}_I$ and $\tilde U^{(p)}= \tilde U^{(p)}_I$, we may write
\begin{align}
    \opr{U} = e^{-i\sum_{k=1}^{mL}\hat{Y}^2_k}, \qquad \tilde U^{(p)} = \prod_{x=1
    }^{p}\prod_{k=1}^{mL}e^{-i\hat{Y}_k^2/p}
\end{align}
For the $p$-th order Trotter expansion, we may bound the norm of the operator difference in terms of the sum over all possible $(p+1)$-th order commutators. Keeping only terms of $O((\sigma_{\max}/p)^{p+1})$ and neglecting the terms $O((\sigma_{\max}/p)^{p+2})$ (valid when $\sigma_{\max}/p\ll1$, i.e., in the limit of small $\hat\tau$ and/or large $p$), we find
\begin{align}
    \|\Delta \opr{U}^{(p)}\| &\leq \sum_{k_1,...,k_{p+1}=1}^{mL}\frac{1}{p^{(p+1)}}\|[\hat{Y}^2_{k_{p+1}},\cdots[\hat{Y}^2_{k_2},\hat{Y}^2_{k_1}]\cdots]\| + O((\sigma_{\max}/p)^{p+2}) \\
    &\leq \left(\frac{1}{2p}\sum_{k=1}^{mL}\text{nnz}(\vec{u}^{(k)})\sigma_k\right)^{p+1} +  O((\sigma_{\max}/p)^{p+2}).
    \label{eq:dUbound}
\end{align}
Note that the estimate of $\|\Delta \opr{U}^{(p)}\|$ in \eq{dUbound} neglects factors of $\frac{1}{n!}$ in the $n$-th terms of the Taylor expansions of $\opr{U}$ and $\tilde U^{(p)}$, of which the combined effect on the expansion of $\Delta \opr{U}^{(p)}$ must be computed explicitly via e.g., the Zassenhaus formula~\cite{magnus1954exponential}, and is difficult to generalize for arbitrary order. The number of distinct commutator terms appearing at $p$-th order may also have been dramatically overestimated, since we have not assumed any cancellations. 

\subsection{Low-rank truncation error}

We now consider the extension of the above result to the case of truncation errors due to the thresholding of the singular values of the cluster operator $\hat T$, prior to the Trotterization. Our exact and approximated unitaries now take the form
\begin{align}
    \opr{U} = \exp\left(-i\sum_{k=1}^{mL}\hat{Y}^2_k\right), \qquad \tilde U = \exp\left(-i\sum_{k=1}^{mR}\hat{Y}^2_k\right),
\end{align}
where $R$ denotes the truncation rank, i.e., $R<L$, such that all singular values beyond $R$ are discarded according to the fixed $\ell^p$-norm threshold $\varepsilon_p$ (see \eq{truncate_L}). Then we can find the corresponding singular-value threshold $\sigma_\text{tol}$ such that for all $i \geq R+1$, $\sigma_i \leq \sigma_\text{tol}$. (Note that for $p=\infty$, we can set $\sigma_\text{tol} = \varepsilon_{\infty}$.) We consider terms of $O(\sigma_\text{tol})$ and neglect terms of $O(\sigma_\text{tol}^2)$ (valid when $\sigma_\text{tol}\ll1$), so that for the Takagi factorization we have
\begin{align}
    \|\Delta \opr{U}\| &\leq \sum_{k=mR+1}^{mL}\|\hat{Y}_k^2\| + \sum_{k'=1}^{mR}\sum_{k=mR+1}^{mL}\|\{\hat{Y}_{k}'^2,\hat{Y}_k^2\}\| + O(\sigma_\text{tol}^2) \\
    &\leq \frac{1}{2} \left( 1 + \sum_{k'=1}^{mR} \text{nnz}(\vec{u}^{(k')})\sigma_{k'}  \right) \left( \sum_{k=mR+1}^{mL} \text{nnz}(\vec{u}^{(k)})\sigma_k \right) + O(\sigma_\text{tol}^2).
\end{align}

\subsection{Multi-reference error propagation}

Having looked at the effect of Trotterization and truncation on the single-reference energies, we now consider the propagation of errors through the generalized eigenproblem to the multi-reference result. First we wish to bound the Trotter error in the matrix elements of $\matr{H}$ and $\matr{S}$. For some operator $\opr{A}$ expressed in the non-orthogonal basis, we find the Trotter error in the matrix element $A_{IJ}$ as
\begin{align}
    \delta A_{IJ} &= A_{IJ}-\tilde A_{IJ} \\ 
    &= \bra{\Phi_I}\left(\opr{U}_I^\dag \opr{A}\opr{U}_J - \tilde{U}_I^\dag \opr{A}\tilde{U}_J\right)\ket{\Phi_J}.
\end{align}
We bound this matrix element by the spectral norm of the operator difference, and multiply by unitaries within the spectral norm to get
\begin{align}
    |\delta A_{IJ}| &\leq \| \tilde{U}_I\opr{U}_I^\dag \opr{A} - \opr{A} \tilde{U}_J\opr{U}_J^\dag\| \label{eq:A_el_norm}\\
    &= \|\tilde{U}_I\Delta \opr{U}_I^\dag \opr{A} + \opr{A}\Delta \opr{U}_J \opr{U}_J^\dag\| \\
    &\leq \|\opr{A}\|(\|\Delta \opr{U}_I\| + \|\Delta \opr{U}_J\|).
\end{align}
Then the maximum possible error in the general off-diagonal elements of the Hamiltonian and overlap matrices is given by
\begin{align}
    |\delta H_{IJ}| &\leq 2\|\opr{H}\|\|\Delta \opr{U}\|_\text{max},\label{eq:hij_bound}\\
    |\delta S_{IJ}| &\leq 2\|\Delta \opr{U}\|_\text{max}. \label{eq:sij_bound}
\end{align}
The error in the Hamiltonian and overlap matrices is then bounded by
\begin{align}
    \|\Delta \matr{H}\| &\leq 2M\|\opr{H}\|\|\Delta \opr{U}\|_\text{max},\label{eq:DelH_bound}\\
    \|\Delta \matr{S}\| &\leq 2M\|\Delta \opr{U}\|_\text{max},\label{eq:DelS_bound}
\end{align}
where $M$ is the number of NOQE reference states. Following the generalized eigenvalue perturbation theory of Mathias and Li~\cite{Mathias2004TheDG}, we define the error in the eigenangles as
\begin{align}
    \Delta\theta_J \equiv |\tan^{-1}\tilde E_J-\tan^{-1} E_J|.
\end{align}
The error in the eigenangles is obtained by applying the Mathias-Li bound~\cite{Mathias2004TheDG},
\begin{align}
    \Delta\theta_J \leq \sin^{-1} \left(Md_J^{-1}\sqrt{\|\Delta \matr{S}\|^2+\|\Delta \matr{H}\|^2}\right),
    \label{eq:mathias-li}
\end{align}
where the condition number $d_J^{-1}$ is defined by $d_J\equiv|\vec{c}_J^{\,\dag}(\matr{H}+i\matr{S})\vec{c}_J| = \sqrt{(1+E_J^2)}$, with $\vec{c}_J$'s being the unit-norm eigenvectors of the generalized eigenvalue problem $\matr{H}\vec{c} = E\matr{S}\vec{c}$. Substituting our bounds \eq{DelH_bound} and \eq{DelS_bound}, as well as easily calculated upper bounds on the condition number and the norm of the Hamiltonian operator,
\begin{align}
    d_J^{-1}&\leq 1,\\
    \|\opr{H}\| &\leq \|\opr{H}\|_\text{u} \equiv \sum_{pq}|h_{pq}|+\sum_{pqrs}|h_{pqrs}|,
\end{align}
we arrive at the final bound
\begin{align}
    \Delta\theta_J \leq \sin^{-1} \left(2M^2\|\Delta \opr{U}\|_{\max}\sqrt{1+\|\opr{H}\|_\text{u}^2}\right).
    \label{final_bound}
\end{align}
Then for a given target eigenangle difference $\Delta\theta$, we may rearrange this bound to obtain a requirement on the Trotter order $p$;
\begin{align}
    \left(\frac{1}{2p}\sum_{k=1}^{mL}\text{nnz}(\vec{u}^{(k)})\sigma_k\right)^{p+1} \lessapprox \frac{\sin(\Delta\theta)}{M^2\sqrt{1+\|\opr{H}\|_\text{u}^2}}.\label{eq:min_p_bound}
\end{align}
The function on the LHS tends rapidly to zero in the limit of large $p$, ensuring a reduction in the Trotter error of the eigenvalues to within chemical accuracy past a finite minimum Trotter order. However, we note that this bound is likely to be a severe overestimate of the true Trotter error, due to the overestimate of $\|\Delta \opr{U}^{(p)}\|$ in \eq{dUbound}. Furthermore, the norm in \eq{A_el_norm} may be too loose a bound on the matrix element, if the UHF reference states happen to have low overlap with the first singular vectors of the quantity inside the norm. The adapted Mathias-Li bound of \eq{mathias-li} is a worst-case estimate of $\Delta\theta$ for a given $\|\Delta \matr{H}\|$ and $\|\Delta \matr{S}\|$, although this bound is still tighter than that of other perturbation theories such as Stewart's bound~\cite{stewart1979pertubation}.

\clearpage
\section{H\texorpdfstring{$_2$}~ results with STO-3G and 6-31G basis sets}\label{app:h2-other-results}

\subsection{STO-3G}
\begin{figure}[h!]
    \begin{minipage}{0.48\columnwidth}
       \includegraphics[width=\linewidth]{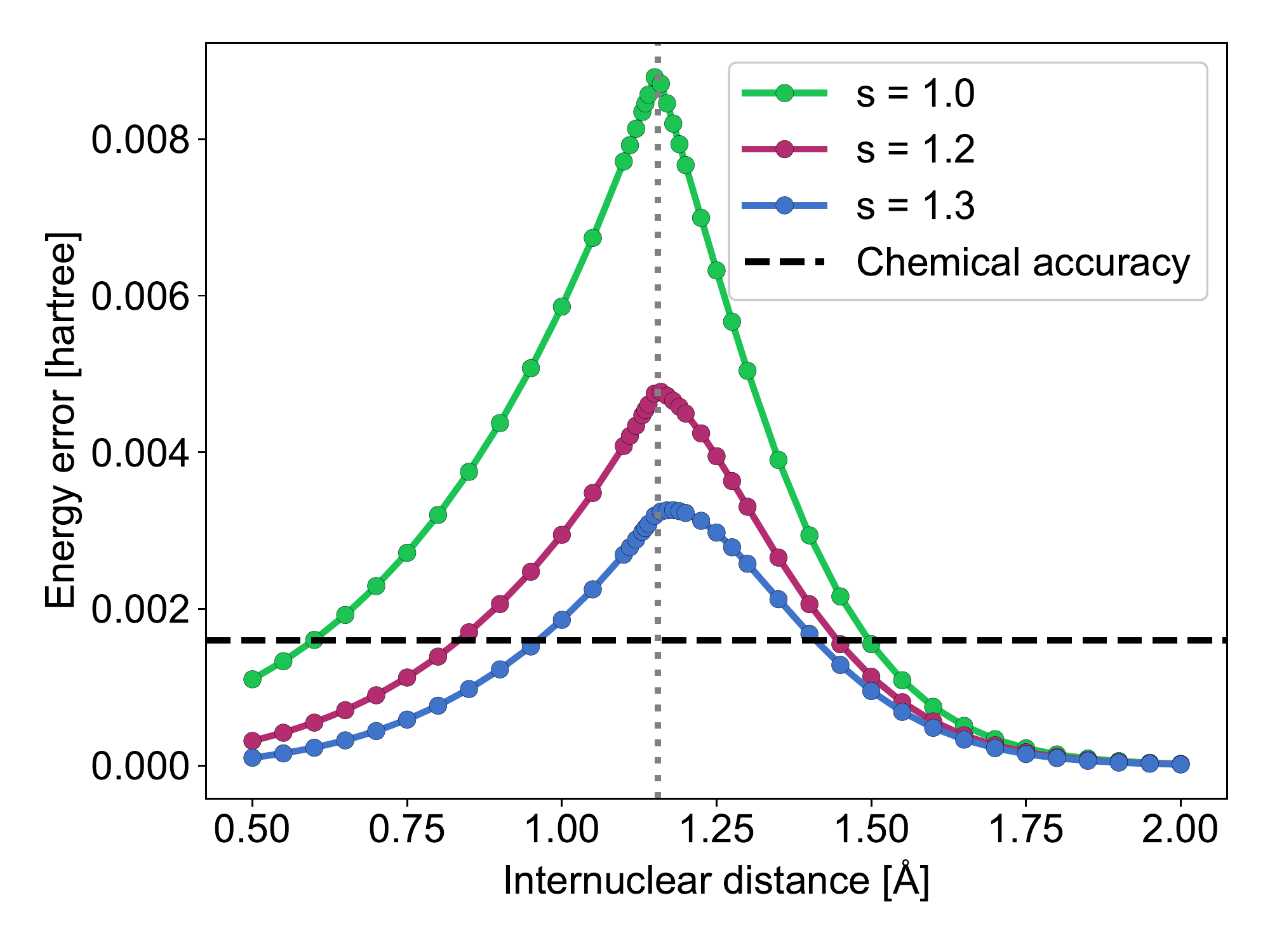}
       \subcaption{S$_0$ energy error relative to FCI.}
       \label{fig:h2-sto3g-S0scale}
    \end{minipage}
    \begin{minipage}{0.48\columnwidth}
       \includegraphics[width=\linewidth]{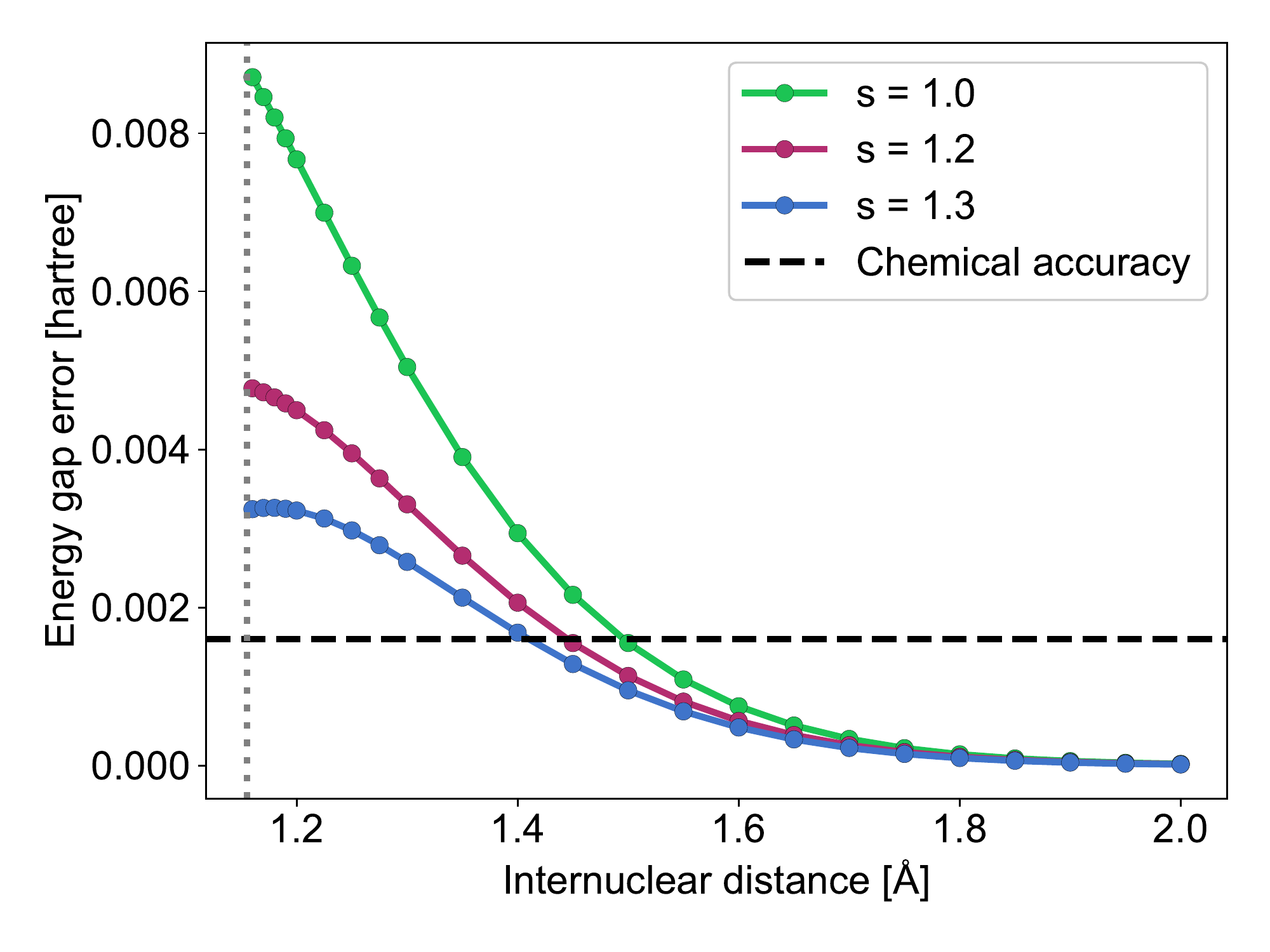}
       \subcaption{Singlet-triplet gap error relative to FCI.}
       \label{fig:h2-sto3g-stgapscale} 
    \end{minipage}
    \caption{Energy errors relative to FCI for singlet and triplet states of H$_2$ from NOQE with NOUCC(2) ansatz states in which the MP2 amplitudes are scaled by $s$ (calculations made with the STO-3G basis). The location of the CF point is marked with a dotted gray line.}
    \label{fig:h2-sto3g-scaled-errors}
     \vspace{-10pt}
\end{figure}

\begin{figure}[h!]
    \begin{minipage}{0.48\columnwidth}
       \includegraphics[width=1\linewidth]{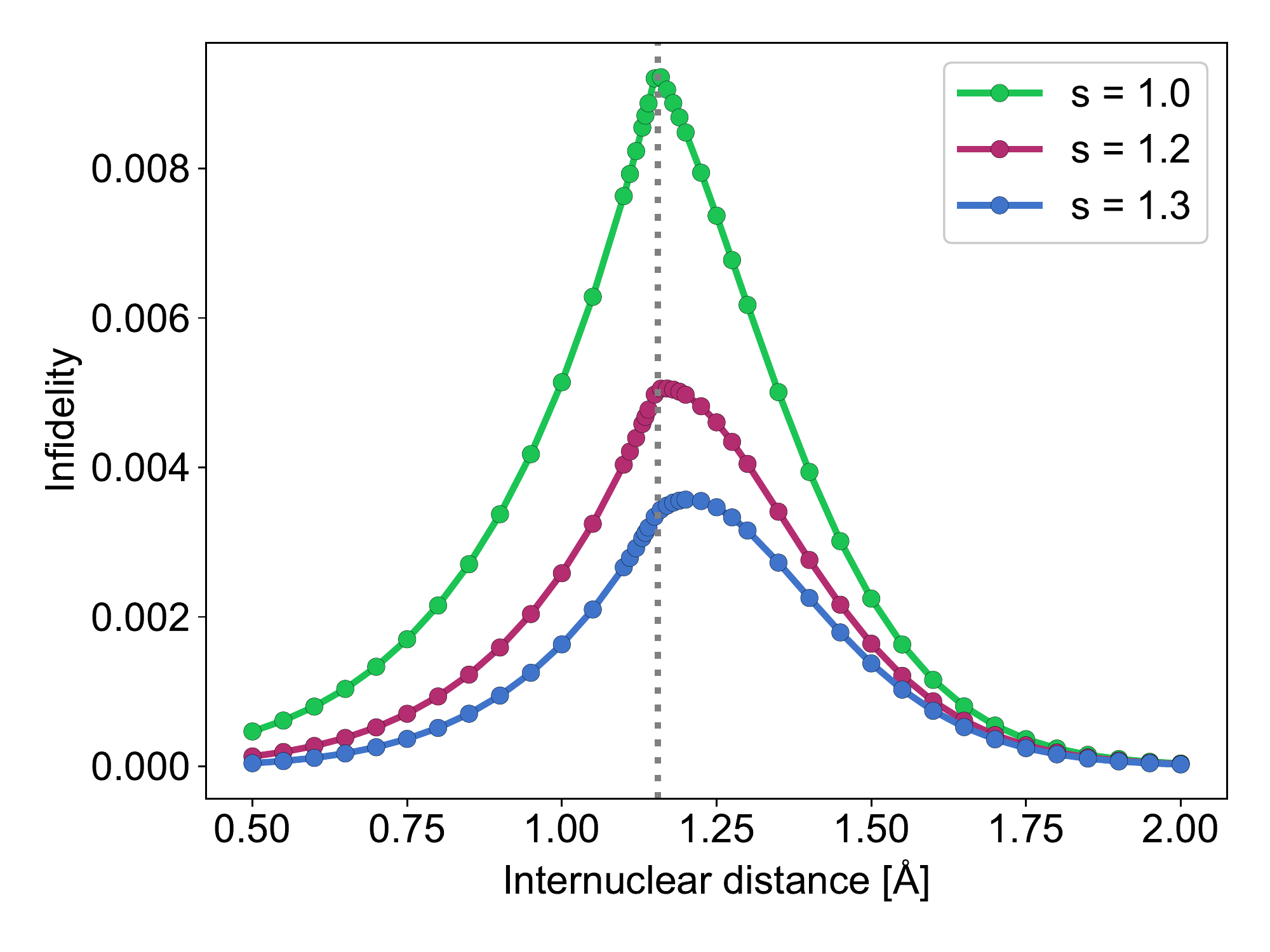}
       \subcaption{Infidelity of NOQE S$_0$ state. }
       \label{fig:h2-sto3g-S0scaleinfidelity}
    \end{minipage}
    \begin{minipage}{0.48\columnwidth}
       \includegraphics[width=1\linewidth]{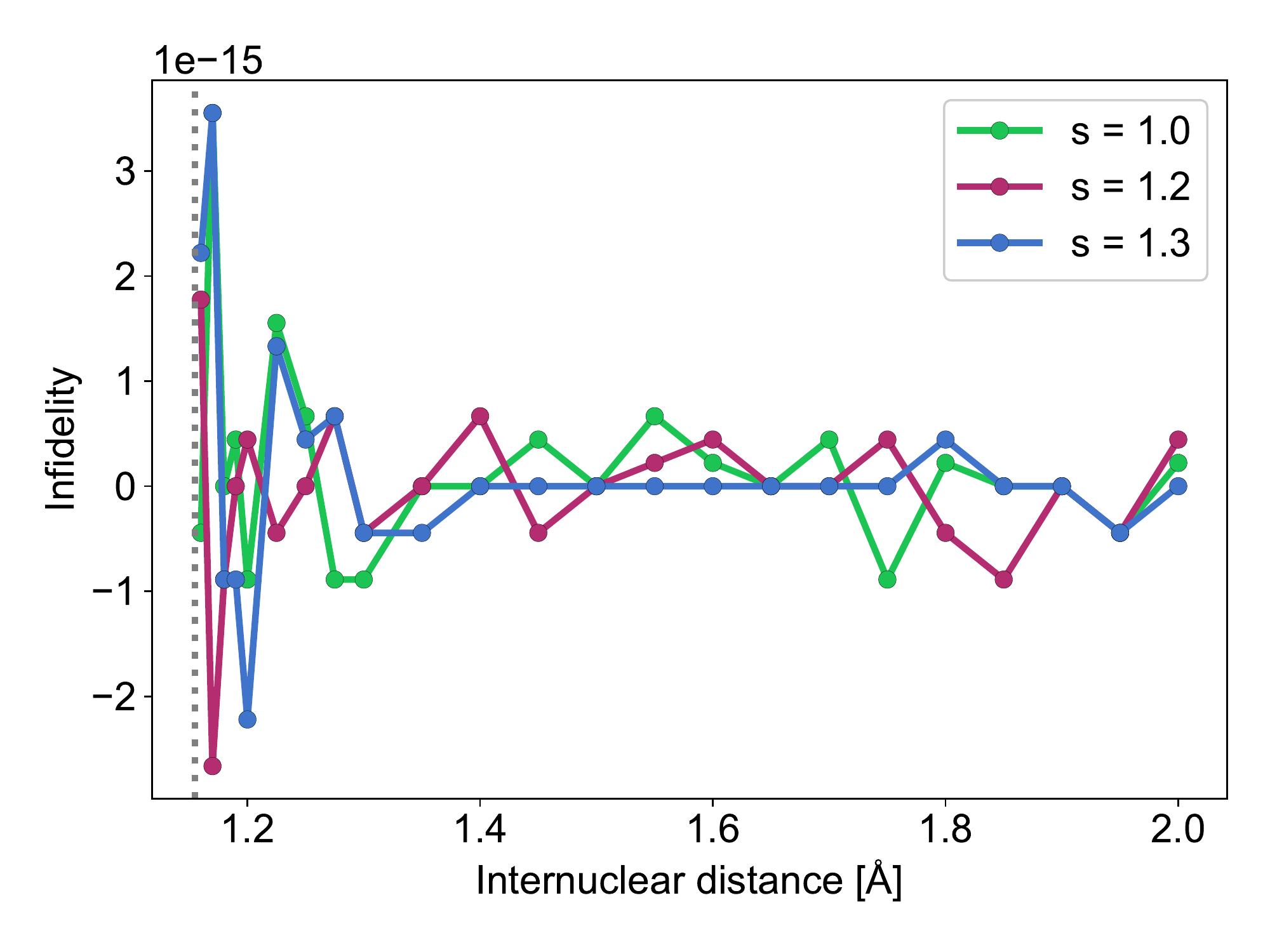}
       \subcaption{Infidelity of NOQE T$_1$ state.}
       \label{fig:h2-sto3g-T1scaleinfidelity} 
    \end{minipage}
    \caption{Infidelity $I_\text{FCI}$ relative to the FCI states of the amplitude scaled NOQE states derived from the NOUCC(2) ansatz, for H$_2$ with the STO-3G basis. See \eq{Infidelity_FCI}. The location of the CF point is marked with a dotted gray line. Note that the extremely low infidelity of the T$_1$ state (which is just numerical noise) demonstrates that this state is predicted exactly by NOQE.}
    \label{fig:h2-sto3g-infidelities}
\end{figure}

\FloatBarrier
\clearpage

\subsection{6-31G}
\begin{figure}[h!]
    \begin{minipage}{0.48\columnwidth}
       \includegraphics[width=\linewidth]{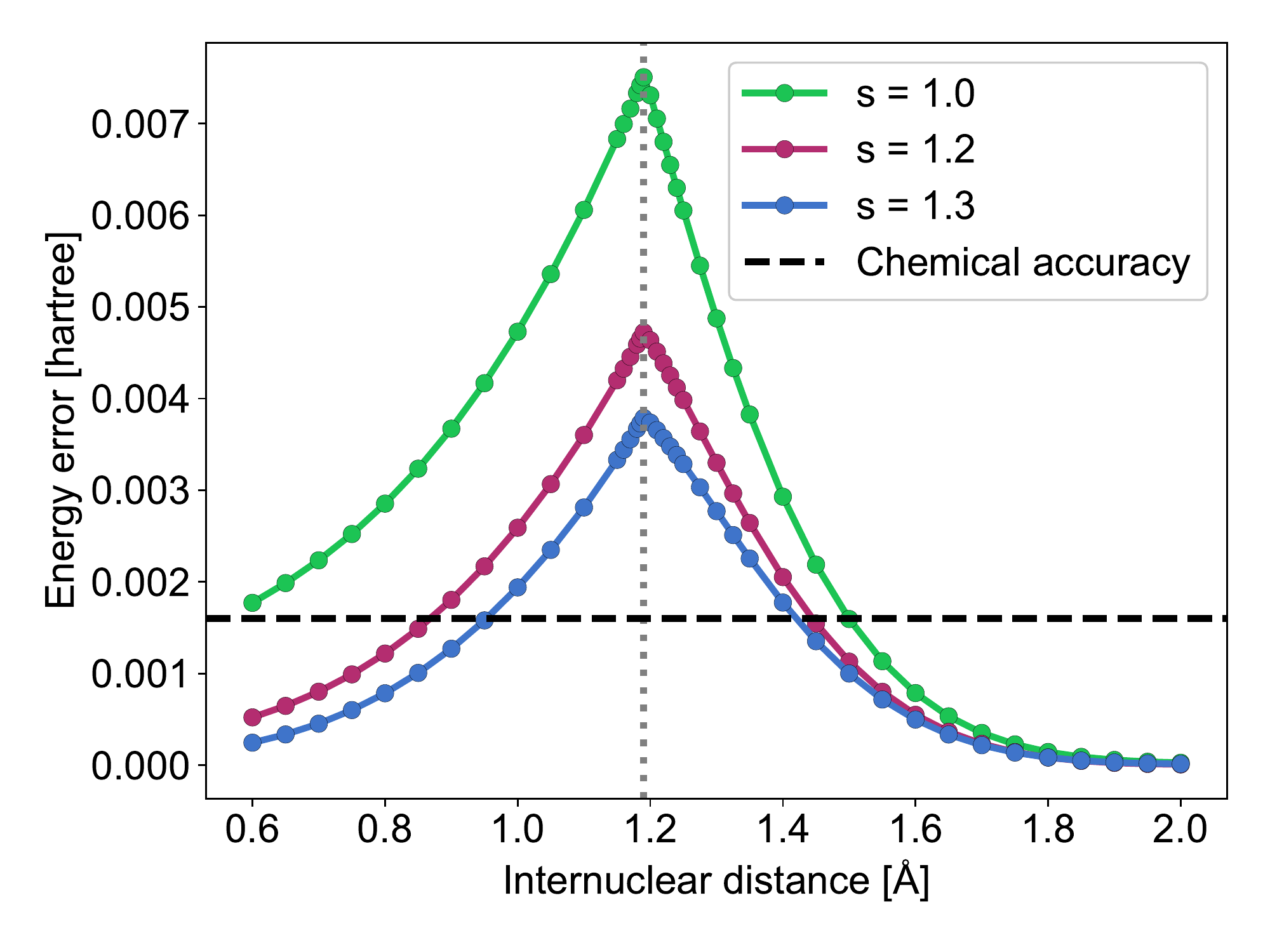}
       \subcaption{S$_0$ energy error relative to FCI.}
       \label{fig:h2-631g-S0scale}
    \end{minipage}
    \begin{minipage}{0.48\columnwidth}
       \includegraphics[width=\linewidth]{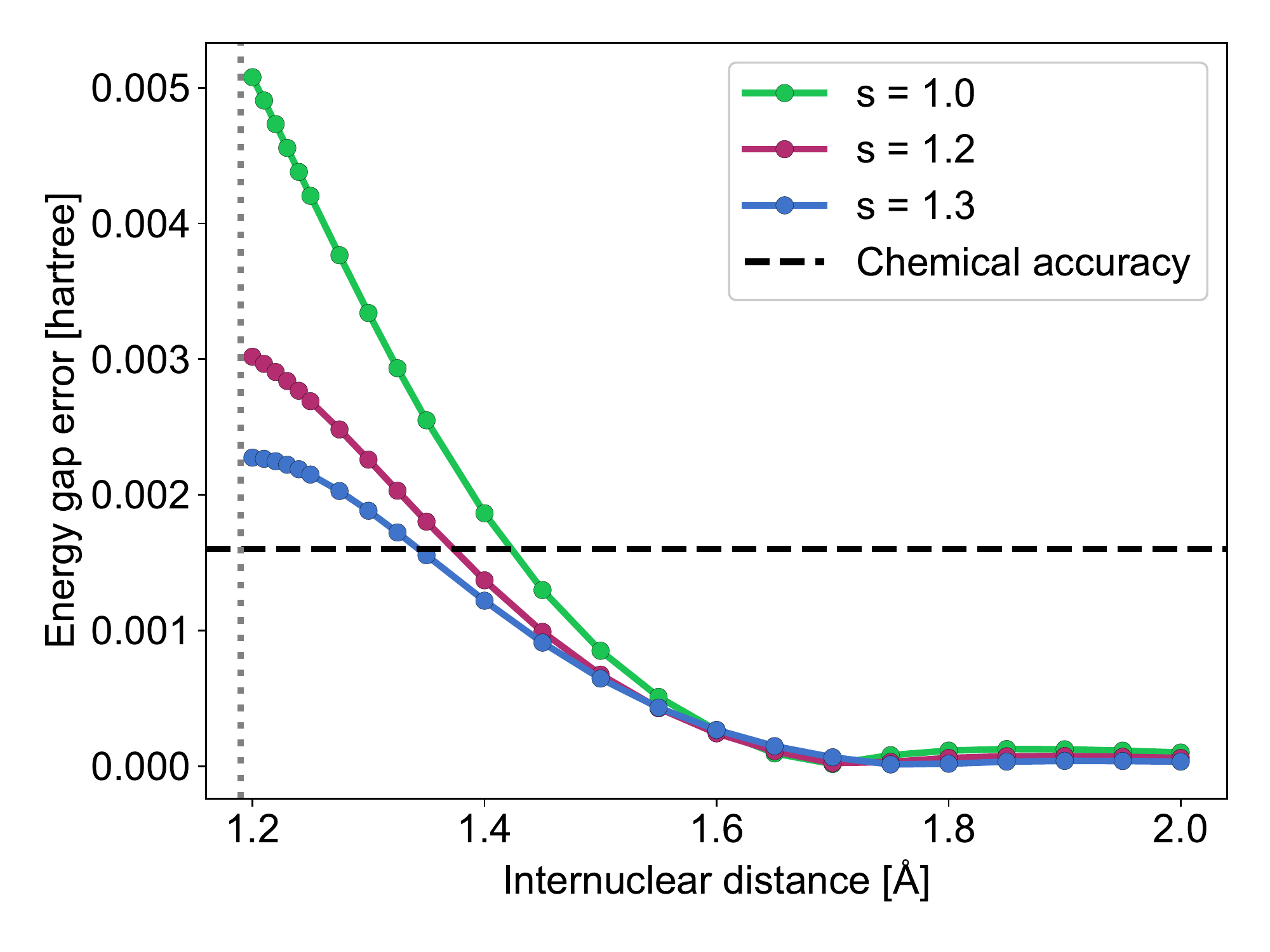}
       \subcaption{Singlet-triplet gap error relative to FCI.}
       \label{fig:h2-631g-stgapscale} 
    \end{minipage}
    \caption{Energy errors relative to FCI for singlet and triplet states of H$_2$ from NOQE with NOUCC(2) ansatz states in which the MP2 amplitudes are scaled by $s$ (calculations made with the 6-31G basis). The location of the CF point is marked with a dotted gray line.}
    \label{fig:h2-631g-scaled-errors}
     \vspace{-10pt}
\end{figure}

\begin{figure}[h!]
    \begin{minipage}{0.48\columnwidth}
       \includegraphics[width=1\linewidth]{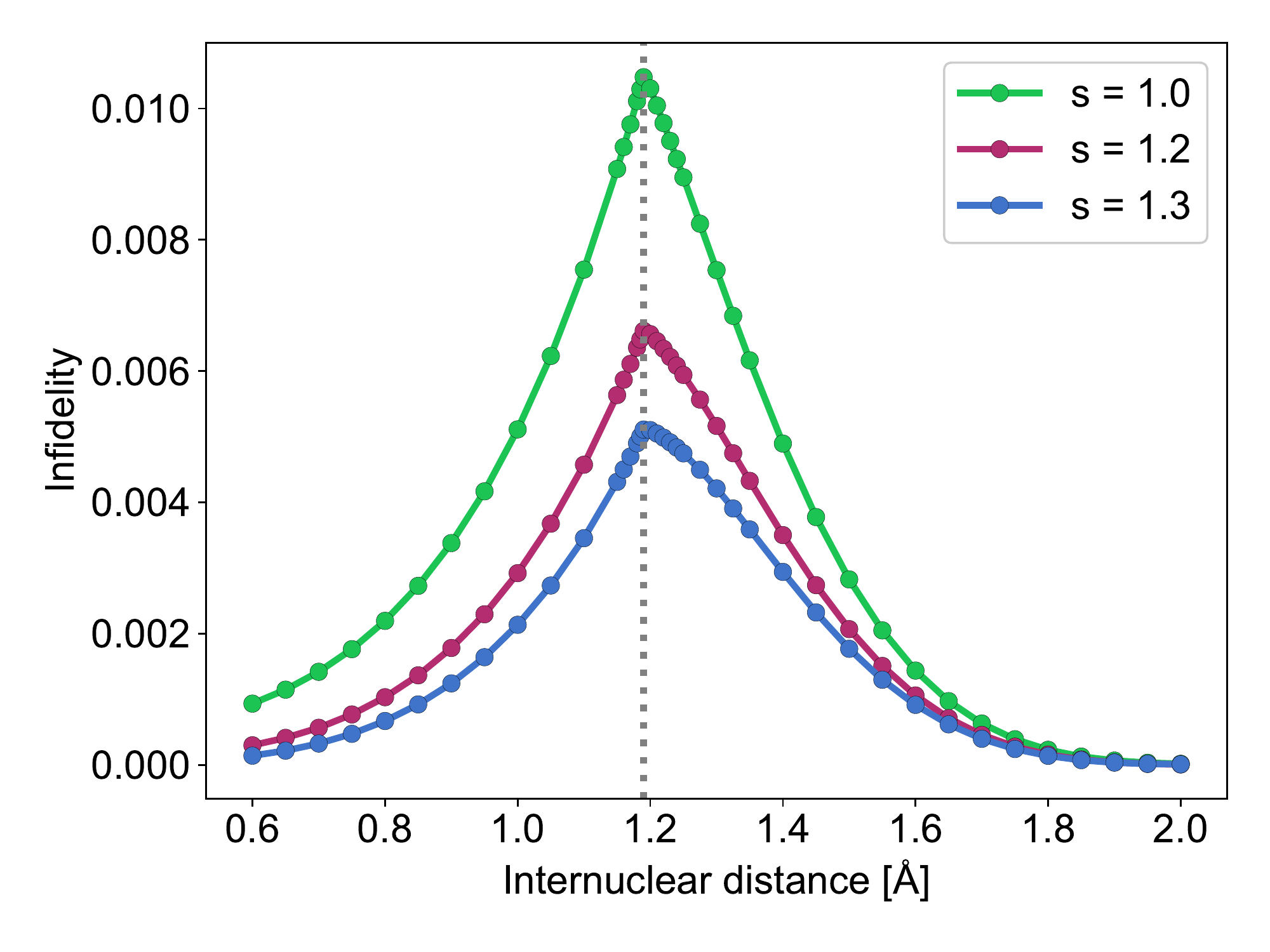}
       \subcaption{Infidelity of NOQE S$_0$ state. }
       \label{fig:h2-631g-S0scaleinfidelity}
    \end{minipage}
    \begin{minipage}{0.48\columnwidth}
       \includegraphics[width=1\linewidth]{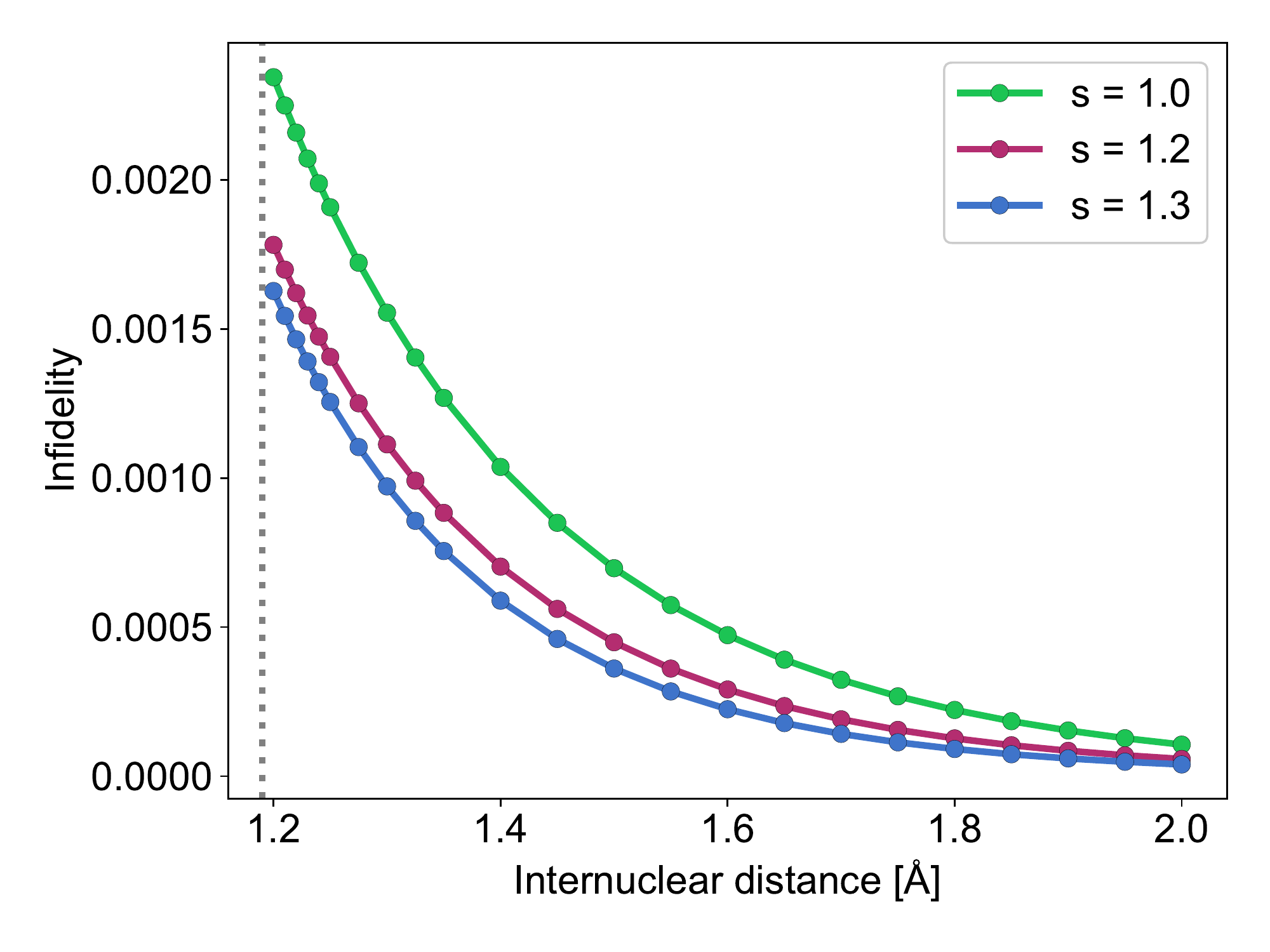}
       \subcaption{Infidelity of NOQE T$_1$ state.}
       \label{fig:h2-631g-T1scaleinfidelity} 
    \end{minipage}
    \caption{Infidelity $I_\text{FCI}$ relative to the FCI states of the amplitude scaled NOQE states derived from the NOUCC(2) ansatz, for H$_2$ with the 6-31G basis. See \eq{Infidelity_FCI}. The location of the CF point is marked with a dotted gray line.}
    \label{fig:h2-631g-infidelities}
\end{figure}

\end{document}